\documentclass[prd,preprintnumbers,twocolumn,eqsecnum,floatfix,a4paper,nofootinbib,superscriptaddress]{revtex4-1}

\usepackage{color}
\usepackage{amsmath,amssymb,graphicx,mathtools}
\usepackage{bm,lipsum}
\usepackage{times}
\usepackage{microtype}
\usepackage[utf8]{inputenc}
\usepackage{booktabs}
\usepackage[caption=false]{subfig}
\usepackage[normalem]{ulem}
\usepackage[varg]{txfonts}
\usepackage{bm,graphics,graphicx,epsfig,xcolor,ulem}
\usepackage[colorlinks, pdfborder={0 0 0}]{hyperref}
\usepackage{multirow}
\usepackage{diagbox}
\definecolor{LinkColor}{rgb}{0.75, 0, 0}
\definecolor{CiteColor}{rgb}{0, 0.5, 0.5}
\definecolor{UrlColor}{rgb}{0, 0, 0.75}
\hypersetup{linkcolor=LinkColor}
\hypersetup{citecolor=CiteColor}
\hypersetup{urlcolor=UrlColor}
\allowdisplaybreaks
\begin{document}
\title{Overtones, mirror modes, and mode-mixing in binary black hole mergers}

\newcommand{\penncosmos}{\affiliation{Institute for Gravitation and the Cosmos, Department of Physics, Pennsylvania State University, University Park, PA 16802, USA}}
\newcommand{\pennastro}{\affiliation{Department of Astronomy \& Astrophysics, Pennsylvania State University, University Park, PA 16802, USA}}
\newcommand{\cardiff}{\affiliation{School of Physics and Astronomy, Cardiff University, Cardiff, UK, CF24 3AA}
}

\author{Arnab Dhani}
\email{aud371@psu.edu}
\penncosmos 
\author{B. S. Sathyaprakash}
\penncosmos
\pennastro
\cardiff

\begin{abstract}
Gravitational waves emitted in the aftermath of a black hole binary coalescence have characteristic complex frequencies called quasinormal modes (QNMs). These can be used to test the nature of the merger remnant, e.g. a test of the black hole ''no-hair" theorem. The relative excitation amplitudes of the QNMs contain information about the progenitor system and can be employed to test the consistency between the pre- and post-merger signals. The post-merger signal from numerical relativity (NR) simulations of binary black holes exhibit a rich structure that has to be understood to fully exploit the information contained therein. In this study, we examine the importance of \emph{overtones} and \emph{mirror} modes for a set of mass ratios of non-spinning black hole binaries and a host of multipole modes corresponding to spherical harmonic indices $l\leq4$ and $m\leq3.$ 
We find that the first overtone is most important in the leading mode for a given $m$ and decreases for sub-dominant modes. The contribution of the fundamental mirror mode is most significant for $m=1$ and increases with mass ratio, although the model systematics are not affected significantly by mirror modes. Mode-mixing is the dominant effect for the sub-leading modes of a given $m$ and minimally affects the leading mode.
\end{abstract}

\maketitle

\section{Introduction}
Gravitational waves with characteristic complex frequencies, called quasinormal mode (QNM) frequencies, are a feature of any dynamical horizon that settles down to an isolated black hole~\cite{Vishveshwara:1970zz}. The QNMs, often referred to as \emph{ringdown} signals, for an isolated Kerr black hole are purely a function of its mass and spin and do not depend on the details of its dynamical past. This property is known as the ``no-hair" theorem~\cite{Misner:1974qy}. The validity of this theorem is an important test for general relativity (GR) and the nature of the final compact object formed in a black hole binary merger \cite{Dreyer:2003bv, Berti:2007zu, Gossan:2011ha}. Any deviation of the measured QNM spectra from its GR prediction would be a tell-tale signature of modified gravity effects or the existence of exotic compact objects~\cite{Cardoso:2019rvt,Carballo-Rubio:2018jzw,Raposo:2018xkf}. 

The detection of a single QNM mode together with an inspiral signal allows for an \textit{inspiral-merger-ringdown} consistency test of GR and the nature of the remnant. In this test, the remnant parameters, namely the mass and spin of the compact object, are independently estimated from the inspiral and the ringdown signals and checked for consistency~\cite{TheLIGOScientific:2016src,Ghosh:2017gfp}. Nevertheless, a direct test of the ``no-hair" theorem involves the detection of at least two ringdown modes~\cite{Dreyer:2003bv,Berti:2018vdi,Isi:2021iql}. Such experimental tests of the no-hair theorem have been carried out using GW150914~\cite{Carullo:2019flw,Isi:2019aib} and GW190521~\cite{Capano:2021etf}. \textcite{Isi:2019aib} report the detection of the first overtone $(l,m,n)=(221)$ in addition to the dominant $(l,m,n)=(220)$ to constrain the fractional deviation of the overtone frequency from its GR value to $-0.05\pm0.2$ while \textcite{Capano:2021etf} uses their secondary detection of the $(l,m,n)=(330)$ mode to constrain the frequency's fractional deviation to $-0.01^{+0.07}_{-0.11}$.

Planned upgrades of current detectors \cite{aPlusWP, AdV_KAGRA_plus, Adhikari:2019zpy} and addition of new ground- \cite{Punturo:2010zz, LIGOScientific:2016wof, Ackley:2020atn} and space-based detectors \cite{LISA:2017pwj} would observe multiple higher-order modes (both overtones and higher angular modes) in addition to the dominant $(l,m,n)=(2,2,0)$ spherical harmonic mode, allowing us to test the ``no-hair" theorem to exquisite precision.  It is, therefore, important to understand the various features of the different modes in a ringdown waveform beyond the dominant quadruple mode. 

There has been a multitude of studies in the literature that have modeled the higher spherical harmonic modes of a ringdown waveform \cite{Kamaretsos:2011um,Baibhav:2017jhs,London:2018gaq,Ota:2019bzl} using the fundamental QNM. Other studies have added overtones and mirror modes~\cite{Giesler:2019uxc,Ota:2019bzl,Bhagwat:2019dtm,Forteza:2020cve,Hughes:2019zmt,Lim:2019xrb,Dhani:2020nik,Forteza:2021wfq} to model the $l=2$ modes. In the effective one-body formalism, the ringdown is modeled using a set of QNMs and pseudo-modes\cite{Buonanno:2006ui,Pan:2013rra,Taracchini:2013rva,Babak:2016tgq}. 

QNMs are the eigenmodes of a perturbed Kerr black hole as the angular part of the perturbation equations separate out in terms of the \emph{spheroidal} harmonic functions \cite{Teukolsky:1973ha, Berti:2005gp}. This is similar in spirit to how the radial wavefunction of a hydrogen atom is given by the Laguerre polynomials as the angular part of the Schrodinger equation separates out in terms of spherical harmonic functions. On the contrary, numerical relativity (NR) simulations of a binary black hole merger are decomposed in the $-2$ spin-weighted \emph{spherical} harmonic basis. However, since the latter are not the angular eigenfunctions of the problem, the radial solution is not given by QNMs. By virtue of the fact that $-2$ spin-weighted spherical harmonics do form a complete set of angular functions, the $-2$ spin-weighted spheroidal harmonics can be decomposed in terms of the former and, hence, the radial solution is a linear combination of QNMs. This is called mode-mixing \cite{Berti:2014fga} because a given spherical harmonic mode is described by a set of QNMs with different angular indices. Specifically, spheroidal harmonic modes with different $l$ but same $m$ contribute to a spherical harmonic mode. 

\textcite{Kelly:2012nd} looked into the effect of mode-mixing in the $(3,2)$ mode of an unequal mass binary black hole merger. \textcite{Mehta:2019wxm} include mode-mixing to construct phenomenological waveform models of non-spinning binary black holes. In both of these studies, mode-mixing is considered only for the higher $l$ modes for a given $m$ which are expected to be affected the most in the case of non-precessing systems. When fitting a set of NR modes with the same $m$ to a given set of QNMs that would account for the effect of mode-mixing, the various NR modes should be fit simultaneously to avoid multiple amplitude estimates for the same QNM. To our knowledge, only~\textcite{London:2014cma} in their \textit{greedy-OLS} algorithm and~\textcite{Cook:2020otn} in their multimode modeling of $m=2$ and $l\leq4$ modes take this into account.

In this work, we model the ringdown waveform of all spherical harmonic modes up to $l=4$ and $m=3$ for non-spinning binaries of mass ratios $q \in \{1.25, 5, 10\}$. 
The salient results of our study are as follows: 
\begin{enumerate}
\item In the leading (or dominant) mode for a given $m$ 
the correction due to the first overtone is more important than that due to other modes but the effect of the overtone decreases for sub-dominant (i.e., $l>m$) modes\footnote{Note that except for $m=1,$ for which the leading mode is $l=2,$ for all other values of $m$ the leading mode is essentially $l=m.$ We do not consider the $m=0$ modes in this study.}.
\item The contribution of the fundamental mirror mode is most significant for the $m=1$ mode and this contribution increases for larger mass ratios becoming the same order of magnitude as mode-mixing for the sub-leading modes of the set $m=1$.
\item The model systematics (e.g., biases in parameter estimation) are not affected significantly by mirror modes. 
\item Most importantly, for a given $m$, mode-mixing is the dominant effect for the sub-leading modes and minimally affects the leading mode corresponding that $m$. 
\end{enumerate}

The outline of the paper is as follows. In Sec.~\ref{sec:rd_model}, we introduce our ringdown waveform models and the various approximations in those models. In Sec.~\ref{sec:fit}, we describe the fits of the various ringdown models to NR waveforms and delineate the features arising from those fits. Sec.~\ref{sec:sys} presents the systematics of each model, specifically how the parameters of the system will be affected due to inaccurate modeling. Finally, in Secs.~\ref{sec:disc} and~\ref{sec:fw} we conclude and lay out the directions for future work.

\section{Ringdown waveform model}
\label{sec:rd_model}
A dynamical horizon settles down to an isolated Kerr black hole of mass $M_f$ and dimensionless spin $a_f$ via the emission of gravitational waves, which for an asymptotic observer at a distance $r$ takes the form\footnote{We use the sign convention in SXS simulations.}
\begin{equation}
    \begin{split}
    \label{eq:PT_waveform0}
        h(t) = h_{+} - i h_{\times} =& \frac{M_f}{r}\sum_{lmn}\Big[ \mathcal{C}_{lmn} e^{-i \omega_{lmn}t} \prescript{}{-2}S_{lm}(\Omega,a_f\omega_{lmn}) \\
        &+ \mathcal{C}_{lmn}' e^{-i \omega_{lmn}'t} \prescript{}{-2}S_{lm}(\Omega,a_f\omega_{lmn}') \Big] \\
    \end{split}
\end{equation}
where $\prescript{}{-2}{}S_{lm}(\Omega,a_f\omega_{lmn})$ is the $-2$ spin-weighted spheroidal harmonic angular eigenfunction, $\Omega$ is the coordinate on a sphere $(\theta,\phi)$, and $t$ is the retarded time at null infinity. The QNMs $\omega_{lmn}$ and $\omega'_{lmn}$ are the positive- and negative-oscillation frequency solutions, respectively, of the radial Teukolsky equation. The negative oscillation frequency solutions are often called \emph{mirror} modes. The azimuthal symmetry of a Kerr spacetime imposes the following relation between the two sets of QNMs:
\begin{equation}
    \label{eq:QNMsym}
    \omega'_{lmn} = -\omega^{*}_{l-mn}.
\end{equation}
In Eq.\,(\ref{eq:PT_waveform0}), $\mathcal{C}_{lmn}=\mathcal{A}_{lmn}\, e^{i\phi_{lmn}}$ and $\mathcal{C}'_{lmn} = \mathcal{A}'_{lmn}\, e^{i\phi'_{lmn}}$ are the complex excitation amplitudes of the regular and mirror modes, respectively. They depend on the nature of the initial perturbation, or when the perturbation is produced by the merger of a pair of compact objects, the parameters and initial configuration of the progenitor binary. The excitation amplitudes cannot be analytically computed from  the initial condition of a black hole binary. We will, therefore, find fits for them using waveforms obtained from NR simulations. 

The symmetry relation in Eq.~(\ref{eq:QNMsym}) can be used to simplify the waveform  in Eq.~(\ref{eq:PT_waveform0}) to
\begin{equation}
    \begin{split}
    \label{eq:PT1}
        h(t) = h_{+} - i h_{\times} =& \frac{M_f}{r}\sum_{lmn}[ \mathcal{C}_{lmn} e^{-i \omega_{lmn}t} \prescript{}{-2}S_{lm}(\Omega,a_f\omega_{lmn}) \\
        &+ \mathcal{C}_{lmn}' e^{i \omega^{*}_{l-mn}t} \prescript{}{-2}S^{*}_{lm}(\Omega',-a_f\omega_{l-mn}^{*}) ] \\
    \end{split}
\end{equation}
where $\Omega'=(\pi-\theta,\phi)$.

The waveform, expressed in terms of spheroidal harmonic angular eigenfunctions, cannot be fit to an NR waveform directly since the latter are decomposed in terms of $-2$ spin-weighted spherical harmonics $\prescript{}{-2}{}Y_{lm}(\Omega)$ and can be written as
\begin{equation}
    h(t) = h_{+} - i h_{\times} = \frac{M_f}{r} \sum_{lm} h_{lm}(t) \prescript{}{-2}{}Y_{lm}(\Omega).
\end{equation}
We will, therefore, expand the spheroidal harmonic angular eigenfunctions in the orthonormal basis of spherical harmonics as 
\begin{equation}
    \label{eq:sph-sphero}
    \prescript{}{-2}S_{lm}(\Omega,a_f\omega_{lmn}) = \sum_{l'} \mu_{lm}^{l'}(a_f\omega_{lmn}) \prescript{}{-2}{}Y_{l'm}(\Omega),
\end{equation}
where $\mu_{lm}^{l'}(a_f\omega_{lmn})$ is called the spherical-spheroidal mixing coefficient. An NR spherical harmonic mode of the form
\begin{equation}
    \begin{split}
    \label{eq:rd_model}
    h_{l'm} &= \sum_{ln} \left [ \mathcal{C}_{lmn}\, \mu_{lm}^{l'}\, (a_f\omega_{lmn})\, e^{-i\omega_{lmn}t} \right . \\
    & \left . + (-1)^{l+l'} \mathcal{C}_{lmn}'\, \mu_{l-m}^{*l'}\, (a_f\omega_{l-mn})\, e^{i\omega_{l-mn}^{*}t}  \right ]
    \end{split}
\end{equation}
can then be arrived at by substituting Eq.~(\ref{eq:sph-sphero}) into Eq.~(\ref{eq:PT1}) and using the symmetry and orthogonality properties of $-2$ spin-weighted spherical harmonics\footnote{For some of the intermediate steps, see~\textcite{Dhani:2020nik}.}. Given the mass and spin of the final black hole, we can now fit for the complex excitation amplitudes to NR waveform. 

We have to now choose a set of QNMs that we want to fit to an NR mode $(l',m)$. In this study, we take that set to comprise of the fundamental mode $(l,m,0)$, the fundamental mirror mode $(l,-m,0)$, and the first overtone $(l,m,1)$. Furthermore, in order to determine the importance of each of these QNMs to the different spherical harmonic modes and physical systems, we define the following ringdown models: (i) Model `F' is a ringdown model that contains only the fundamental QNM, (ii) model `FM' contains both the fundamental mode and the fundamental mirror mode, (iii) model `FO' contains the first overtone together with the fundamental mode, and (iv) model `FMO' contains all the three QNMs. The various models are enumerated in Table.~\ref{tab:models}.

\begin{table}[ht]
    \centering
    \begin{ruledtabular}
        \begin{tabular}{l l}
            Model Name & Mode Content\\
            \hline
            F & Fundamental mode \\
            FM & Fundamental mode + mirror mode \\
            FO & Fundamental mode + overtone \\
            FMO & Fundamental mode + mirror mode + overtone \\
        \end{tabular}
    \end{ruledtabular}
    \caption{The different ringdown models defined in this study and their symbols. We build four different models with varied mode content. Starting from the simplest model with just the fundamental mode, FM, FO, and FMO represent progressively greater mode content.}
    \label{tab:models}
\end{table}

A spherical harmonic mode $(l',m)$ has contributions not only from the corresponding spheroidal harmonic quantum number $l=l'$ but from all $l$ for a given $m$, though the dominant contribution of mode-mixing is from $l=l'$ as can be seen from~\textcite{Berti:2014fga}. Consequently, we do a simultaneous fit of all the modes with the same $m$. For the modes considered in this paper ($\forall l\leq4, m\leq3$), this implies a simultaneous fit of the following three sets of NR modes: 
\begin{itemize}
    \item $(2,2),(3,2),(4,2)$
    \item $(2,1),(3,1),(4,1)$
    \item $(3,3),(4,3)$.
\end{itemize}
The inclusion of mode-mixing entails that, for the `FMO' model, the first two sets that contain three modes are fit to $3\times3=9$ QNMs and the last set containing two modes is fit to $3\times2=6$ QNMs. Additionally, we want to ascertain the impact of mode-mixing to the different modes under consideration. For that, we establish two additional classes of models, the first of which would include the effect of mode-mixing in the above defined models and the second class containing our four models would disregard mode-mixing in the ringdown waveform. The latter involves setting $\mu_{lmn}^{l'}=\delta_l^{l'}$ in Eq.~(\ref{eq:rd_model}). In this case, the general ringdown model simplifies to 
\begin{equation}
    \label{eq:rd_model_nomm}
    h_{l'm} = \sum_{n} \left[ \mathcal{C}_{l'mn}e^{-i\omega_{lmn}t} +  \mathcal{C}_{lmn}'e^{i\omega_{l-mn}^{*}t} \right].
\end{equation}

\section{Waveform fits}
\label{sec:fit}
We consider three non-spinning black hole binaries from the publicly available SXS\footnote{Simulating eXtreme Spacetimes\cite{SXS}} catalog of NR simulations corresponding to three different mass ratios. The SXS IDs, mass ratios, final masses, and dimensionless final spins of the simulations are give in Table~\ref{tab:sims}. We have chosen three simulations corresponding roughly to mass ratios 10:8, 5:1, and 10:1. For the sake of convenience we call them S10:8, S10:2, and S10:1, respectively. The chosen systems span the range of mass ratios of interest to ground-based detectors. We set the complex QNM frequencies to their GR values and fit the complex amplitudes in each of the ringdown models introduced in the previous section to the NR waveform via a linear least-squares method. To calculate the goodness of the fit, we define the mismatch between the fit and the NR waveform as
\begin{equation}
    \label{eq:mismatch}
    \mathcal{M} = 1 - \frac{\langle h_{\rm NR}|h_{\rm fit} \rangle}{\sqrt{\langle h_{\rm NR}|h_{\rm NR} \rangle \langle h_{\rm fit}|h_{\rm fit} \rangle}}
\end{equation}
where the inner product is defined as
\begin{equation}
    \langle a(t)|b(t) \rangle = \int_{t_0}^{t}dt' \oint d\Omega a^{*}(t',\Omega)b(t',\Omega). 
\end{equation}
Since we fit modes with the same $m$ simultaneously, the combined mismatch for a given set of modes takes the form
\begin{equation}
    \label{eq:mismatch_comb}
    \mathcal{M}_m = 1 - \frac{\sum_l \langle h^{lm}_{\rm NR}|h^{lm}_{\rm fit} \rangle}{\sqrt{\sum_l \langle h^{lm}_{\rm NR}|h^{lm}_{\rm NR} \rangle \sum_l \langle h^{lm}_{\rm fit}|h^{lm}_{\rm fit} \rangle}}.
\end{equation}
Additionally, to measure the effect of the individual modes, we define
\begin{equation}
    \label{eq:mismatch_indv}
    \mathcal{M}_{lm} = 1 - \frac{\langle h^{lm}_{\rm NR}|h^{lm}_{\rm fit} \rangle}{\sqrt{\langle h^{lm}_{\rm NR}|h^{lm}_{\rm NR} \rangle \langle h^{lm}_{\rm fit}|h^{lm}_{\rm fit} \rangle}}
\end{equation}
for the individual modes of the same fits.

We vary the start time of the ringdown waveform between $t_0=t_{\rm peak}$ and $t_0=t_{\rm peak}+50M,$ where the reference time $t_{\rm peak}$ is defined as the peak of the $h_{22}$ mode. The stop time is taken to be $t=t_{\rm peak}+90M$. 

\begin{table}
    \centering
    \begin{ruledtabular}
        \begin{tabular}{c c c c c}
            SXS ID & Proxy & $q$ & $M_f$ & $a_f$ \\
            \hline
            SXS:1143 & S10:8 & 1.250 & 0.9528 & 0.6795 \\
            SXS:0187 & S10:2 & 5.039 & 0.9825 & 0.4148 \\
            SXS:1107 & S10:1 & 10.00 & 0.9917 & 0.2605 \\
        \end{tabular}
    \end{ruledtabular}
    \caption{The SXS IDs, proxies, mass ratios, final masses, and dimensionless spins of the binary black hole simulations considered in this study from the publicly available catalog of SXS simulations.}
    \label{tab:sims}
\end{table}

In the rest of the section, we will discuss the different effects and QNMs that affect the mismatch curves for the ringdown signal of an NR waveform. Specifically, we will discuss the effects of mode-mixing, overtones, and mirror modes to the various sets of modes considered. The mismatch curves as a function of the ringdown start time $t_0$ for the four different models and the two classes of models for the set of modes with $m=2,1, 3,$ and the three simulations considered, are given in Fig.\,\ref{fig:mism_m2}, Fig.\ref{fig:mism_m1} and Fig.\,\ref{fig:mism_m3}, respectively. In Appendix.~\ref{sec:fits}, we plot the mode amplitudes $|h_{lm}|$ and mode frequencies $f_{lm}=-\Im(\Dot{h}_{lm}/h_{lm})$ as a function of time $t-t_{\rm peak}$. We additionally show our four best-fit ringdown models that include mode-mixing in Fig.~\ref{fig:amp} for the mode amplitudes and Figs.~\ref{fig:freq_m2}, Fig.\,\ref{fig:freq_m1}, and~Fig.\,\ref{fig:freq_m3} for the mode frequencies. The ringdown start time is taken to be $t_0-t_{\rm peak}=20M$ for these plots. 

\begin{figure*}
    \centering
        \includegraphics[width=\textwidth]{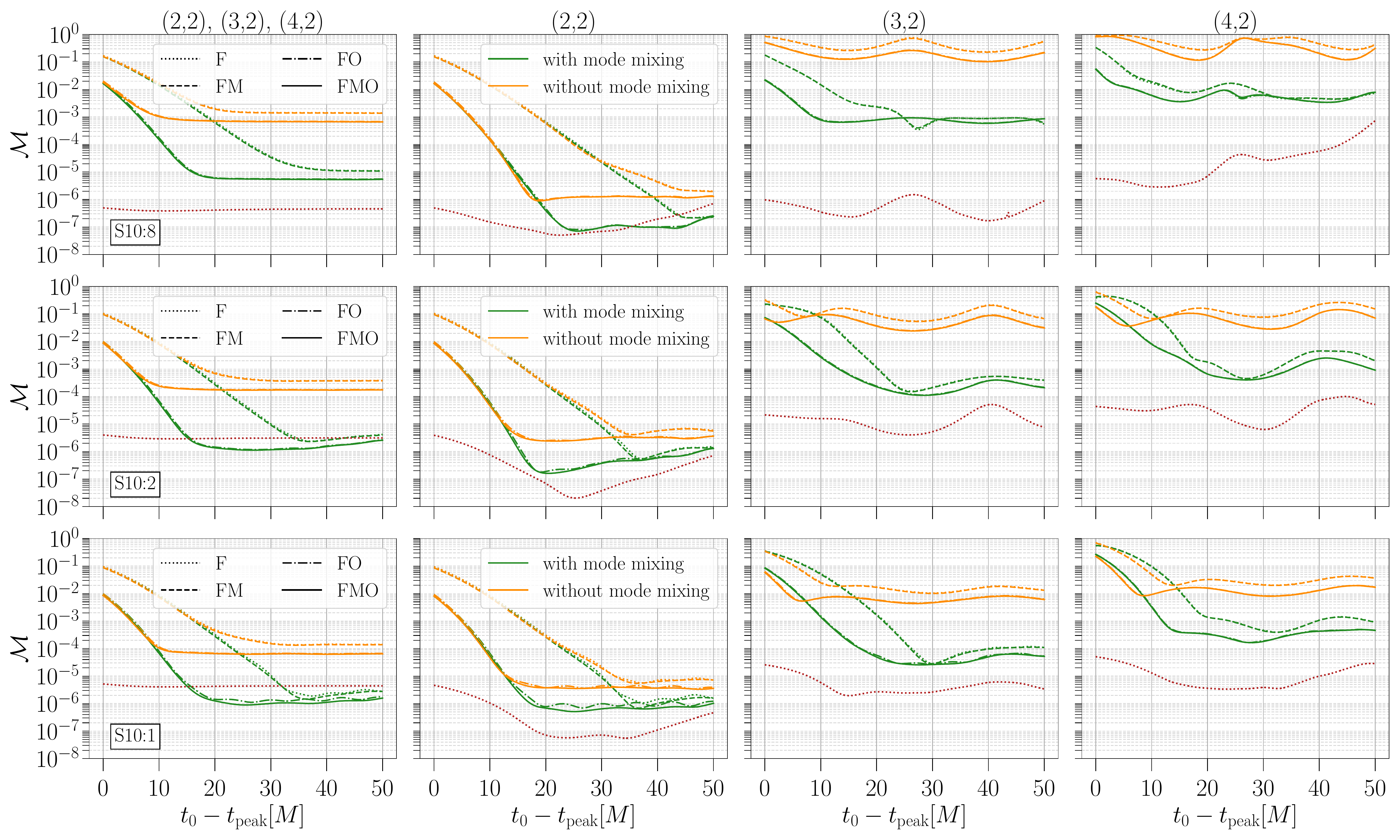}
    \caption{Comparison of the mismatches $\mathcal{M}$ corresponding to the $m=2$ mode, for the four models enumerated in Table~\ref{tab:models} and three mass ratios $q=1.25$ (S10:8, top row), $q=5$ (S10:2, middle row) and $q=10$ (S10:1, bottom row),  considered in this study as a function of the ringdown start time $t_0$. The reference time for $t_0$ is taken to be the peak of the $h_{22}$ mode. The green curves denote the class of models that include mode-mixing while mode-mixing is ignored for the orange curves. The left most panel in each row shows the mismatch $\mathcal{M}_m$ for the combined set of three modes with the same $m$ given in Eq.~(\ref{eq:mismatch_comb}), the other panels of the row show the mismatches $\mathcal{M}_{lm}$ given in Eq.~(\ref{eq:mismatch_indv}) and depict the contribution of the individual modes to the total mismatch. In all cases the fits were obtained simultaneously for all the modes.  The red, dotted line shows the mismatch between the two highest resolution NR simulations and is a proxy for the NR noise floor. We note that the mirror modes do not improve the mismatch and, hence, the curves for `F' and `FM' models essentially overlay on top of each other as do `FO' and `FMO' models. On the contrary, the overtone model essentially captures most of the features in the NR waveform and closely follow the `FMO' model. See text for other details.}
    \label{fig:mism_m2}
\end{figure*}

\subsection{Mode-mixing}
We describe the mixing of various spheroidal to spherical modes described above in this subsection. 

The effect of mode-mixing in a given spherical harmonic mode is determined by two independent factors: (a) the relative excitation amplitudes of the QNMs with the same $m$ and (b) the magnitude of the mixing-coefficients between them. For the physical systems considered (non-spinning binaries), the mode with a smaller $l$ for a given $m$ is more strongly excited and, therefore, the corresponding $l'$ is affected less by mode-mixing from a larger $l$. This is accentuated by the fact that the mixing coefficient is close to unity for $l=l'$ and at most a few percent otherwise~\cite{Berti:2014fga}. Let's consider the set $\{(2,2),\, (3,2)\}$ as an example and for simplicity of argument consider the ringdown model `F'. Our fits show that $\mathcal{C}_{220}$ is more strongly excited than $\mathcal{C}_{320}.$ This can be inferred from the fact that for the $(3,2)$ mode (the bottom left sub-figure for each mass ratio), the green curves (which include mode-mixing) show considerably lower mismatches over orange curves (which don't). On the contrary, for the $(2,2)$ mode orange curves show similar mismatches to green curves. Both of these excitation amplitudes contribute to the spherical harmonic modes $(2,2)$ and $(3,2)$ with their relative contribution governed by their mixing-coefficients as can be seen from Eq.~(\ref{eq:rd_model}). 

The coefficients with $l'=l$, $\mu^2_{220}$ and $\mu^3_{320}$, are close to unity while the ones with $l'\neq l$, $\mu^2_{320}$ and $\mu^3_{220}$, are a few percent. Consequently, the $(2,2)$ mode doesn't have a considerable effect due to mode mixing but the effects of the same on $(3,2)$ can be easily seen in the mismatch curves in Fig.~\ref{fig:mism_m2}, the mode amplitudes $|h_{lm}|$ in Fig.~\ref{fig:amp}, and the mode frequencies in Fig.~\ref{fig:freq_m2}. Similar arguments follow for the mode $(4,2),$ which is affected by both of the stronger modes, as can be seen in the two distinct modulations of the mode amplitudes and mode frequencies in Fig.~\ref{fig:amp} and Fig.~\ref{fig:freq_m2}, respectively, for mass ratio $q=1.25$. For the other mass ratios, the subdominant modulation due to $(3,2)$ is not visible. On the other hand $(3,2)$ is modulated only by the frequency of the sole stronger mode. In the absence of mode-mixing, the $(3,2)$ and $(4,2)$ modes would be described by a single damped sinusoid for the `F' model and these are shown as a straight line whose slope corresponds to the damping time and a horizontal line corresponding to the oscillation frequency in the mode amplitude and mode frequency plots, respectively.

\begin{figure*}
    \centering
        \includegraphics[width=\textwidth]{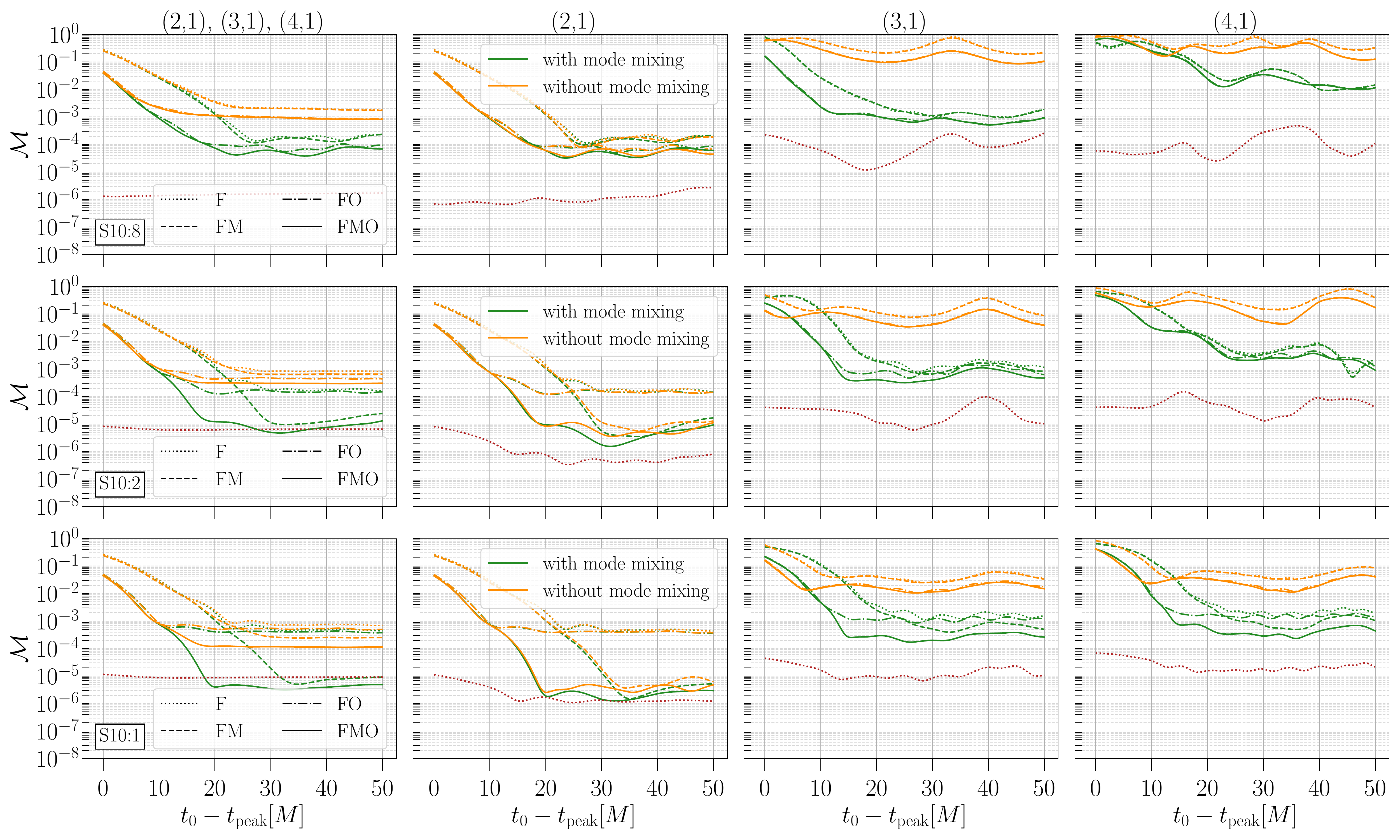}
    \caption{Same as Fig.\,\ref{fig:mism_m2} but for $m=1$ modes. The left most panels show that a mere overtone (FO) or a mirror model (FM) is not enough to obtain the best fit model. The second column of panels show that mode-mixing is unimportant for in the leading mode (2,1), while the third and fourth columns show that sub-leading modes (3,1) and (4,1) are significantly affected by mode-mixing.}
    \label{fig:mism_m1}
\end{figure*}

The effect of mode-mixing is similar for all the sets of modes considered in this study. The modes with the smallest $l$ for a given $m$ are least affected by mode-mixing. The mode amplitudes for $m=1$ and $m=3$ are given in~\ref{fig:amp} and the mode frequencies for the same modes are given in Figs.~\ref{fig:freq_m1} and~\ref{fig:freq_m3}, respectively.

We additionally observe from the mode amplitude and mode frequency plots that the mass ratio least affected by mode-mixing is the largest mass ratio we consider, $q=10$. As one goes to larger mass ratios, the final black hole tends towards a Schwarzschild black hole, because we started with an initially non-spinning binary, and, hence, the angular eigenfunctions tend towards $-2$ spin-weighted spherical harmonics making the NR modes the eigenmodes of the system. Note that we stave off making a similar statement comparing the smaller two mass ratios because even though the mixing coefficients are monotonically decreasing between the final spins of the $q=1.25$ and $q=5$ binary, the relative excitation amplitudes are not and, therefore, the contribution of mode-mixing is maximum for some mass ratio between them.

\begin{figure*}
        \includegraphics[width=0.75\textwidth]{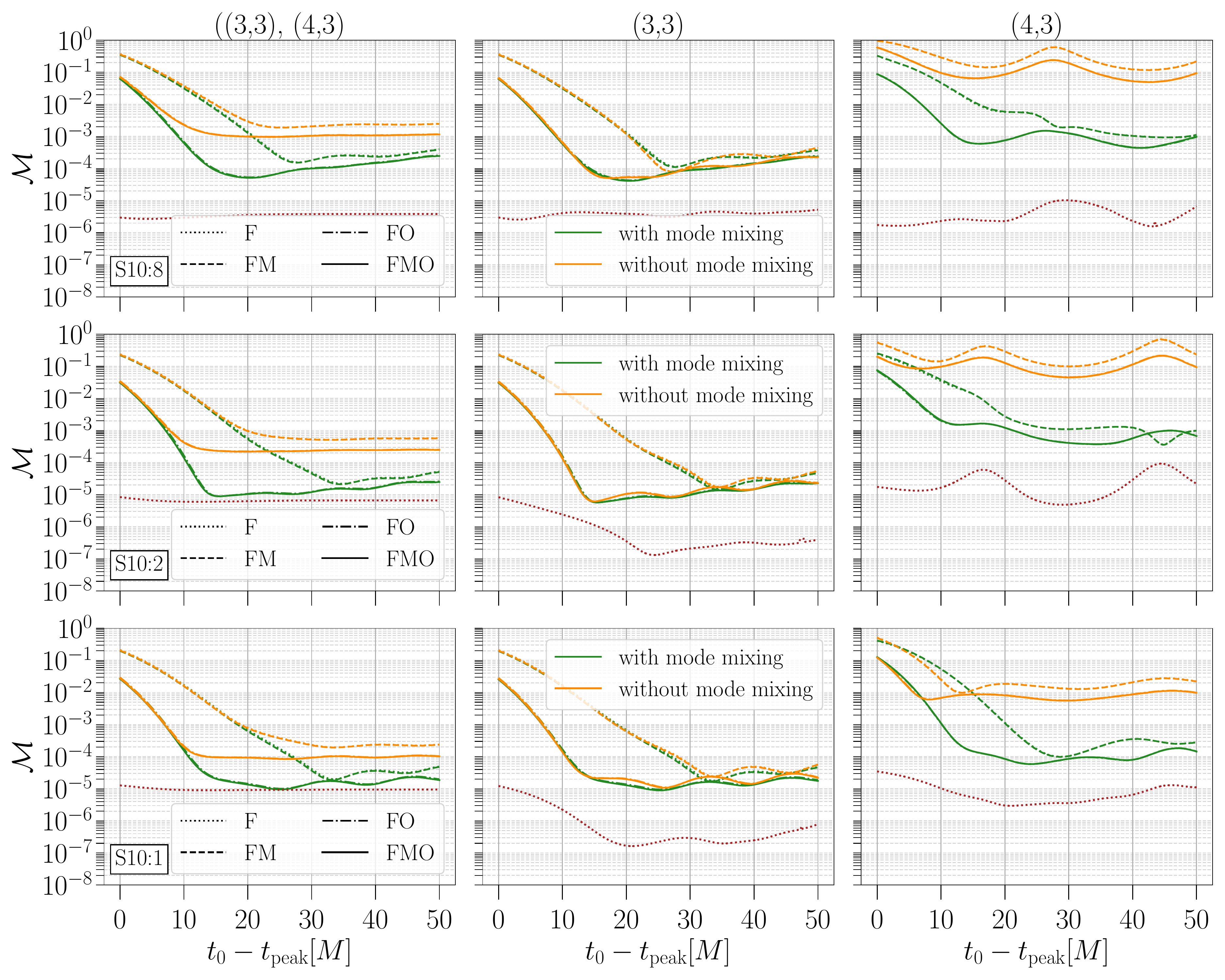}
    \caption{Same as Fig.~\ref{fig:mism_m3} but for $m=3$. Since $l \ge m,$ there are only two columns showing ${\cal M}_{lm},$ in addition to the left most column showing ${\cal M}_m,$ for each mass ratio, in contrast to three columns for $m=2,1$ in Figs.\,\ref{fig:mism_m2} and \ref{fig:mism_m1}. As in the case of $m=2,$ here too we see that the contribution of the mirror modes is sub-dominant compared to the overtones. This can be seen from the fact that the `FO' (dot-dashed curves) and `FMO' (solid curves) models overlay on top of each other as do `F' (dotted curves) and `FM' models (dashed curves). The middle and right columns once again shows that mode-mixing is less of an issue for the leading (3,3) mode but is significant for the sub-leading (4,3) mode, respectively.}
    \label{fig:mism_m3}
\end{figure*}

\subsection{Overtones}
We next turn to the effect of overtones (dot-dashed curves) and look at the $m=2$ modes first. Some features of the mismatch curves are qualitatively similar as a function of the mass ratio. We observe that adding the first overtone improves the match of the fit waveform with the NR waveform for early times. Indeed, the dot-dashed curves closely follow the solid curves corresponding to the FMO model for all mass ratios for $m=2.$ Our results are consistent with previous studies which either only considered the $(l,m)=(2,2)$ mode \cite{Giesler:2019uxc,Bhagwat:2019dtm} or included higher modes but only showed the combined mismatch $\mathcal{M}_m$ \cite{Cook:2020otn}. Although the improvement is most prominent for the $(2,2)$ mode, we report similar conclusions for the individual modes $(3,2)$ and $(4,2)$ of this set for the mode-mixing class of models. We note, however, that the level of improvement with the addition of the first overtone decreases for higher angular modes. 

If mode-mixing is ignored, we see that adding the first overtone doesn't improve the match much for the $(3,2)$ and $(4,2)$ modes, presumably because mode-mixing is the dominant effect over overtones even at early times. We further note that the higher modes of smaller mass ratio systems perform poorer than larger mass ratios. This is most stark in the $m=1$ modes where the first overtone of the $(4,1)$ mode barely improves the mismatch especially for $q=1.25$. The match of the NR waveform with the leading $(2,1)$ mode of this set sees a huge improvement with the addition of the first overtone dominating the combined mismatch curves. The $(4,3)$ mode responds much better to the first overtone with significant improvement over excluding the QNM but still the leading mode of the set $(3,3)$ governs the combined mismatch curves. 

The performance of the overtone for the higher angular modes of a set lead us to caution against their use for spectroscopic tests.

\subsection{Mirror modes}
For the modes of the sets $m=2$ and $m=3$, we see that adding the fundamental mirror mode does not improve the mismatch across mass ratios $q=1$ to $q=10.$ This is seen in Figs.\,\ref{fig:mism_m2} and \ref{fig:mism_m3} where the mismatch curves for the `F' and `FM' models and that of `FO' and `FMO' models, respectively, essentially coincide with each other and is again consistent with the results of~\textcite{Forteza:2020cve} and~\textcite{Dhani:2020nik} but where too only the $(2,2)$ mode was considered and the fundamental mirror mode didn't improve the mismatch significantly. 

For the $m=2$ set of modes the effect of adding mirror modes to a ringdown model can be clearly seen from Fig.~\ref{fig:freq_m2} where we plot the mode frequencies as a function of time $t-t_{\rm peak}$. In the left panels, the plot for the `F' model is represented by a horizontal line corresponding to the oscillation frequency of the fundamental QNM (this is true for the $(2,2)$ mode because mode-mixing is negligible in this case) while the `FM' model shows characteristic modulations on top of the fundamental mode frequency corresponding to the mirror mode oscillation frequency. For the $(2,2)$ mode, across mass ratios, the mirror modes are not visible to the eye in the numerical waveform and the corresponding features in the best-fit model are a result of over-fitting. Particularly, for the $(2,2)$ mode, notice that the mirror mode modulations vanish for the `FMO' model though they are visible for the `FM' model. We believe this is because the fitting procedure tries to fit some of the overtone features at early times using the mirror mode QNM. In fact, in the case of $m=2$ the high-frequency mirror mode modulations first become visible for mass ratio $q=10$ and the $(4,2)$ mode, less so for the $(3,2)$ mode. However, these features are not picked up in the mismatch plot in Fig.~\ref{fig:mism_m2}, which shows the same mismatch for the models `FO' and `FMO' for the said mode(s).

For the $m=3$ modes, even though mirror modes do not seem to play an important role, we cannot quantify this due to the presence of spurious oscillations in the $(3,3)$ mode. These are easily visible in the mode frequency plot, Fig.~\ref{fig:freq_m3}. The `FM' model shows a beat pattern which is dissimilar to the typical mirror mode modulations that we have seen. We are not able to assign any physical reasoning to these oscillations. These features are though not present in the mismatch curves for this set.

We find that mirror modes are most strongly excited for $m=1$ out of the three sets we consider in this study. Furthermore, the excitation amplitudes of the mirror modes increase with mass ratio $q$ as can be readily seen from the mode frequency plot Fig.~\ref{fig:freq_m1}, where the high frequency modulations on top of the fundamental QNM is due to the mirror modes while the low frequency ones are due to mode mixing. We further see that the excitation amplitudes weighed by the mixing coefficients, and therefore, the effective contribution of a QNM to an NR mode, are the same order of magnitude for the mirror mode and the mode-mixing mode for the $(3,1)$ and $(4,1)$ modes of the mass ratio $q=10$ case.

\section{Final mass and spin: ringdown model systematics}
\label{sec:sys}
Until now, we have fixed the final mass and spin of the black hole and fit for the complex excitation amplitudes. In this section, we look at the systematic uncertainties in each of our models arising from their estimation of the remnant parameters. For this, we will vary the remnant mass an spin on a grid and calculate the mismatch across the grid. We will then associate the mass-spin pair that minimizes the mismatch on this grid as the estimated remnant parameters for our model. We will then estimate the systematic uncertainty of a model by the deviation of the best-fit remnant parameters from the values determined by NR. We quantify this deviation as the Euclidean distance in the remnant parameter space given by
\begin{equation}
\label{eq:eps}
    \epsilon = \sqrt{\left(\frac{\delta M_f}{M}\right)^2 + \left(\delta a_f\right)^2}
\end{equation}
where $\delta M_f$ and $\delta a_f$ are the differences between the best-fit values and the NR values. Such a measure will help us in distinguishing models with the same or a similar mismatch. For example, as we discussed in the previous section, the `FM' model for $m=2$ gives rise to spurious mirror mode modulations that cannot be seen in the NR waveform. Even so, it produces the same mismatch as the `F' model. These spurious modulations have no physical reason to produce the minimum mismatch for the true remnant values and as such we expect the systematic uncertainties for such models to be larger.

In the following, we will depict the mismatches on a grid of remnant parameters for a ringdown start time of $t_0-t_{\rm peak}=20M$. We choose this time because it is approximately the latest time when the mismatch curves bottom out for the models with the best mismatches. We will later look at the effect of the start time on the model systematics by calculating $\epsilon$ for different choices of start times. 

\subsection{Modes: (2,2), (3,2), (4,2)}
\begin{figure*}
    \centering
    \subfloat{
        \includegraphics[width=2\columnwidth, trim={0 2.5cm 0 0}, clip]{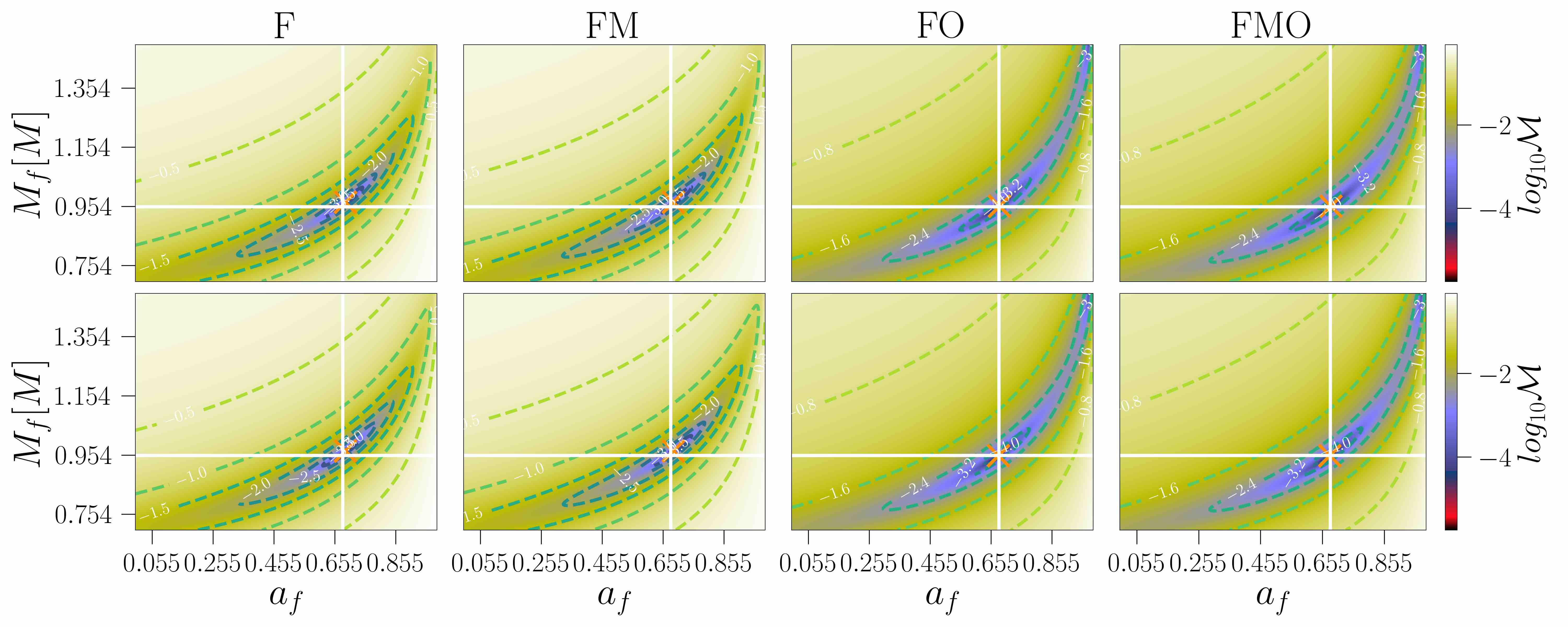}
    }
    
    \subfloat{
        \includegraphics[width=2\columnwidth, trim={0 2.5cm 0 1.3cm}, clip]{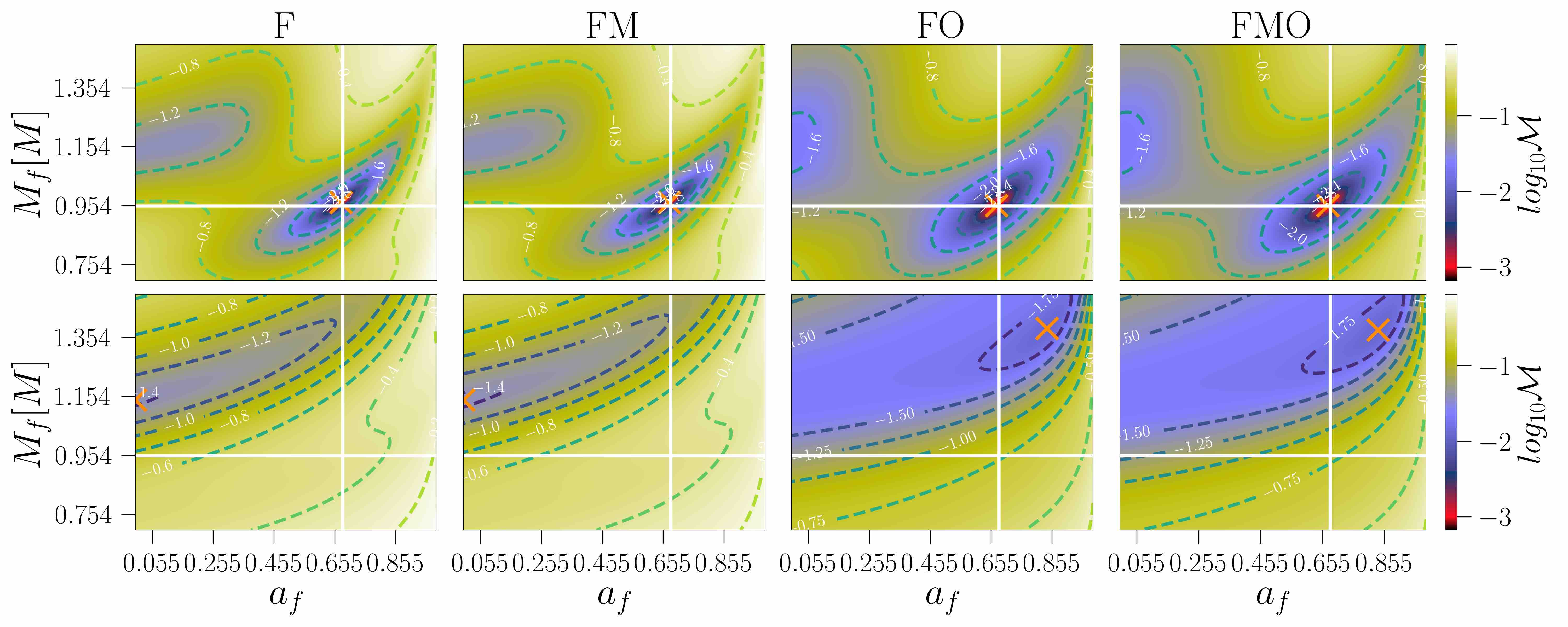}
    }
    
    \subfloat{
        \includegraphics[width=2\columnwidth, trim={0 0 0 1.3cm}, clip]{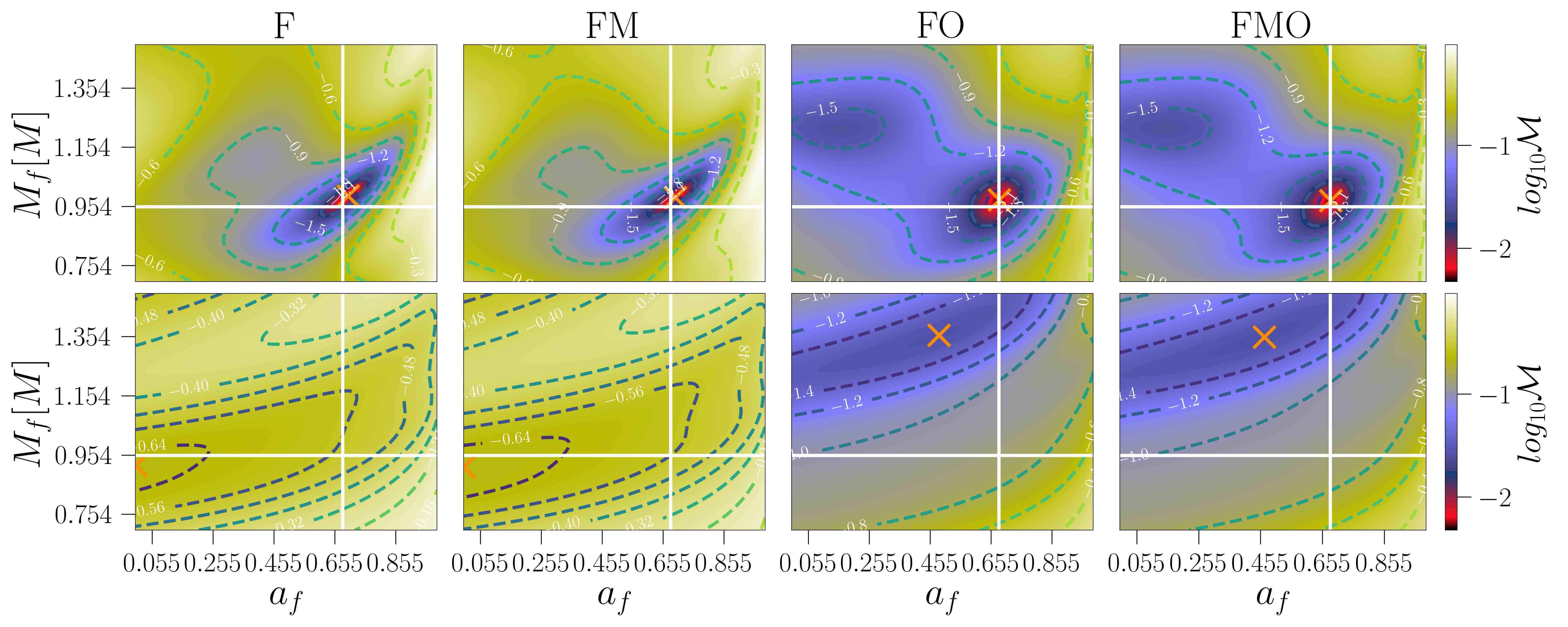}
    }
    \caption{Mismatch $\mathcal{M}_{lm}$ for the best-fit ringdown model on a grid of $M_f$--$a_f$ for a ringdown start time of $t_0-t_{\rm peak}=20M$. The white cross hairs show the value of the remnant determined by NR and the orange crosses show the best-fit value. The different sub-figures correspond to the angular modes of the set $m=2$, namely, $l=2$, $l=3$, and $l=4$, respectively. The columns plot the mismatches on the remnant parameter space for the four different models considered in this study (see Table.~\ref{tab:models}). The top and bottom rows of each sub-figure plot the mode-mixing and the non-mode-mixing classes of models, respectively. The scale for the color bars are the same for each sub-figure. These set of plots correspond to the simulation S10:8.}
    \label{fig:grid_m2_q1}
\end{figure*}

\begin{figure*}
    \centering
    \subfloat{
        \includegraphics[width=2\columnwidth, trim={0 2.5cm 0 0}, clip]{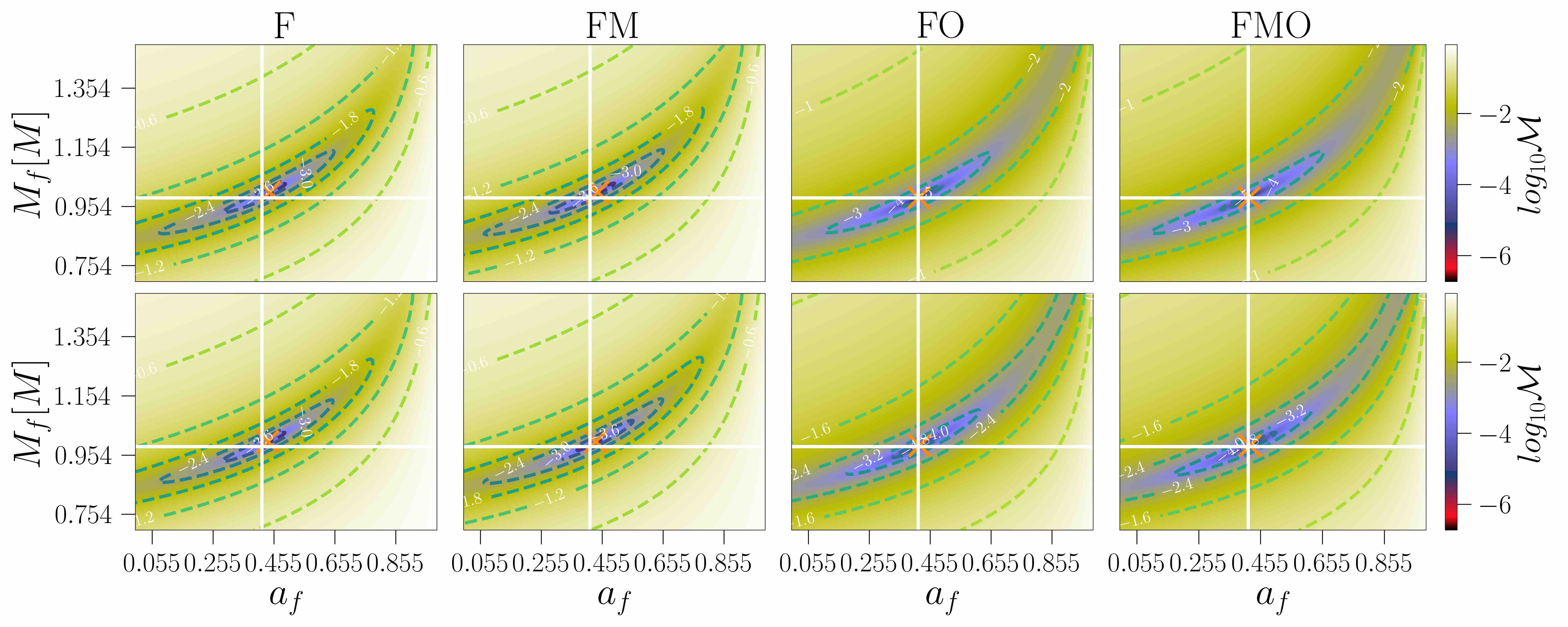}
    }
    
    \subfloat{
        \includegraphics[width=2\columnwidth, trim={0 2.5cm 0 1.3cm}, clip]{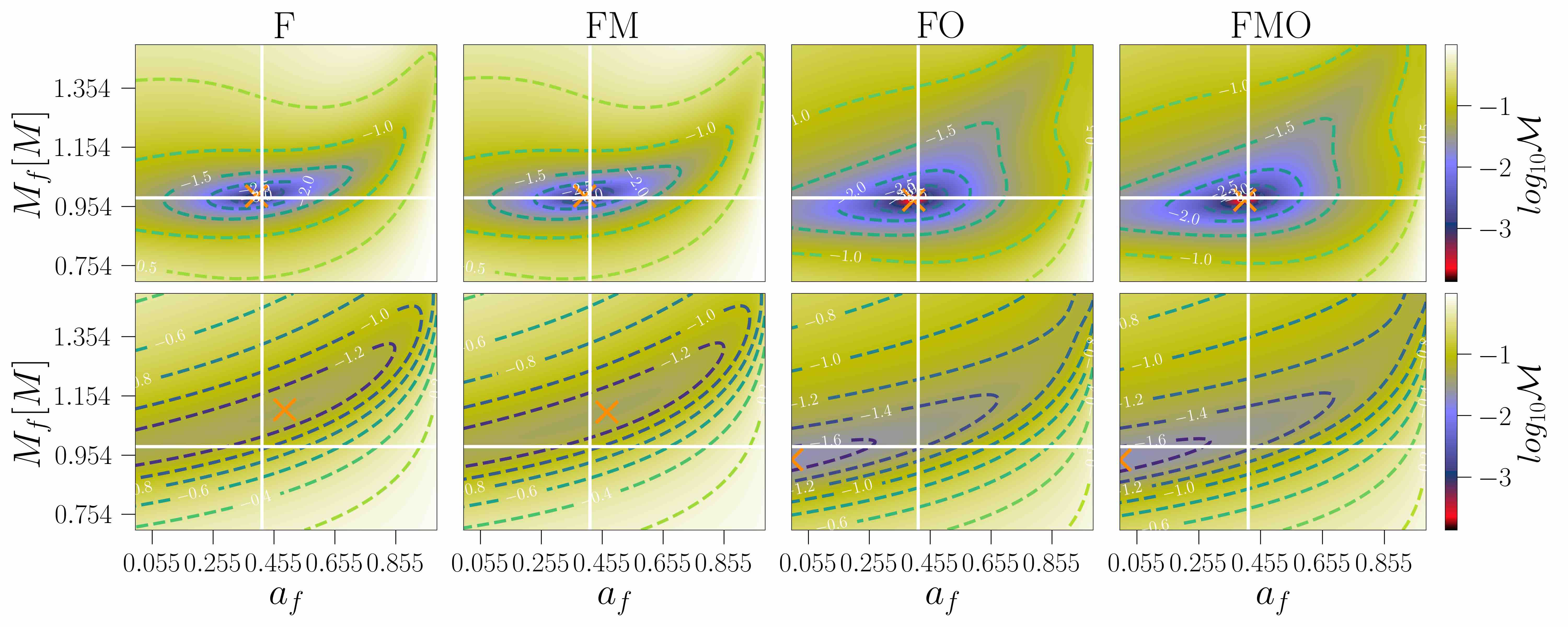}
    }
    
    \subfloat{
        \includegraphics[width=2\columnwidth, trim={0 0 0 1.3cm}, clip]{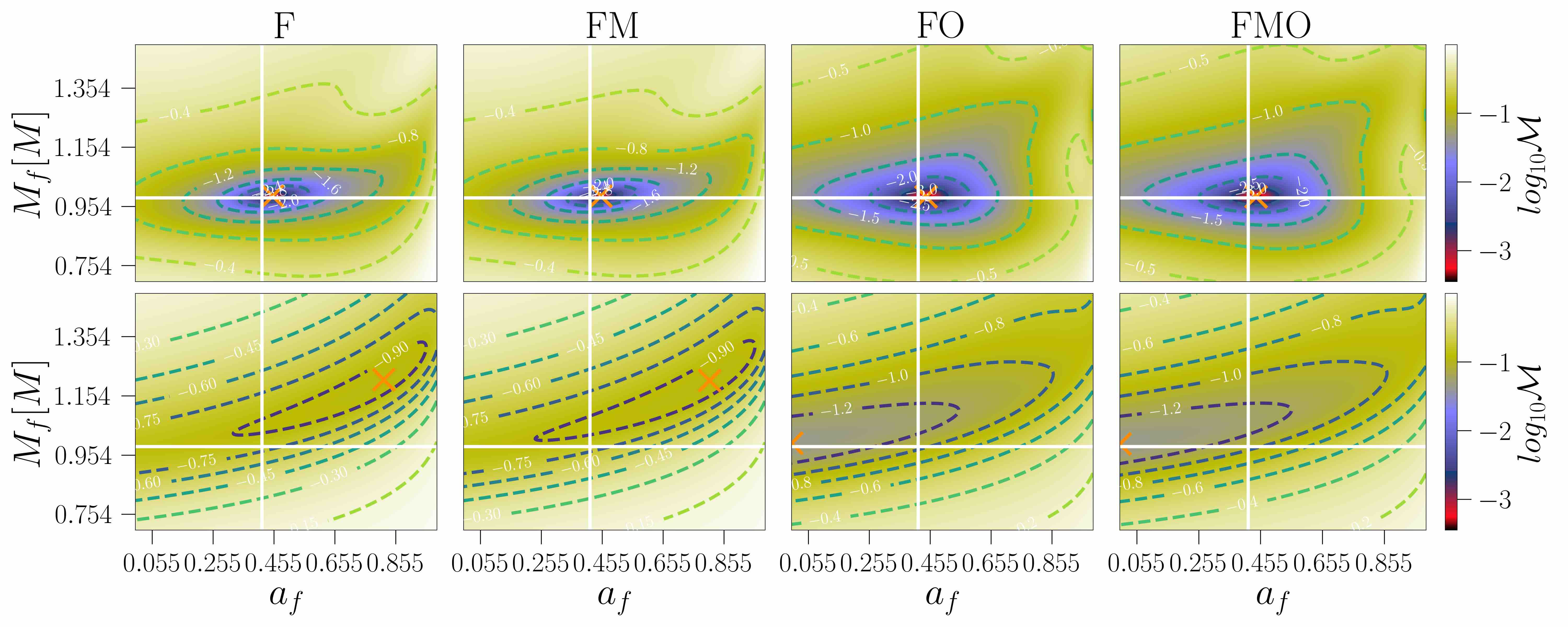}
    }
    \caption{Same as Fig.~\ref{fig:grid_m2_q1} but for the simulation S10:2.}
    \label{fig:grid_m2_q5}
\end{figure*}

\begin{figure*}
    \centering
    \subfloat{
        \includegraphics[width=2\columnwidth, trim={0 2.5cm 0 0}, clip]{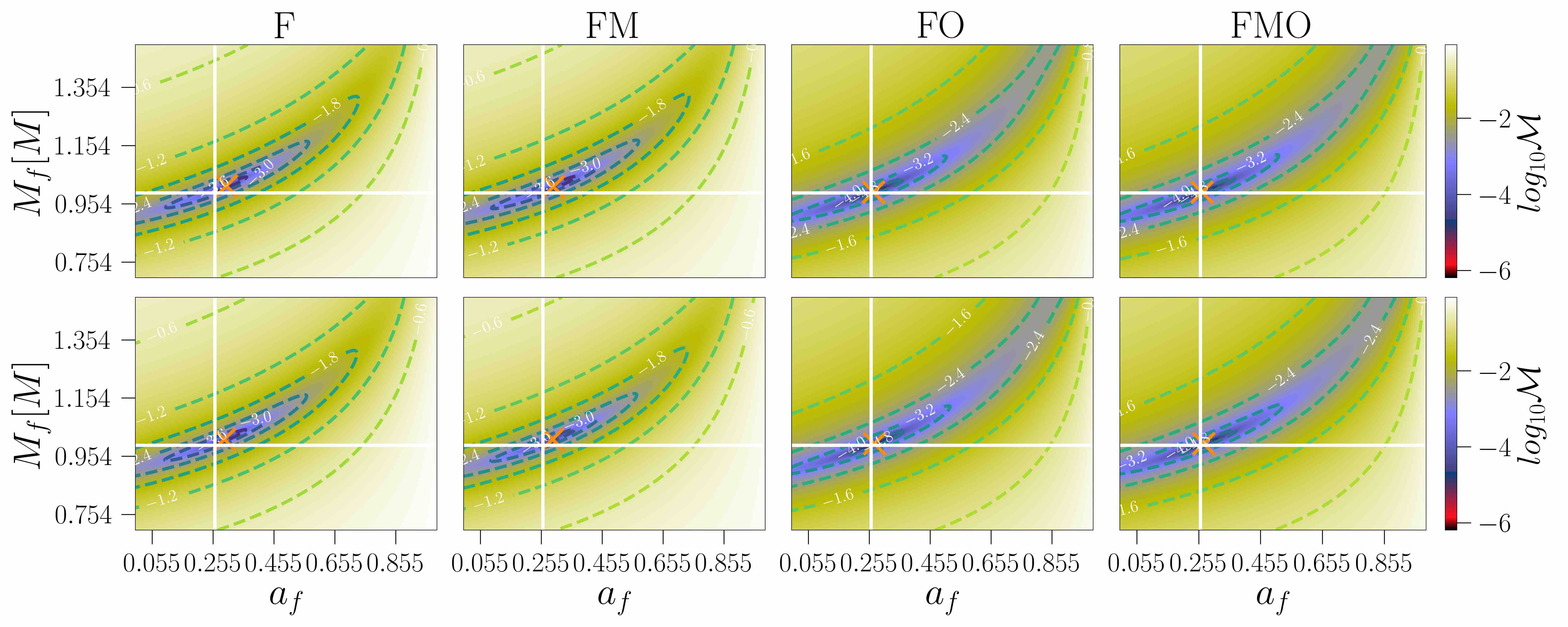}
    }
    
    \subfloat{
        \includegraphics[width=2\columnwidth, trim={0 2.5cm 0 1.3cm}, clip]{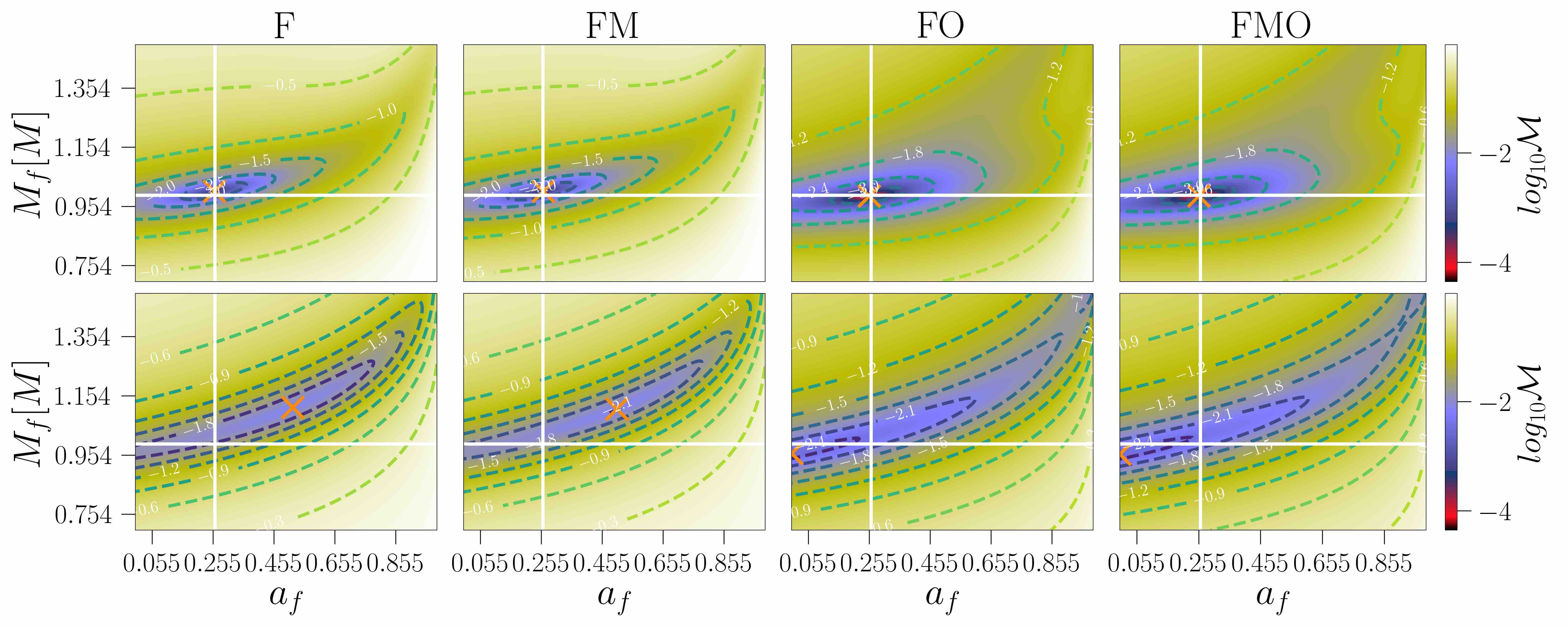}
    }
    
    \subfloat{
        \includegraphics[width=2\columnwidth, trim={0 0 0 1.3cm}, clip]{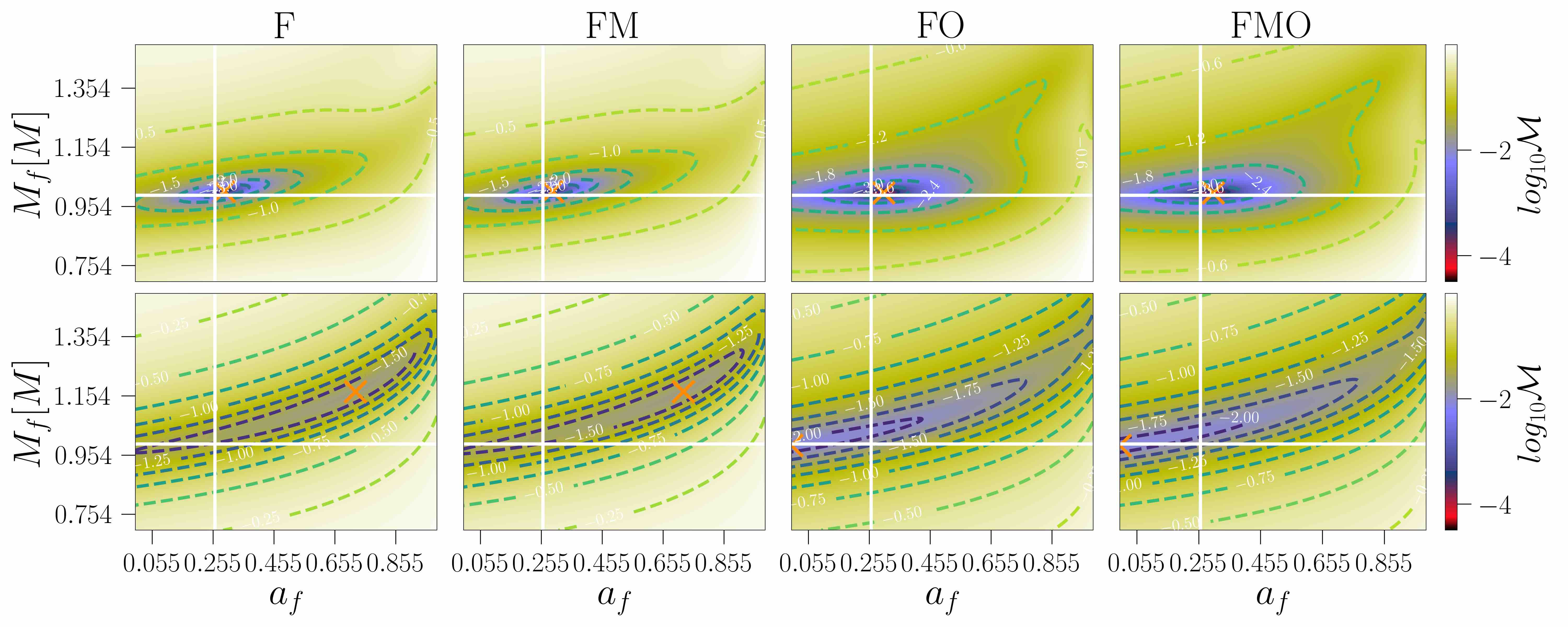}
    }
    \caption{Same as Fig.~\ref{fig:grid_m2_q1} but for the simulation S10:1.}
    \label{fig:grid_m2_q10}
\end{figure*}

We will look at the $m=2$ modes first shown in Figs.~\ref{fig:grid_m2_q1}, \ref{fig:grid_m2_q5}, and \ref{fig:grid_m2_q10}, corresponding to mass ratios $q=1.25$, $q=5$, $q=10$, respectively. Each pair of rows from the top depict the angular modes $l=2$, $l=3$, and $l=4$, respectively; the top panels of each pair show the class of models that include the effect of mode mixing while the bottom panels take no mode-mixing into account. 

For all mass ratios, the most important effect that stands out is mode-mixing. In line with our earlier observation that mode-mixing does not improve the mismatch for the $l=2$ mode of this set, we find that it does not notably affect the remnant parameter estimates either across all mass ratios. This is not the case for the higher $l$ modes of this set, which are significantly affected by mode mixing. We additionally note that the class of models with mode-mixing produces a deeper mismatch minima, which would assist in more accurate parameter estimation.

The impact of the first overtone to model the systematics is detailed in Fig.~\ref{fig:eps_m2_m1}, where one can see that the best-fit remnant parameters for the `FO' model is closer to the true values determined by the NR waveform than both the `F' and `FM' models for $l=2,$ which shows the greatest improvement in mismatch. The case for higher $l$ values is more nuanced, with clear improvement in the remnant parameter estimate for the smallest mass ratio $q=1.25$ and for start times before $t_0-t_{\rm peak}=20M$. For the two larger mass ratios, the first overtone does not affect $\epsilon$ much. The `FMO' model closely follows the `FO' model reinforcing that the fundamental mirror mode does not play an important role for $m=2$. 

\subsection{Modes: (2,1), (3,1), (4,1)}
\begin{figure*}[h]
    \centering
    \subfloat{
        \includegraphics[width=2\columnwidth, trim={0 2.5cm 0 0}, clip]{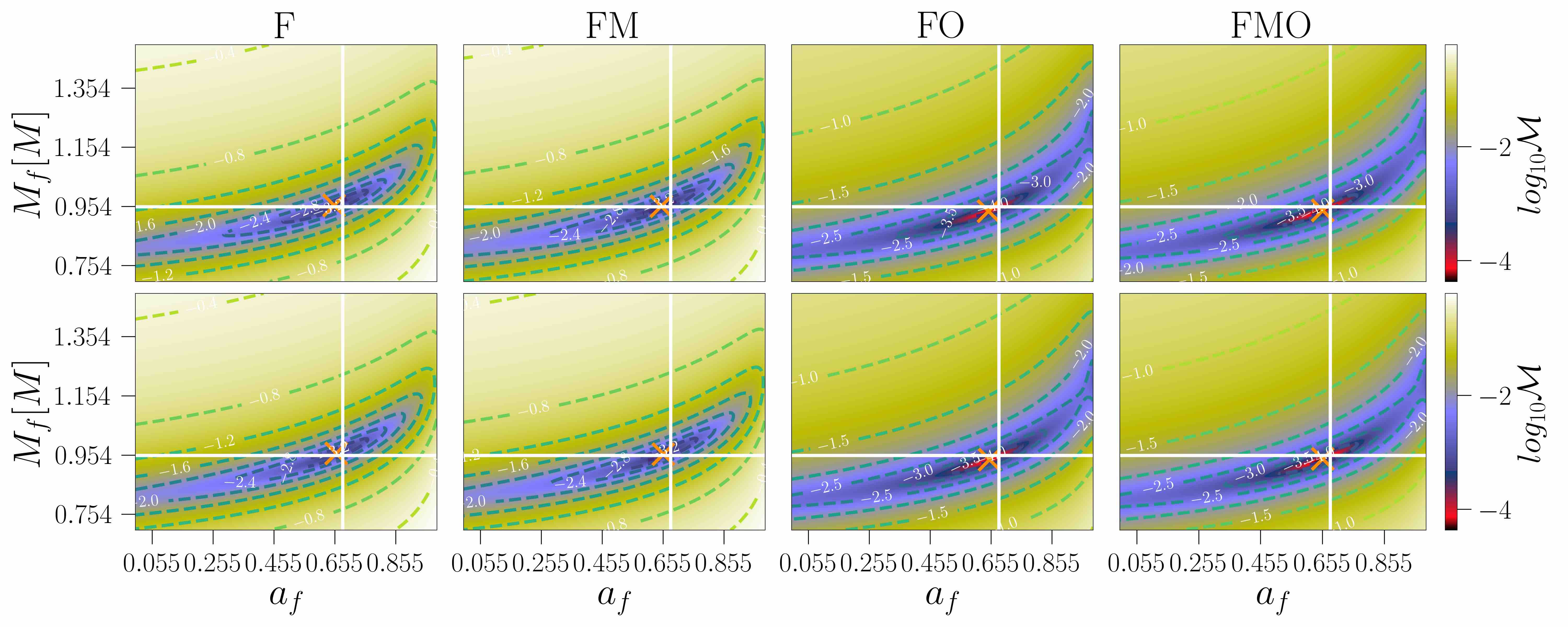}
    }
    
    \subfloat{
        \includegraphics[width=2\columnwidth, trim={0 2.5cm 0 1.3cm}, clip]{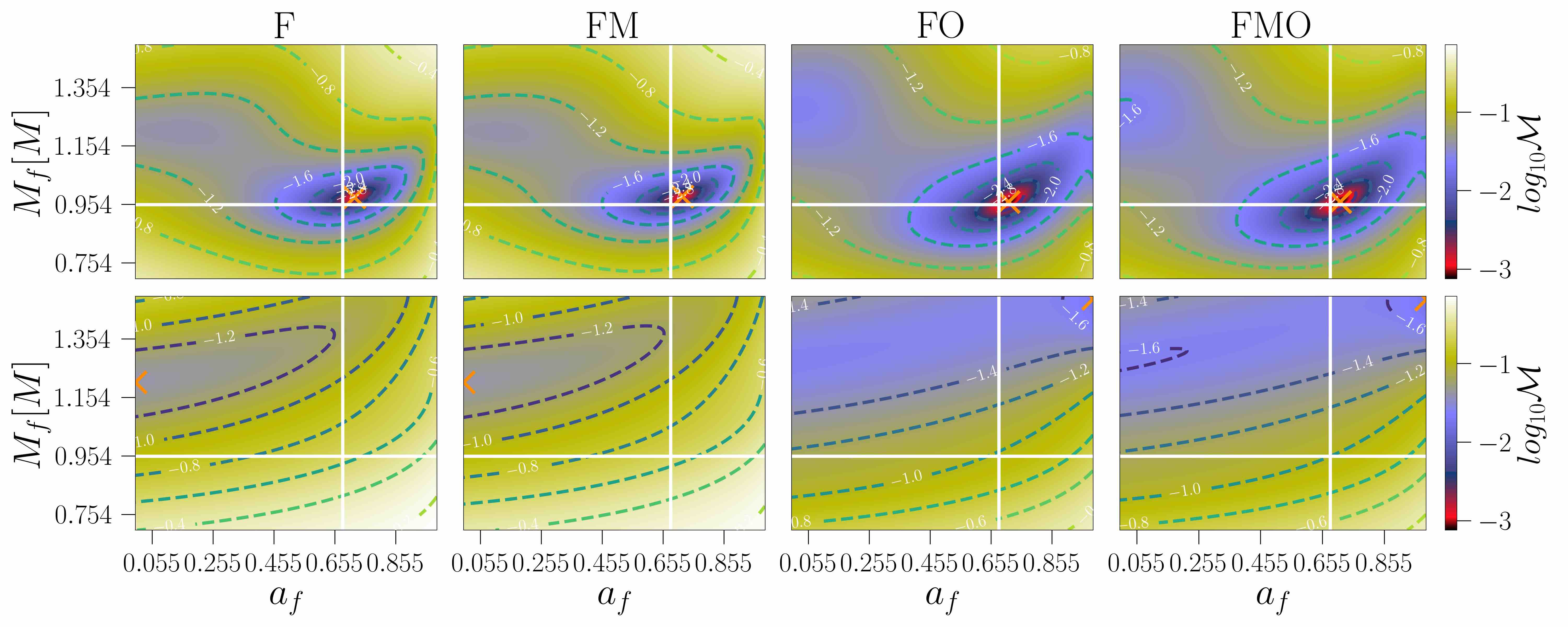}
    }
    
    \subfloat{
        \includegraphics[width=2\columnwidth, trim={0 0 0 1.3cm}, clip]{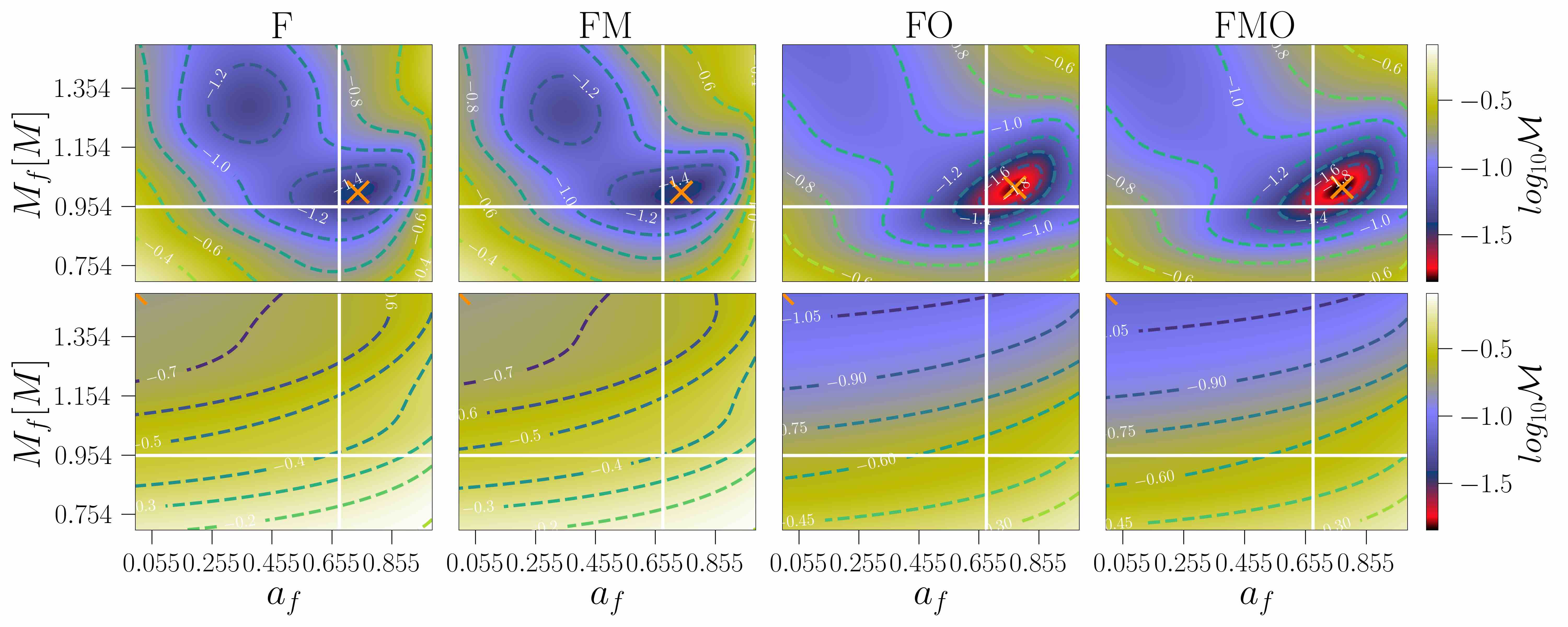}
    }
    \caption{This figure plots the mismatch $\mathcal{M}_{lm}$ in the remnant parameter space for a ringdown start time of $t_0-t_{\rm peak}=20M$ as in Fig.~\ref{fig:grid_m2_q1} but for $m=1$.}
    \label{fig:grid_m1_q1}
\end{figure*}

\begin{figure*}[h]
    \centering
    \subfloat{
        \includegraphics[width=2\columnwidth, trim={0 2.5cm 0 0}, clip]{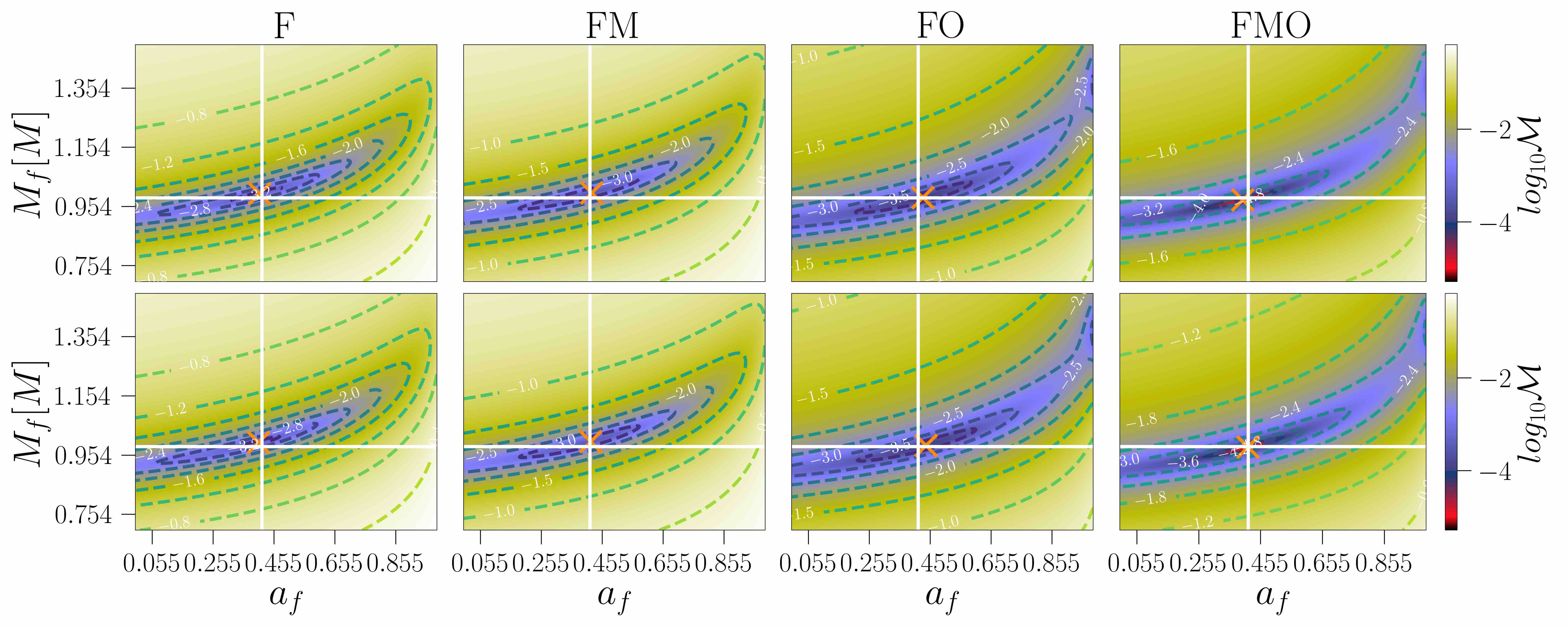}
    }
    
    \subfloat{
        \includegraphics[width=2\columnwidth, trim={0 2.5cm 0 1.3cm}, clip]{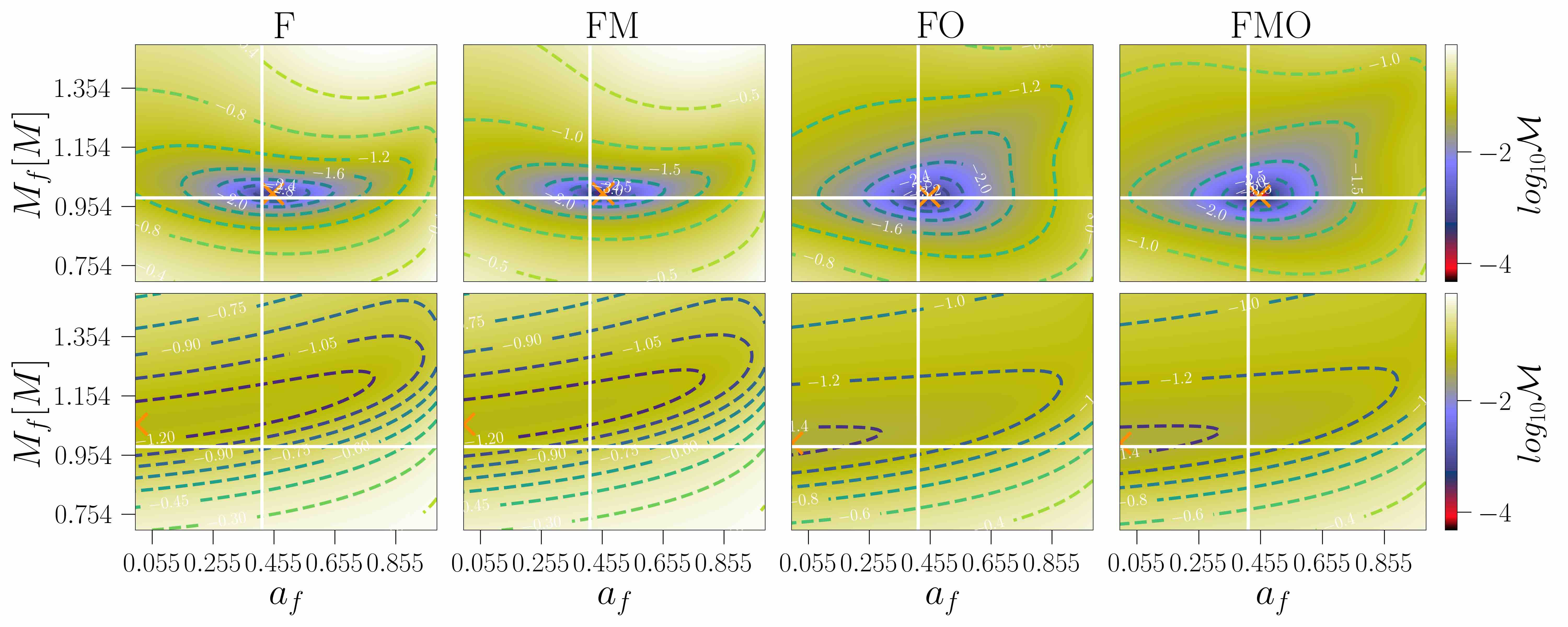}
    }
    
    \subfloat{
        \includegraphics[width=2\columnwidth, trim={0 0 0 1.3cm}, clip]{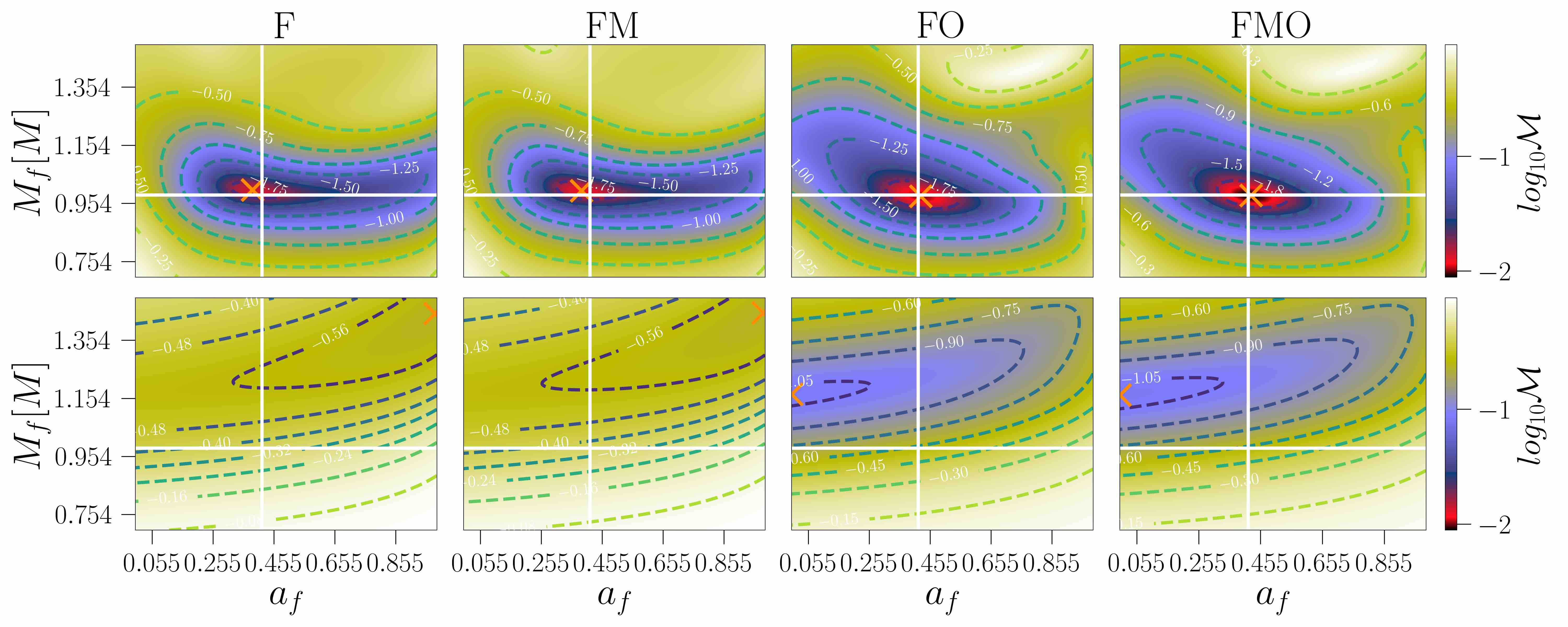}
    }
    \caption{Same as Fig.~\ref{fig:grid_m1_q1} but for the simulation S10:2.}
    \label{fig:grid_m1_q5}
\end{figure*}

\begin{figure*}[h]
    \centering
    \subfloat{
        \includegraphics[width=2\columnwidth, trim={0 2.5cm 0 0}, clip]{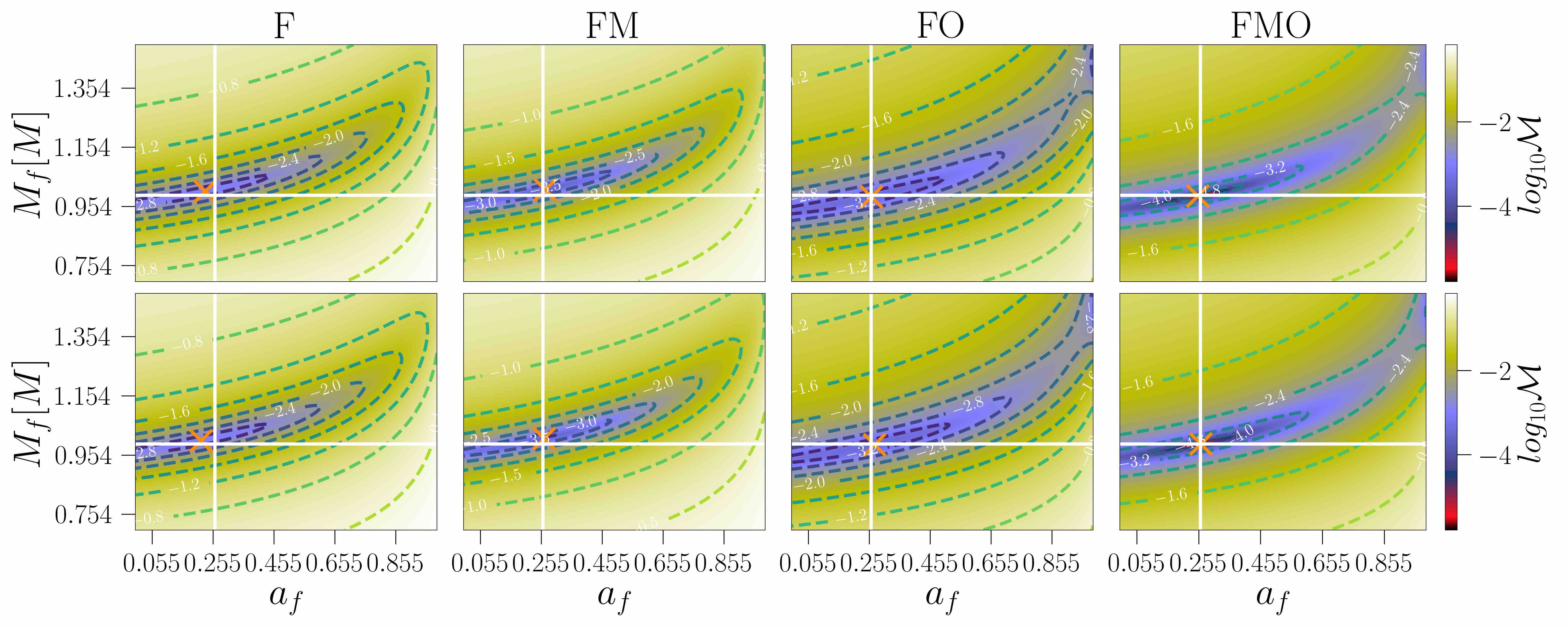}
    }
    
    \subfloat{
        \includegraphics[width=2\columnwidth, trim={0 2.5cm 0 1.3cm}, clip]{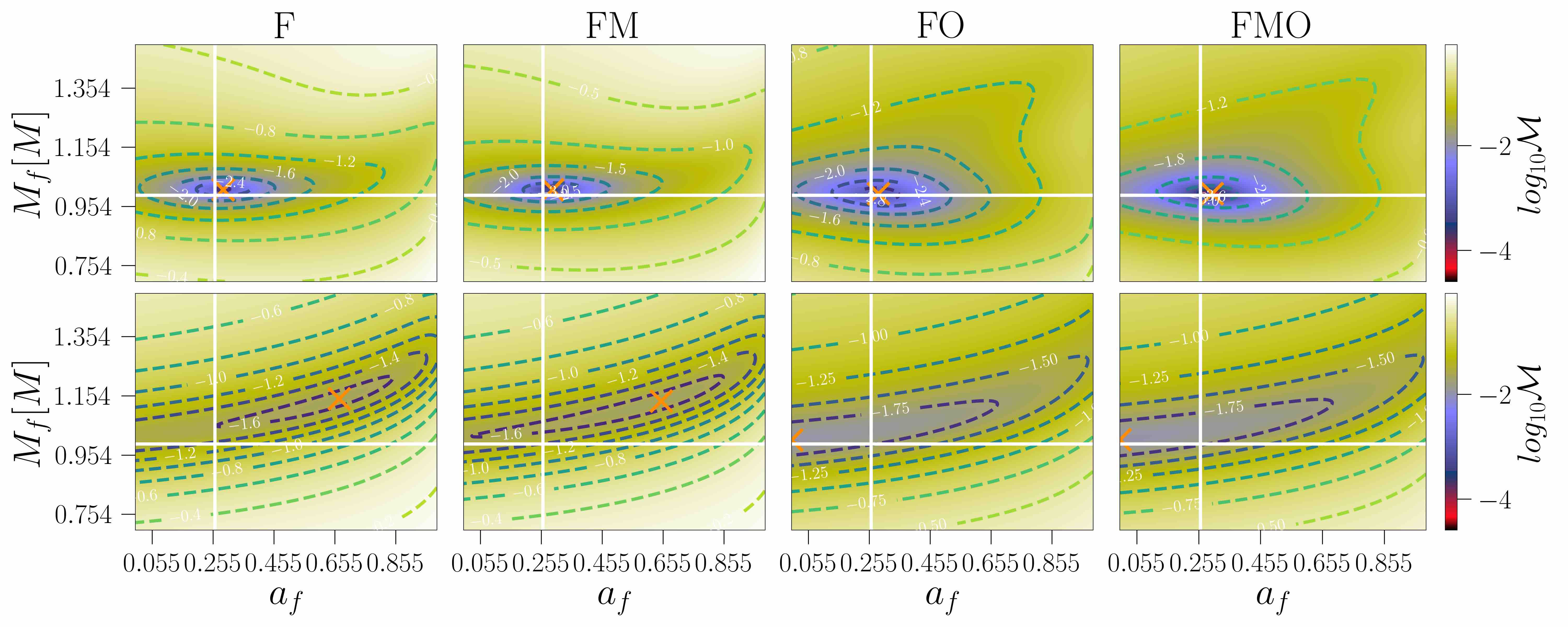}
    }
    
    \subfloat{
        \includegraphics[width=2\columnwidth, trim={0 0 0 1.3cm}, clip]{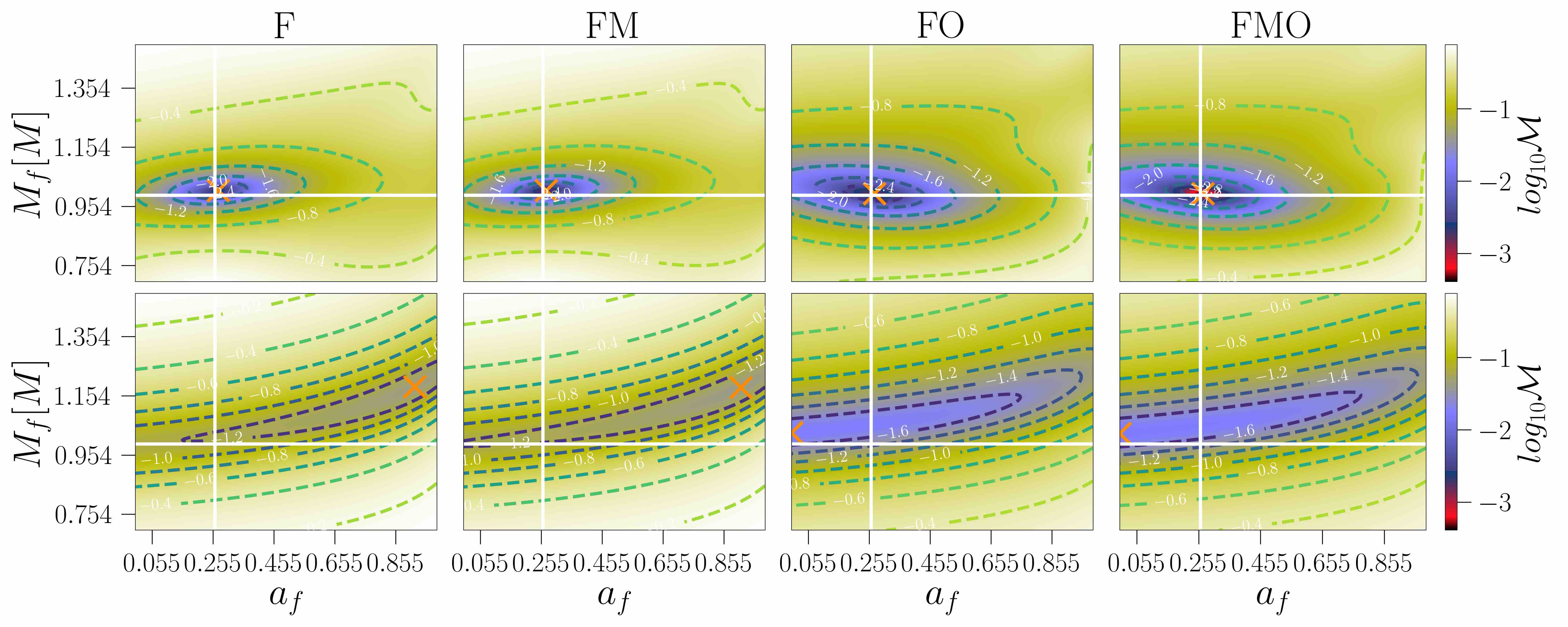}
    }
    \caption{Same as Fig.~\ref{fig:grid_m1_q1} but for the simulation S10:1.}
    \label{fig:grid_m1_q10}
\end{figure*}

The mismatch plots in the remnant parameter space can be seen in Figs.~\ref{fig:grid_m1_q1},~\ref{fig:grid_m1_q5}, and~\ref{fig:grid_m1_q10} for mass ratios $q=1.25$, $q=5$, $q=10$, respectively. We report similar conclusions for the $m=1$ modes as we did for the $m=2$ modes with regard to mode-mixing. It is the most important effect for modes with $l>2$ and when ignored gives wildly different remnant parameter estimates. The $l=2$ mode of this set is again unaffected by mode-mixing as expected. 

Overtones are important for certain cases of this set at early times as can be seen from Fig.~\ref{fig:eps_m2_m1}. 
The $\epsilon$ values for the modes of this set fluctuates across start times and mass ratios and no clear pattern can be made out though there is a general improvement in $\epsilon$ for the `FO' model compared to the `F' or `FM' models. It is not clear whether this is a feature of these modes or if there are hidden numerical artifacts causing this behaviour. 

A surprising feature of these modes is the effect of the mirror modes. Although the addition of mirror modes improves the mismatch curves significantly, as we saw in the previous section, and its effect can be clearly seen in the mode frequency plots in Fig.~\ref{fig:freq_m1}, they do not seem to improve the model systematic in the remnant parameter space. In fact, the amplitude of mirror mode modulations in the modes $(3,1)$ and $(4,1)$ for the $q=10$ case is of the same order of magnitude as the amplitude of mode-mixing modulations. Yet, where mode-mixing improves the remnant parameter estimates by an order of magnitude, the inclusion of mirror modes do not. 
This indicates that there are other effects, either numerical or physical, that dominate the deviation of the best-fit parameters from their true values. At this point, we cannot rule out either. The mismatch curves for the $m=1$ modes too indicate that we are missing some features since the mismatches of our best models fall way short of the NR error floor, suggesting that either some physical structures are being missed or there are systematic numerical artifacts across resolutions. 

\subsection{Modes: (3,3), (4,3)}
\begin{figure*}[h]
    \centering
    \subfloat{
        \includegraphics[width=2\columnwidth, trim={0 2.5cm 0 0}, clip]{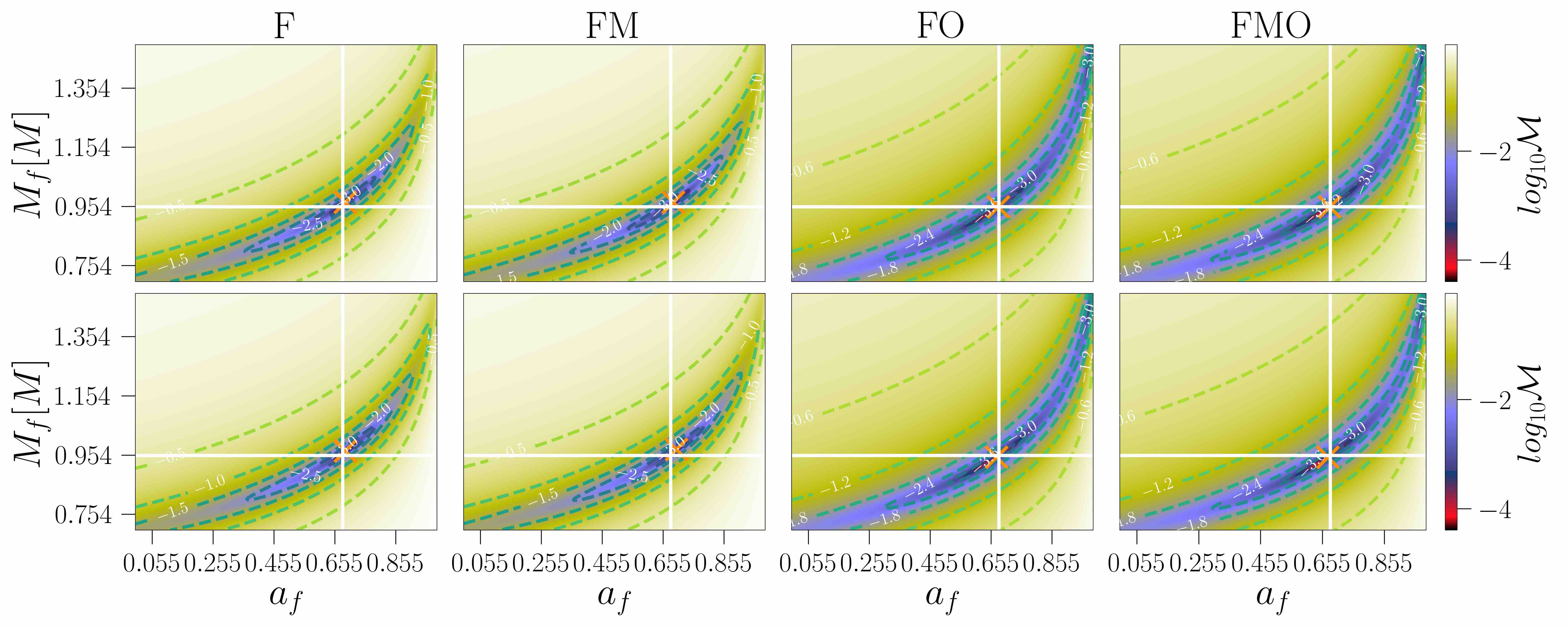}
    }
    
    \subfloat{
        \includegraphics[width=2\columnwidth, trim={0 0 0 1.3cm}, clip]{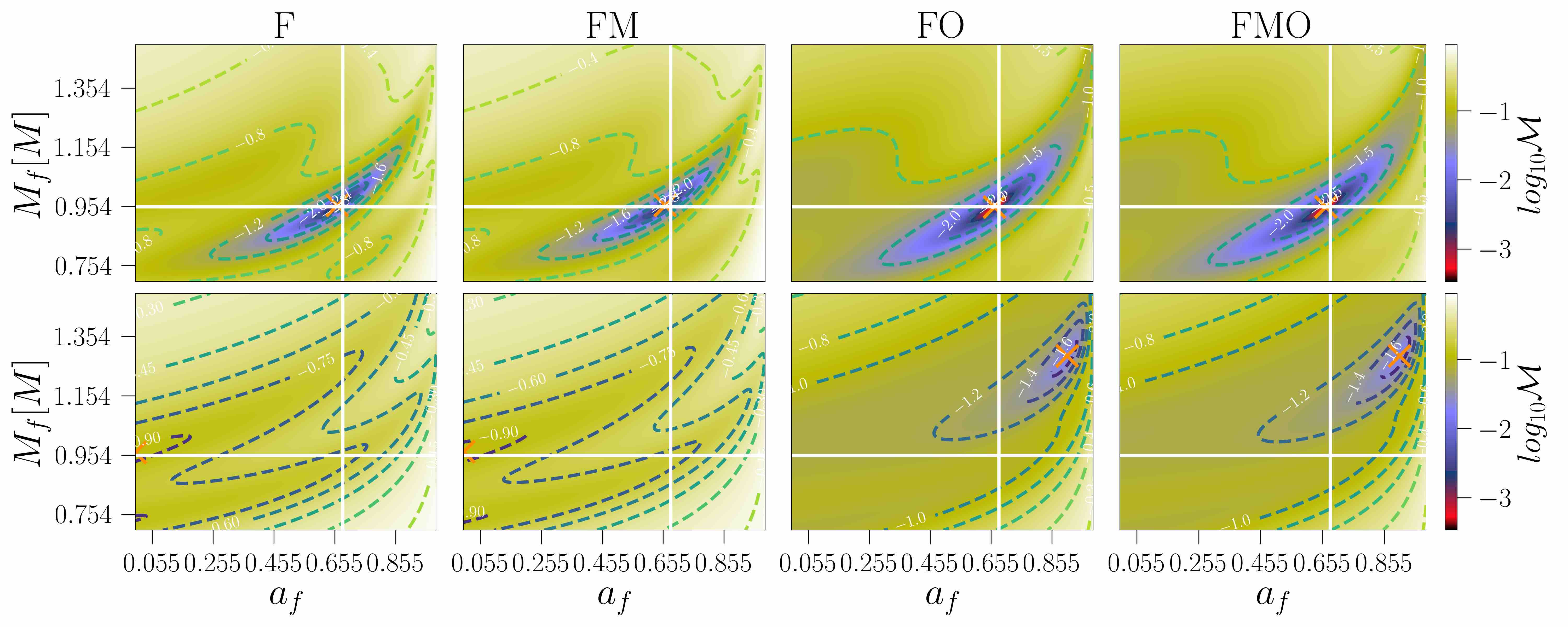}
    }
    \caption{This figure plots the mismatch $\mathcal{M}_{lm}$ in the remnant parameter space for a ringdown start time of $t_0-t_{\rm peak}=20M$ as in Fig.~\ref{fig:grid_m2_q1} but for $m=3$.}
    \label{fig:grid_m3_q1}
\end{figure*}

\begin{figure*}[h]
    \centering
    \subfloat{
        \includegraphics[width=2\columnwidth, trim={0 2.5cm 0 0}, clip]{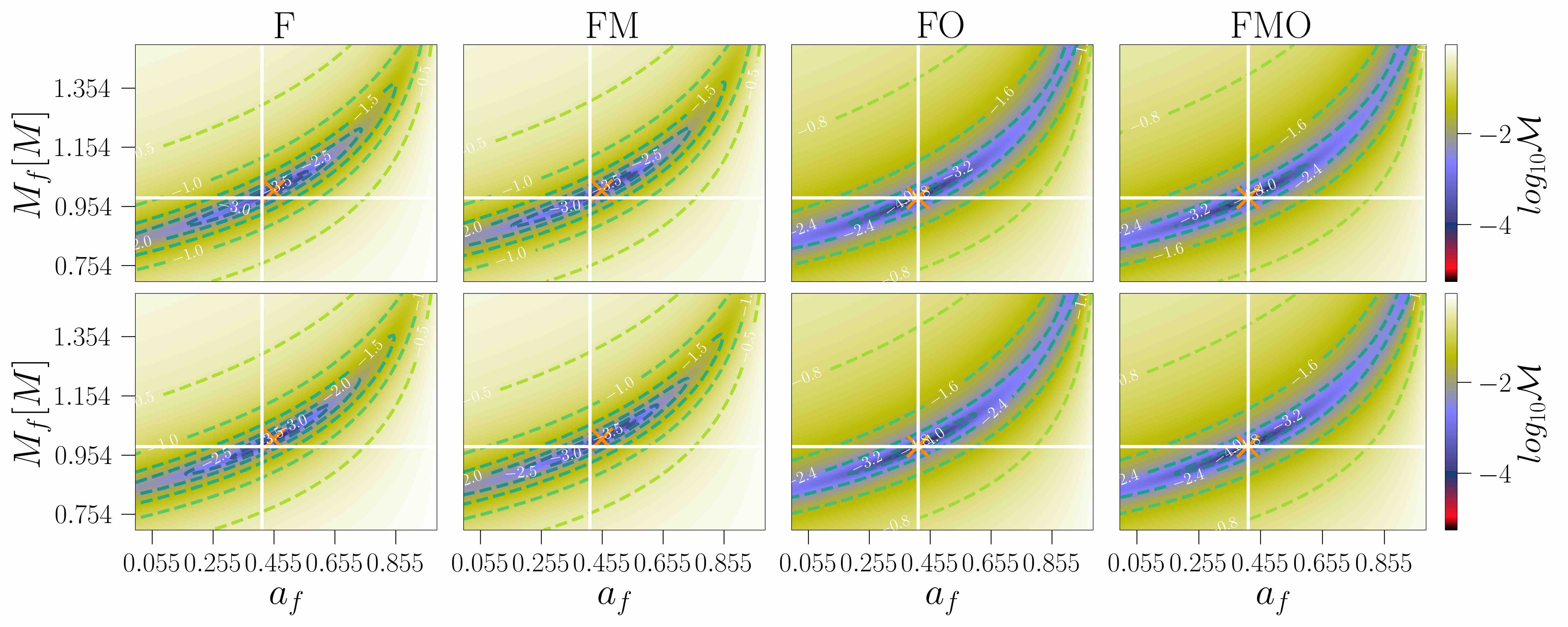}
    }
    
    \subfloat{
        \includegraphics[width=2\columnwidth, trim={0 0 0 1.3cm}, clip]{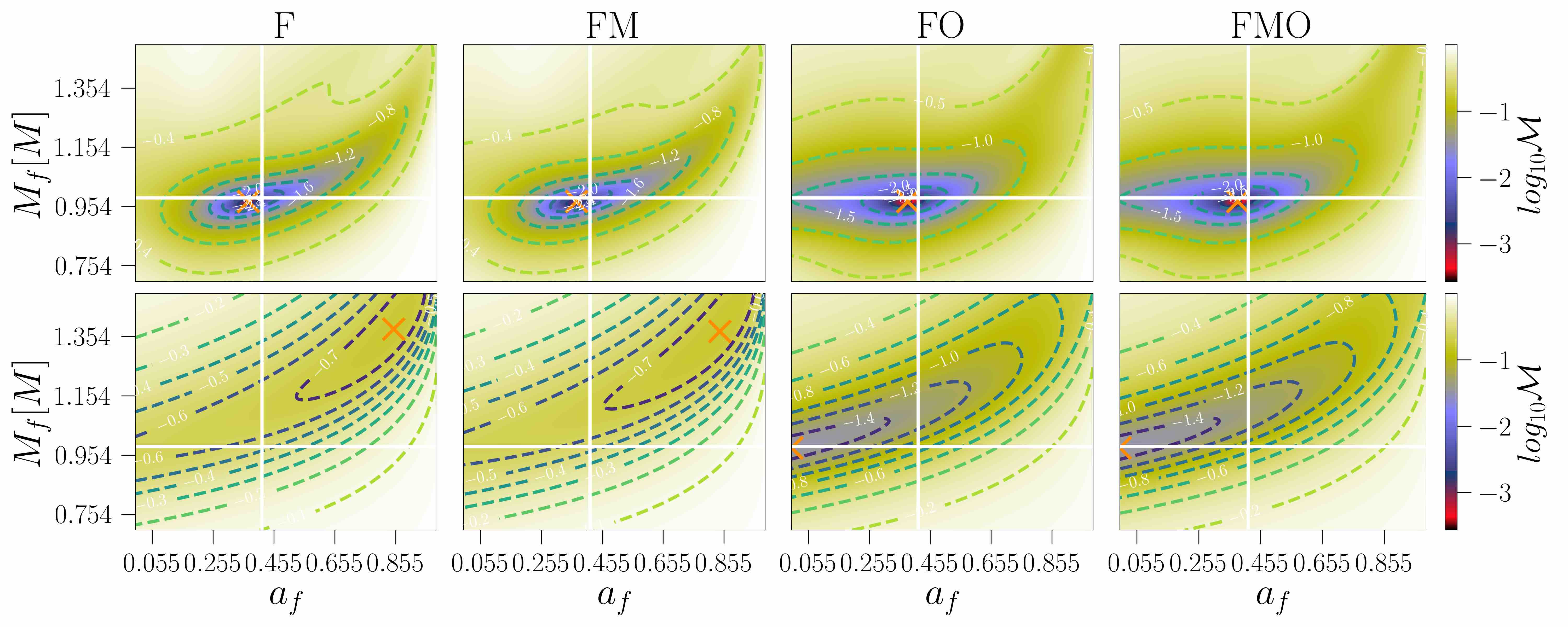}
    }
    \caption{Same as Fig.~\ref{fig:grid_m3_q1} but for the simulation S10:2.}
    \label{fig:grid_m3_q5}
\end{figure*}

\begin{figure*}[h]
    \centering
    \subfloat{
        \includegraphics[width=2\columnwidth, trim={0 2.5cm 0 0}, clip]{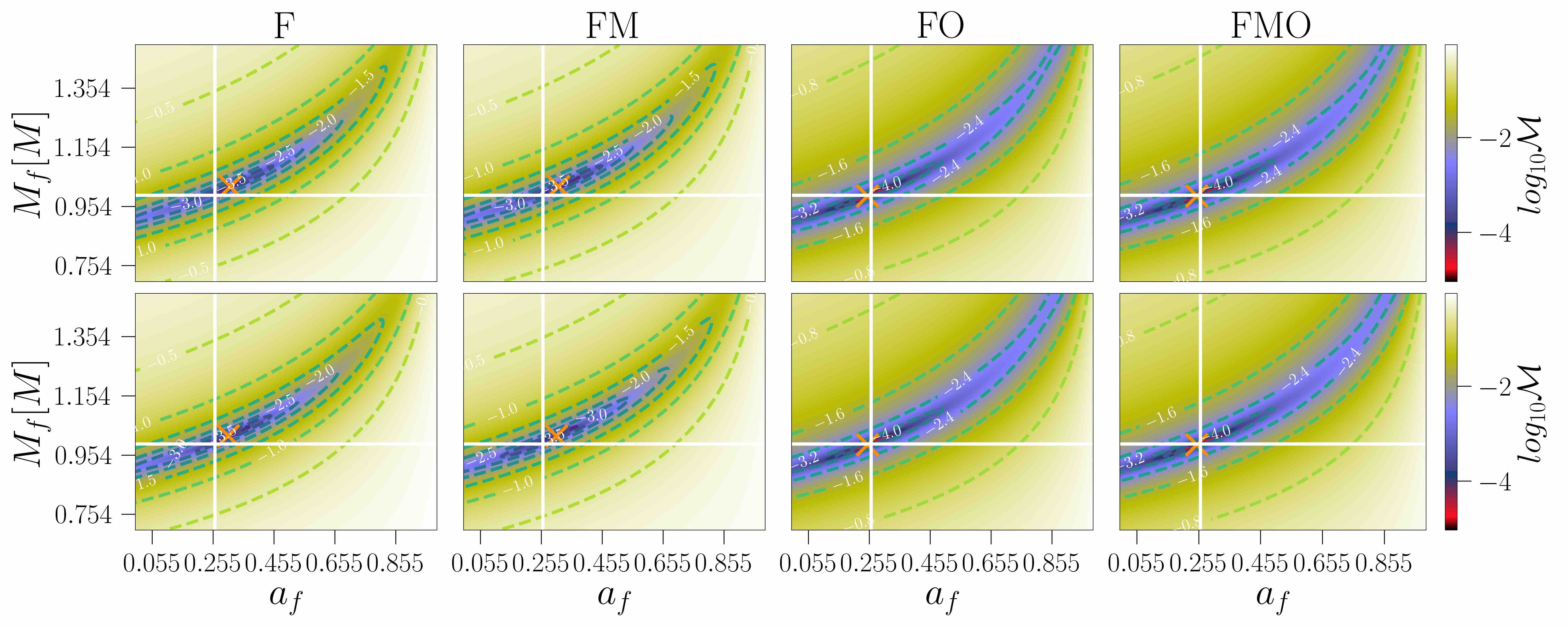}
    }
    
    \subfloat{
        \includegraphics[width=2\columnwidth, trim={0 0 0 1.3cm}, clip]{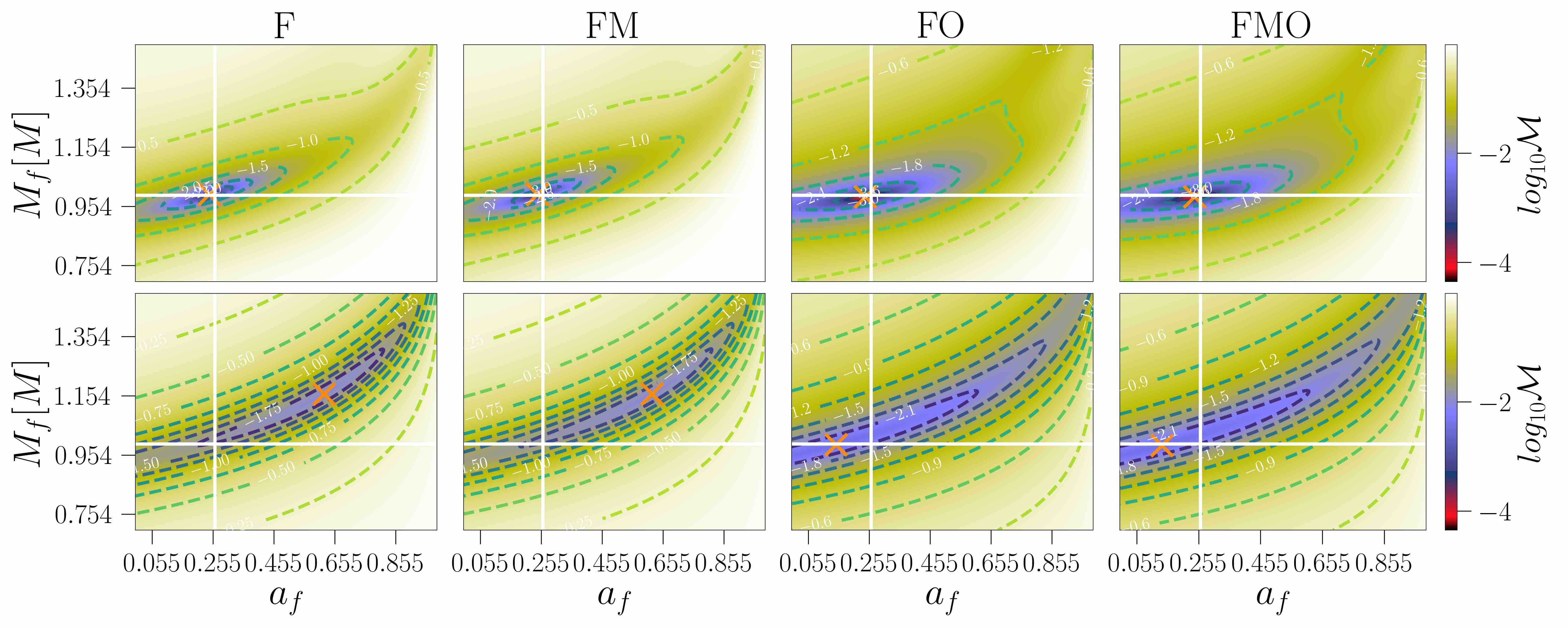}
    }
    \caption{Same as Fig.~\ref{fig:grid_m3_q1} but for the simulation S10:1.}
    \label{fig:grid_m3_q10}
\end{figure*}

The best-fit mismatches on an $M_f$--$a_f$ grid for the set of modes with $m=3$ are given in Figs.~\ref{fig:grid_m3_q1}, \ref{fig:grid_m3_q5}, and \ref{fig:grid_m3_q10}, for the three mass ratios under consideration. We report similar findings for mode-mixing, namely, the $(3,3)$ mode is unaffected though the $(4,3)$ mode gets significant contribution across all mass ratios. 

The `FO' model that includes the first overtone performs significantly better than the `F' model for the $(3,3)$ mode for all mass ratios but the overtone does not play much of a role for the $(4,3)$ mode as can be seen in Fig.~\ref{fig:eps_m3}. The fundamental mirror mode is again unimportant as can be seen by the overlapping $\epsilon$ curves for the `F' and `FM' models as well as the `FO' and `FMO' models.

\section{Discussion and conclusions}
\label{sec:disc}
In this work, we examined the features of ringdown waveforms using NR simulations for the modes $l\leq4$ and $m\leq3$. We constructed four ringdown models by adding the fundamental mirror mode, the first overtone, or both to a base model model consisting of only the fundamental QNM to assess the importance of each of these QNMs in a ringdown waveform. 

We found that overtones improve the mismatch to an NR waveform for all modes, though the improvement is most prominent for the modes with the smallest $l$ for a given $m$. The refinement is insignificant for $l=4,$ which is the largest $l$ considered in this study. We also found that adding the first overtone improves the model systematic in the remnant parameter space for the $(2,2)$ and $(3,3)$ modes across all mass ratios. The $(3,2)$, $(4,2)$, and $(4,3)$ modes see refinement only for mass ratio $q=1.25$ and times between $t_0-t_{\rm peak}=10M$ and $t_0-t_{\rm peak}=15M$. The other mass ratios for these modes do not see any significant improvement with the inclusion of the first overtone. The case for the set of modes with $m=1$ is not very clear because the epsilon values for the various models fluctuate though a general improvement can be seen with the addition of the overtone.

Moreover, we found that mirror modes are most significantly excited for the $m=1$ modes with their relative strengths growing for higher $l$ modes and larger mass ratios. Curiously, we observed that even though the mirror modes improve the mismatches by orders of magnitude and are clearly visible in the mode frequency plots of the waveforms, they do not improve the model systematics in the remnant parameter space. This indicates that the departure of the best-fit values of the remnant parameters from their true values is dominated by other features, either physical or numerical, though we cannot rule out either.

The angular part of a gravitational waveform in an NR simulation of a binary black hole merger is decomposed in $-2$ spin-weighted spherical harmonics whereas the Teukolsky equation, which describes the perturbations of a Kerr black hole, separates out the angular part in terms of $-2$ spin-weighted spheroidal harmonics. As a result, an NR spherical harmonic mode has contributions from multiple spheroidal harmonic angular quantum numbers though the azimuthal quantum numbers $m$ stays the same. This is known as mode-mixing. We estimated the effect of mode-mixing for the various spherical harmonic modes under consideration by further classifying the above four models into a class that accounts for mode-mixing and one that does not. To account for mode-mixing, we fit the set of spherical harmonic modes with the same $m$ simultaneously to prevent multiple fit amplitudes from being associated to the same QNM. We note that mode-mixing affects the dominant mode of a set, or the mode with the smallest $l$ for a given $m$, the least. This is true for both the mismatch curves and the model systematic in the remnant parameter space. For the higher $l$ modes for a given $m$, we find that mode-mixing plays a significant role, with the $\epsilon$ value that provides a measure of the deviation of the best-fit remnant parameters from the true values superior by an order of magnitude or better for the class of models that account for mode-mixing.

\section{Future work}
\label{sec:fw}
A natural extension of this work would be to fit the amplitudes as a function of the mass ratio for the best models of a given $m$ as determined in this study. LISA is expected to observe binary black hole mergers with total mass greater than $10^8\,M_{\odot}$ that would have little to no inspiral part. Such an amplitude model can be used to infer the progenitor properties as has already been done in~\textcite{Capano:2021etf} for GW190521. Additionally, such amplitude fits can be used for testing the strong field dynamics as proposed by~\textcite{Bhagwat:2021kfa}. 

In this study, we focused on non-spinning binaries. An interesting continuation would be to consider initially spinning binaries which are expected to play a role for modes with odd $l+m$ and look at their effects on these modes. 

Finally, we plan on simulating real gravitational wave signals for the various current, planned, and proposed gravitational wave detector networks to determine the bias in parameter estimation due to ignoring mode-mixing for the relevant modes considered in this study.

\acknowledgements
This research was supported by National Science Foundation, grant number PHYS-2012083, PHY-1836779 and
AST-2006384.

\appendix
\renewcommand\thefigure{\thesection \arabic{figure}}

\setcounter{figure}{0}
\section{Waveform fits}
\label{sec:fits}
In this section, we plot the mode amplitudes $|h_{lm}|$ and mode frequencies, $f_{lm}=-\Im(\Dot{h}_{lm}/h_{lm})$, as a function of time $t-t_{\rm peak}$ for our four best-fit ringdown models. We only consider the models from the mode-mixing class. We overlay the models on top of the corresponding NR mode to depict the features of the NR mode that each QNM takes into account. We take the ringdown start time to be $t_0-t_{\rm peak}=20M$ for these plots. 

Let us now consider a simplified example to make apparent the various features seen in the mode amplitude and mode frequency plots. Consider that we model a given NR mode $h$ using two QNMs with complex frequencies $\omega=\omega_o-i/\tau$ and $\omega'=\omega'_o-i/\tau'$ and their respective mixing coefficients are $\mu$ and $\mu'$. Let us further assume that $\mu\mathcal{C}\gg\mu'\mathcal{C}'$ where $\mathcal{C}$ and $\mathcal{C}'$ are the complex excitation amplitudes of the two QNMs, respectively. The NR mode can then be written as
\begin{equation}
    h = \mathcal{C}\mu e^{-i\omega t} + \mathcal{C}'\mu' e^{-i\omega' t}
\end{equation}
The above formula is valid for any pair of QNMs and, therefore, the relevant indices $(l,m)$ can be introduced as necessary.

When the above approximation for the effective excitation amplitudes holds, the mode amplitude $|h|$ takes the form
\begin{equation}
\label{eq:mode_amp}
    |h| \approx |\mathcal{C}\mu|e^{-t/\tau} \left[1 + \Re\left(\frac{\mu'}{\mu} \frac{\mathcal{C}'}{\mathcal{C}} e^{i(\omega-\omega')t}\right)\right]
\end{equation}
and the mode frequency $f$ is given by
\begin{equation}
\label{eq:mode_freq}
    f \approx \omega \left[ 1- \Re\left(\left(1-\frac{\omega'}{\omega}\right)\frac{\mu'}{\mu} \frac{\mathcal{C}'}{\mathcal{C}} e^{i(\omega-\omega')t}\right) \right].
\end{equation}

\subsection{Mode amplitudes}
\begin{figure*}[ht]
    \centering
    \includegraphics[width=2\columnwidth]{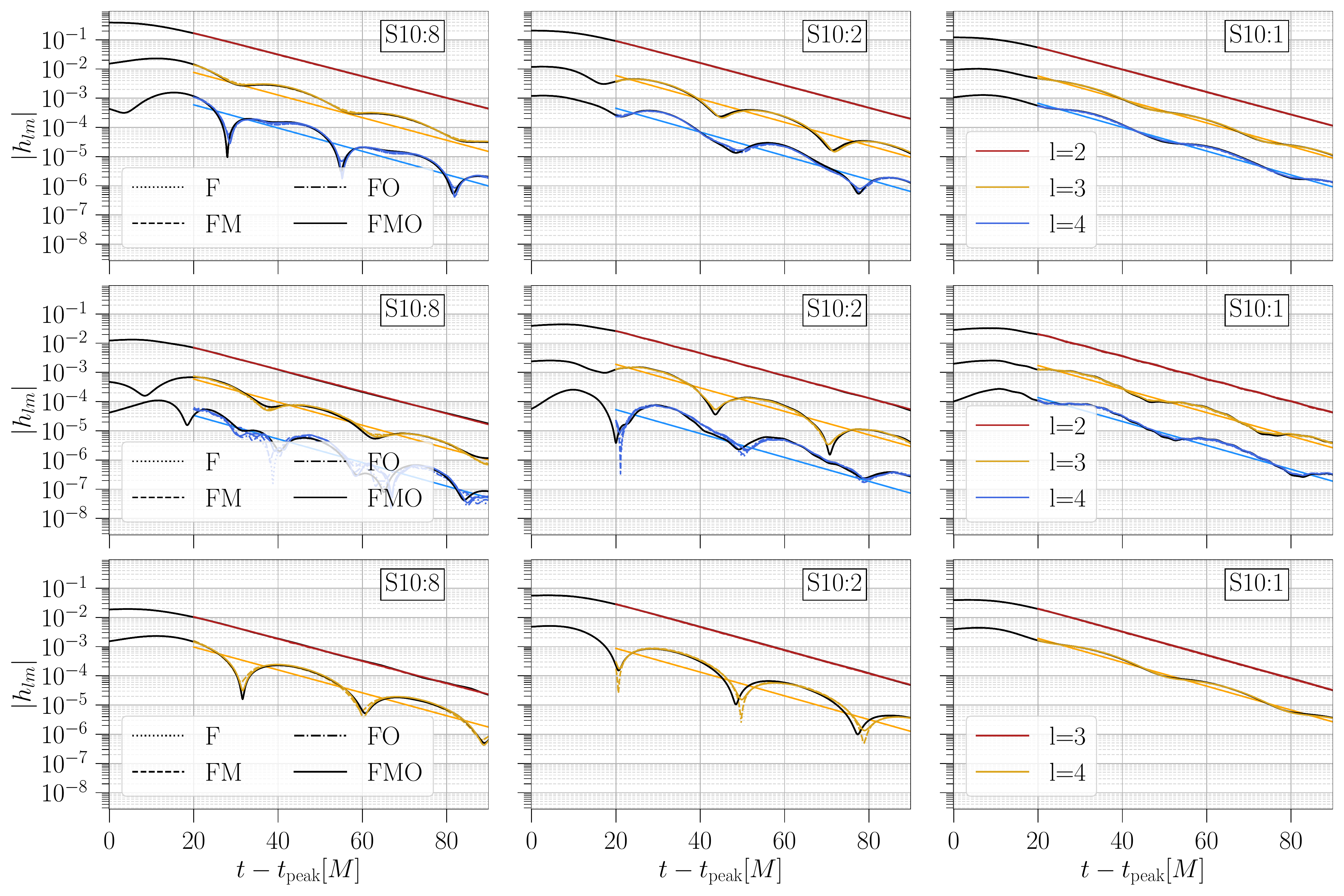}
    \caption{The mode amplitudes (black) for $m=2$, $m=1$, and $m=3$ (top to bottom) are plotted as a function of time $t-t_{\rm peak}$ for the three simulations in this study (left to right). The four different best-fit mode-mixing models are overlaid on top for comparison. The slanted straight curves are fits to their respective modes ignoring mode-mixing and using only the fundamental mode. The start time for the fits are taken to be $t_0-t_{\rm peak}=20M$. The effect of mode-mixing is negligible for $l=2$ but becomes more important as $l$ increases.}
    \label{fig:amp}
\end{figure*}

In Fig.~\ref{fig:amp}, the mode amplitudes for $m=2$, $m=1$, and $m=3$ are plotted. The straight lines are the best-fits using the corresponding fundamental QNM without the inclusion of mode-mixing. The foremost feature of these plots is the presence of characteristic modulations about the straight line for the higher modes of a set. More specifically, the second mode of a set sees a single modulating frequency whereas the third mode, if present, sees an interference between two modulating frequencies. The secondary modulations are visible to the eye for the smaller mass ratios only. 

Let us take the mode $(3,2)$ as a case study to describe the behavior of the second mode of a set and use the above defined model. The two QNMs we use to model are the fundamental modes $(3,2,0)$ ($\omega=\omega_{320}$) and $(2,2,0)$ ($\omega'=\omega_{220}$). The mixing coefficients are then $\mu\equiv\mu^3_{320}\sim1$ and $\mu'\equiv\mu^2_{320}\sim10^{-2}$, respectively, for the two modes. The relative excitation amplitudes are $\mathcal{C}'/\mathcal{C}\sim10^{1}$ from our fits. Therefore the amplitude of modulations are $\sim10^{-1}$ with the modulating frequency $\omega_{320}-\omega_{220}$. 

If we carry out the same exercise for the $(3,2)$ mode using the fundamental QNMs $(3,2,0)$ ($\omega=\omega_{320}$) and $(4,2,0)$ ($\omega'=\omega_{420}$), we find the mixing coefficients to be $\mu\equiv\mu^3_{320}\sim1$ and $\mu'\equiv\mu^4_{320}\sim10^{-2}$, respectively, with the relative excitation amplitudes being $\mathcal{C}'/\mathcal{C}\sim10^{-1}$. Therefore, the amplitude of modulations are $\sim10^{-3}$ and, hence, not visible in the figures. 

The above model can be easily extended to describe the third mode of a set, if present. Note that till this point we are not describing the high frequency modulations seen in the mode $(4,1)$ for the mass ratio $q=10$. We will describe this in the next section.



\subsection{Mode frequencies}
\begin{figure*}[ht]
    \centering
    \includegraphics[clip,width=2\columnwidth]{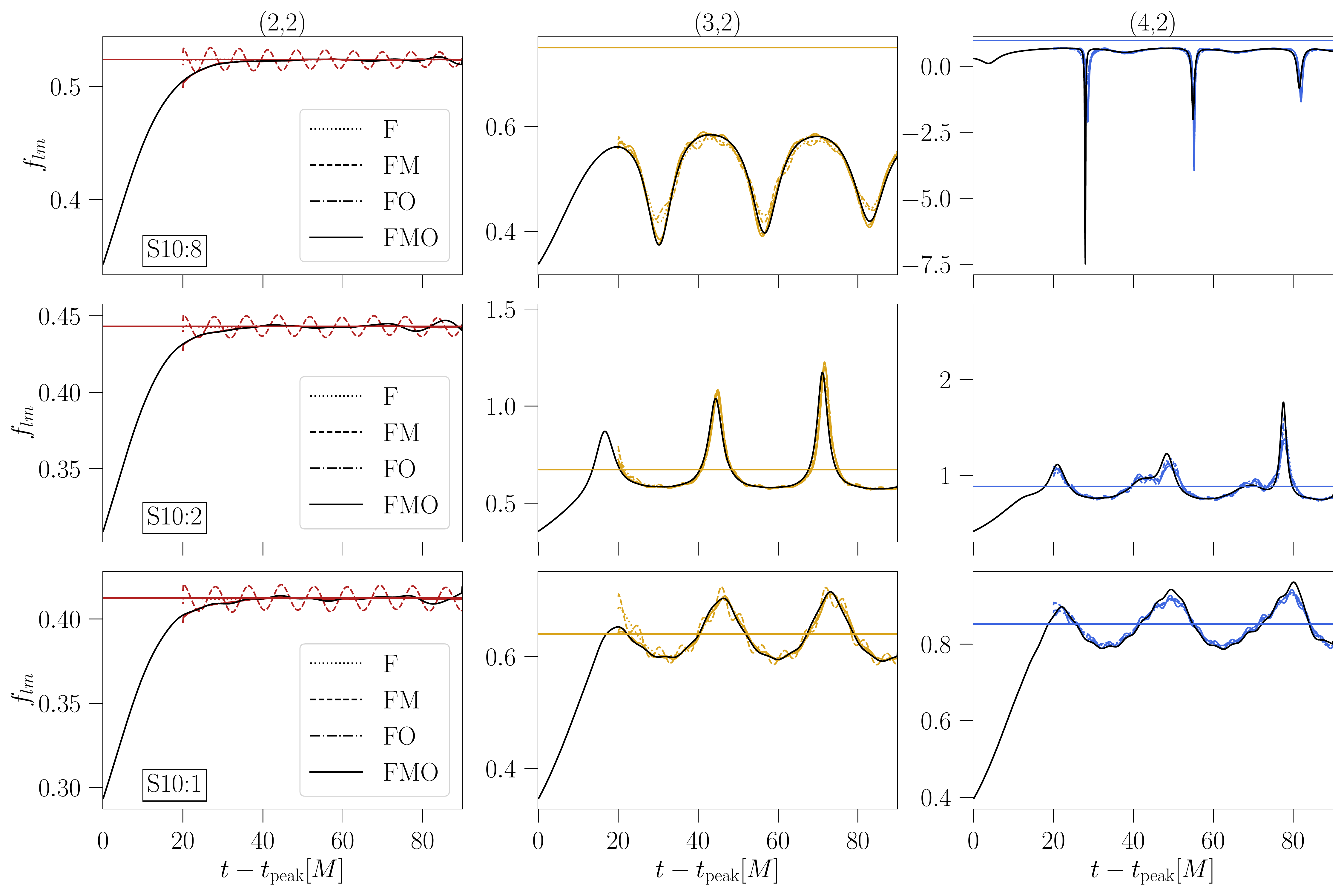}
    \caption{The mode frequencies (black) for $m=2$ as a function of time $t-t_{\rm peak}$ for the mass ratios examined in this study. The four different best-fit mode-mixing models are overlaid on top for comparison. The horizontal curves show the fundamental mode frequency for the respective modes. The start time for the fits are taken to be $t_0-t_{\rm peak}=20M$. Note that for modes $(3,2)$ and $(4,2)$ and mass ratio $q=1.25$, the fundamental QNM grossly under estimates the true mode frequency. Note, further, that for mode $(4,2)$ and mass ratio $q=10$, we start seeing the modulations due to the fundamental mirror mode on top of mode-mixing for the first time for $m=2$.}
    \label{fig:freq_m2}
\end{figure*}

\begin{figure*}[ht]
    \centering
        \includegraphics[clip,width=2\columnwidth]{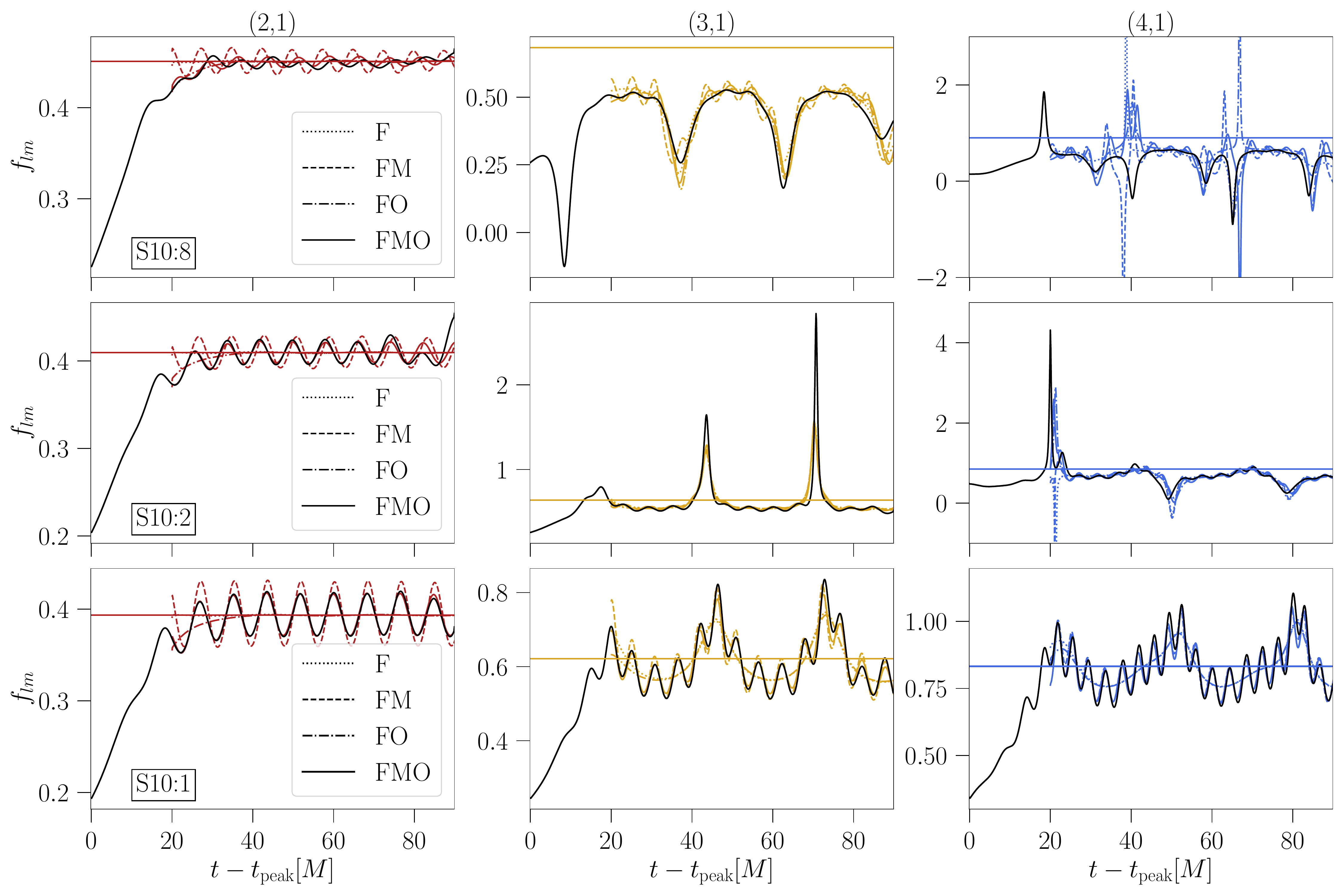}
    \caption{Same as Fig.~\ref{fig:freq_m2} but for $m=1$.}
    \label{fig:freq_m1}
\end{figure*}

\begin{figure}[ht]
    \centering
        \includegraphics[clip,width=\columnwidth]{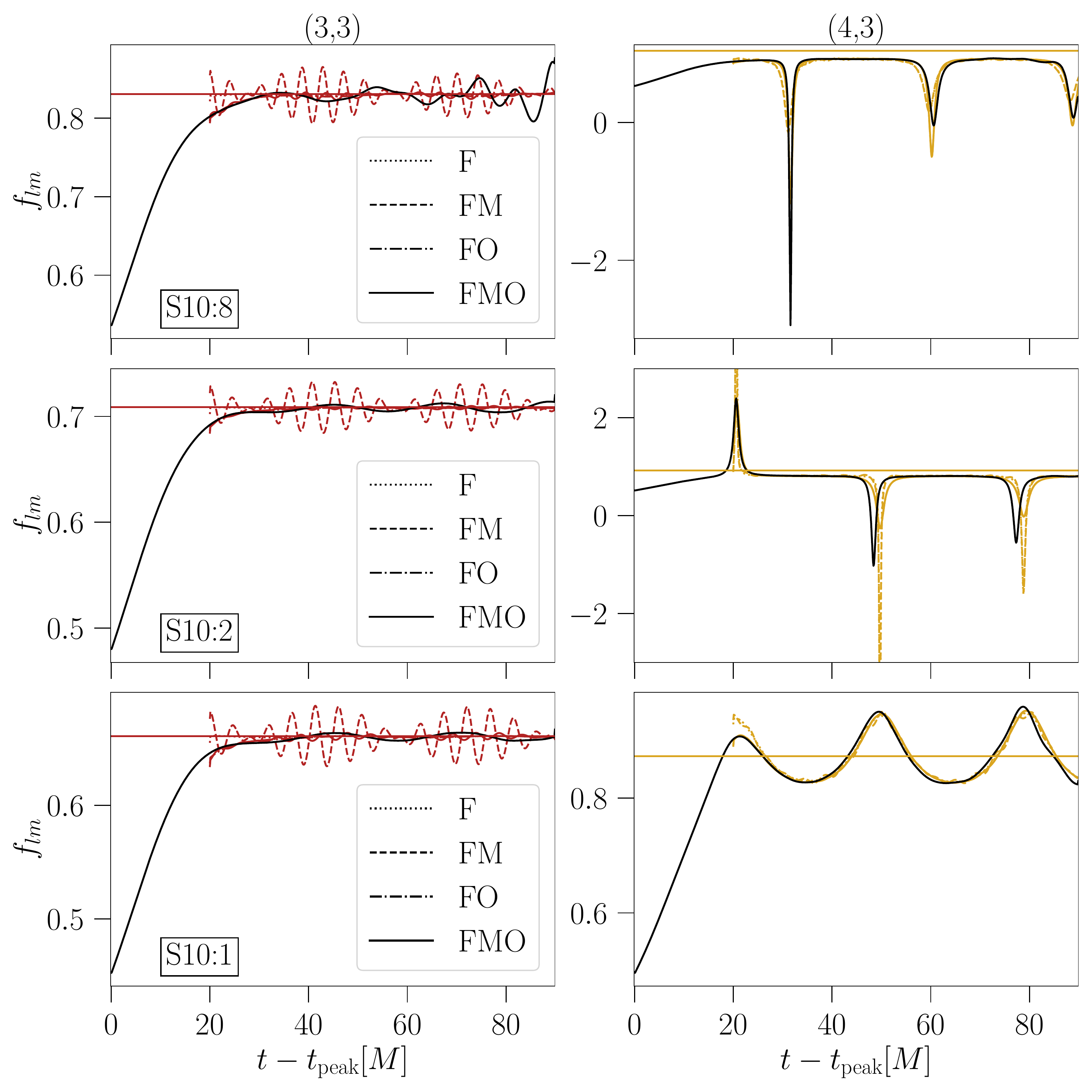}
    \caption{Same as Fig.~\ref{fig:freq_m2} but for $m=3$.}
    \label{fig:freq_m3}
\end{figure}

The modulations due to mode-mixing can also be seen in the mode frequency plots in Figs.~\ref{fig:freq_m2},~\ref{fig:freq_m1}, and~\ref{fig:freq_m3} for the sets of modes with $m=2$, $m=1$, and $m=3$, respectively. The horizontal line in the plots for each mode corresponds to the oscillation frequency of the corresponding fundamental QNM. 

The foremost difference from the mode amplitude plots is the clear presence of high frequency modulations on top of the mode-mixing modulations. These are due to the fundamental mirror mode. We see from Eq.~(\ref{eq:mode_freq}) that the modulating frequency is $\Re(\omega-\omega')$. For mode-mixing, this is the difference in the oscillation frequencies of the fundamental QNM both of which are positive whereas the oscillation frequency of mirror modes are negative and, hence, the modulating frequency is typically higher. Further, note that for the $(2,1)$ mode, even though the `FM' model generally traces out the mirror mode modulations, it is off in the amplitude and the phase whereas the `FMO' model reproduces the mode frequency curves exactly indicating the role played by the first overtone. 

Finally, we point to the $(3,3)$ mode plots where the `FM' model shows beat patterns which are atypical of mirror mode modulations we have seen so far. These are due to the spurious oscillatory features we see for this mode and are not able to assign any physical reasoning for the presence of these features.

\setcounter{figure}{0}
\section{Effect of start time on model systematics}
\label{sec:eps_t0}
Throughout the main text of this paper, we showed the results for the model systematics on the remnant parameter space taking the ringdown start time to be $t_0-t_{\rm peak}=20M$ because that is approximately the time when the mismatch curves for the model with the lowest mismatch bottom out for all the modes considered in this study. In this section we look at the effect of varying the start time of ringdown as this is a phenomenologically defined quantity. Specifically, we look at its effect on $\epsilon$ (see Eq.~(\ref{eq:eps})) which provides a measure of the deviation of the best-fit remnant parameters from its true value as determined from NR simulations. 

\begin{figure*}[ht]
    \centering
    \subfloat{
        \includegraphics[width=1.7\columnwidth, trim={0 1.9cm 0 0}, clip]{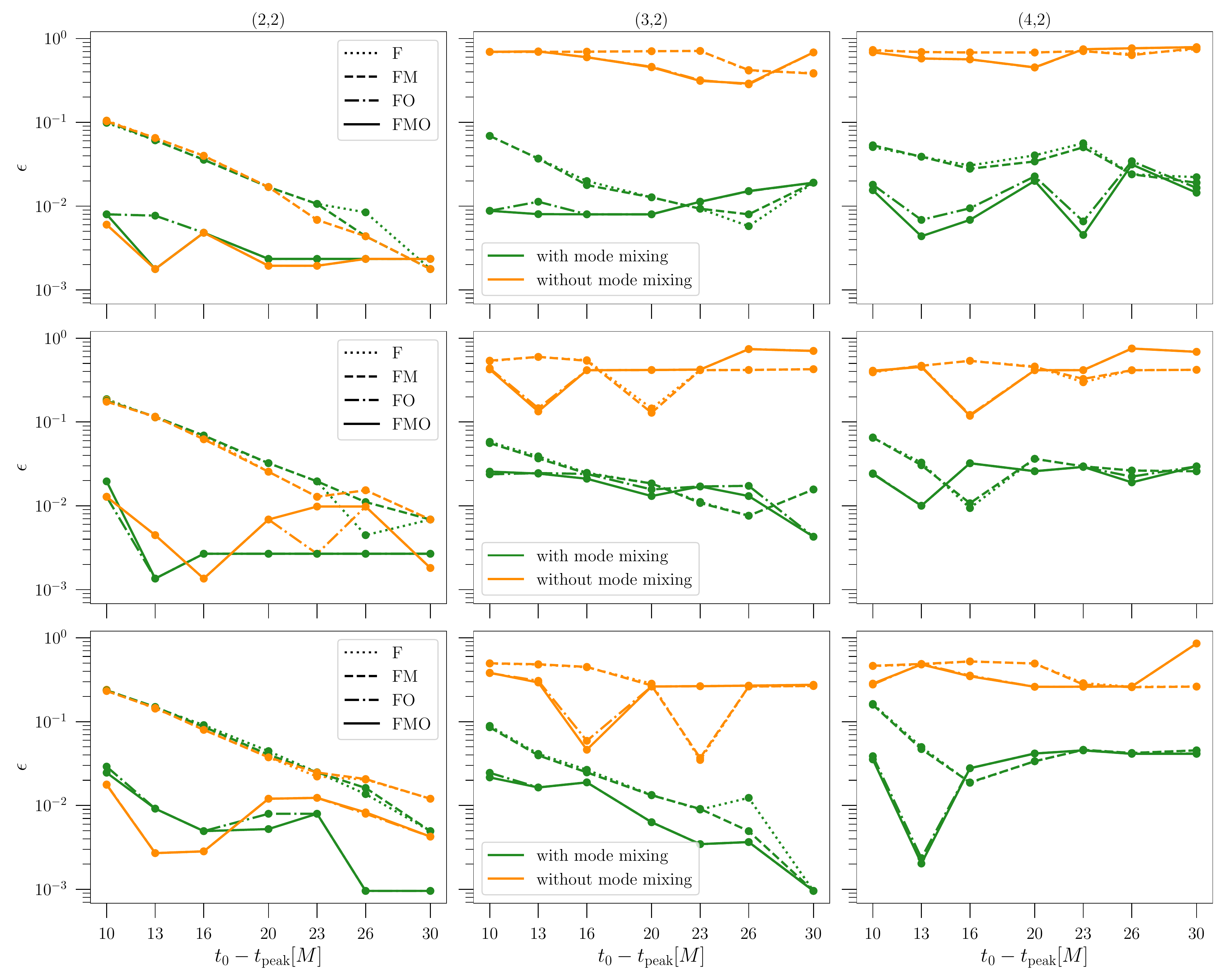}
    }
    
    \subfloat{
        \includegraphics[width=1.7\columnwidth, trim={0 0 0 0}, clip]{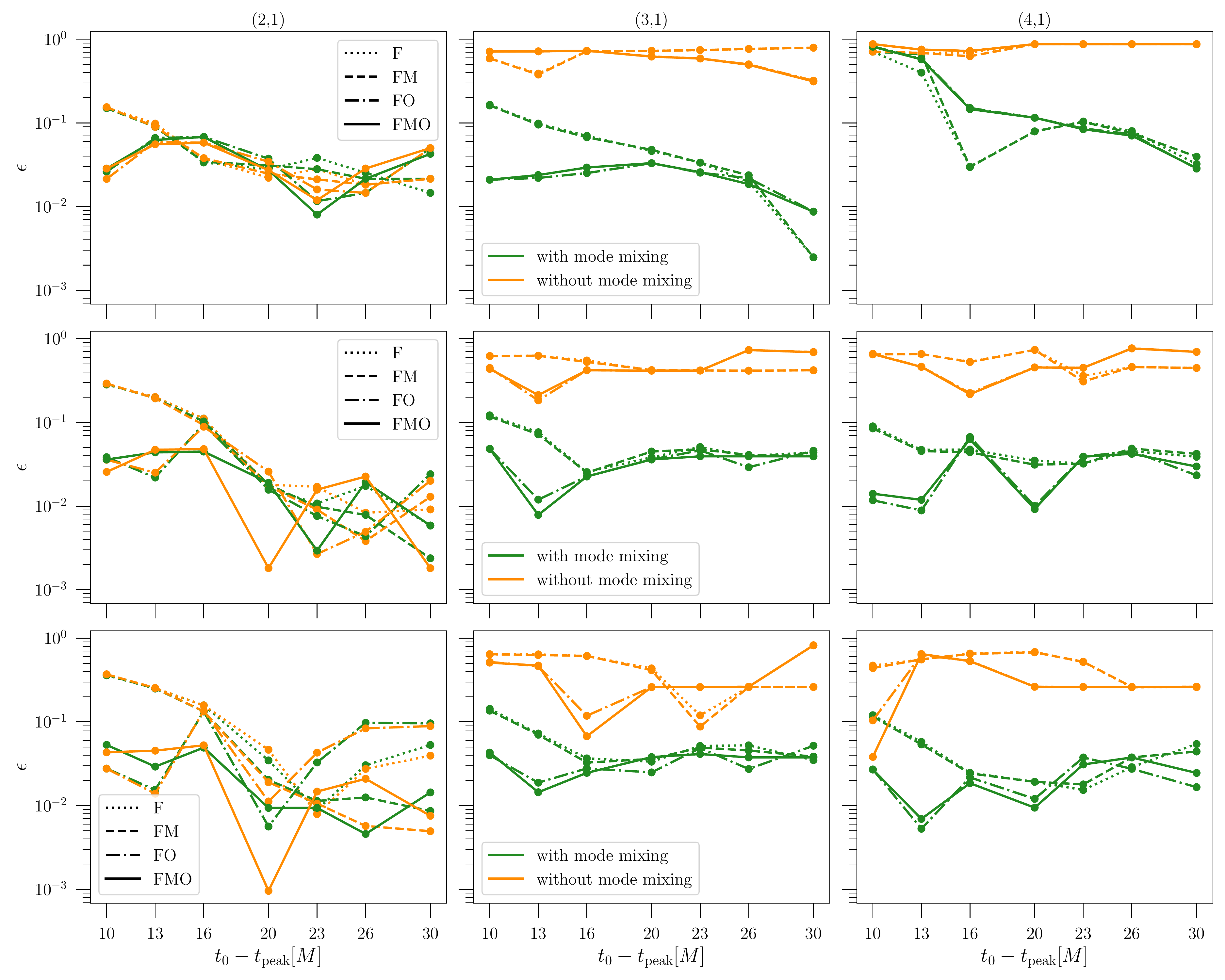}
    }
    \caption{The error in the estimation of the remnant parameter $\epsilon$ as a function of the start time of ringdown $t_0-t_{\rm peak}$ for the four individual models and the two classes of models including (green) and excluding (orange) mode-mixing. This figure corresponds to the set $m=2$ (top) and $m=1$ (bottom) for the three mass ratios (rows in a sub-figure) considered in this study.}
    \label{fig:eps_m2_m1}
\end{figure*}

\begin{figure}[ht]
    \centering
    \includegraphics[width=\columnwidth]{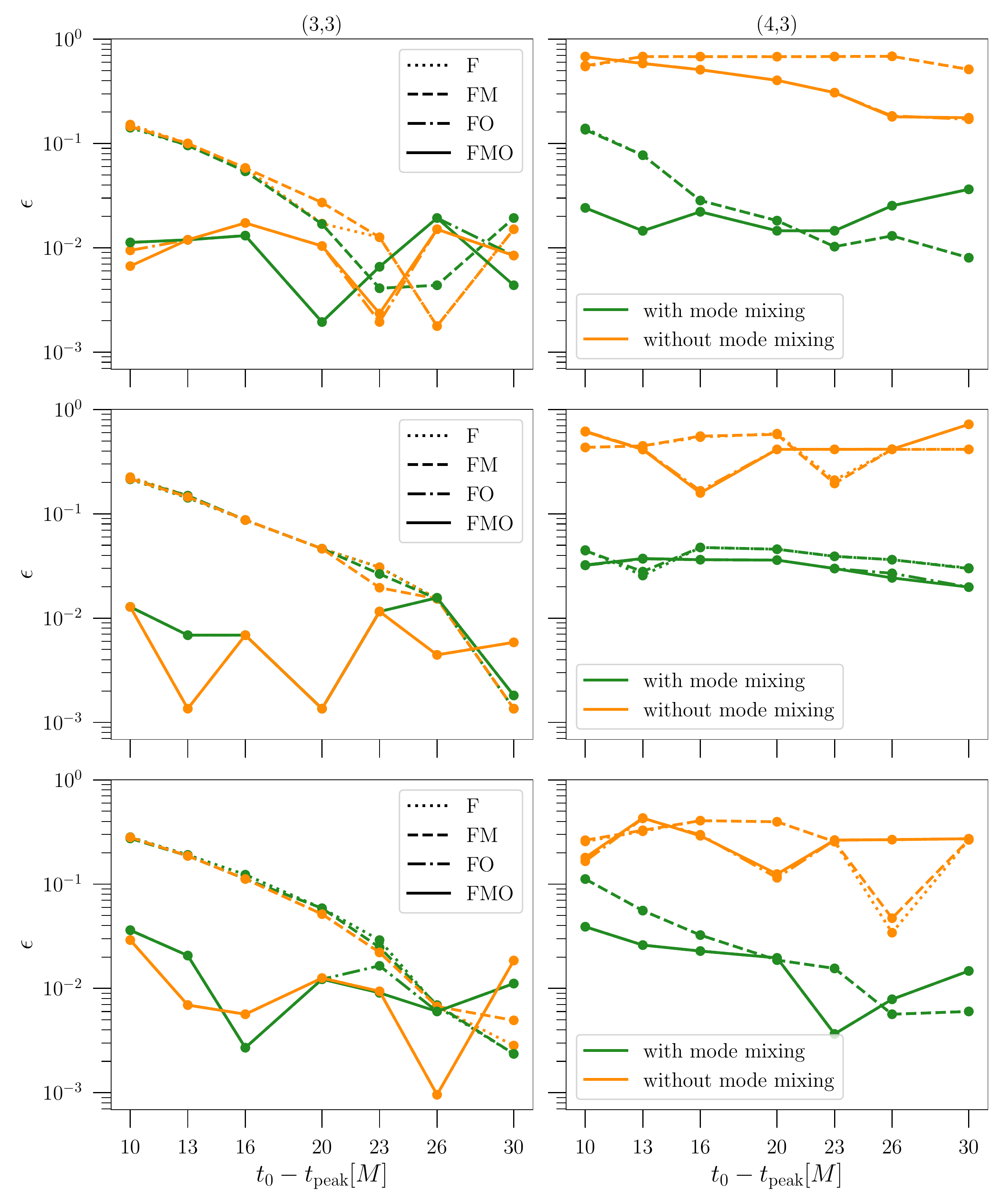}
    \caption{Same as Fig.~\ref{fig:eps_m2_m1} but for $m=3$.}
    \label{fig:eps_m3}
\end{figure}

In Figs.~\ref{fig:eps_m2_m1} and~\ref{fig:eps_m3}, we plot $\epsilon$ for the sets $m=2$, $m=1$, and $m=3$, respectively. The different panels in each figure correspond to the different mass ratios considered. We find that, qualitatively, the conclusions we drew from choosing the ringdown start time to be $t_0-t_{\rm peak}=20M$ remain valid across the a range of start times. The first overtone improves $\epsilon$ for the dominant mode of a set but not for the subdominant modes. Mode-mixing is of prime importance for the subdominant modes of a set but not the dominant mode, as expected. Ignoring mode-mixing gives wildly different remnant values compared to the true value. In fact, the $\epsilon$ values for the subdominant modes of the non-mode-mixing class of models are a conservative estimate as the best-fit values rail against the boundaries of our remnant space for most cases meaning that their correct estimates are much worse than what is indicated here. The fundamental mirror mode do not seem to improve the remnant estimates even though these model the modulations seen in the waveform and improve the mismatches significantly indicating that there are other features in the waveform dominating the error budget. These conclusions are true across mass ratios. 

\bibliography{multimode_ringdown}

\begin{thebibliography}{46}%
\makeatletter
\providecommand \@ifxundefined [1]{%
 \@ifx{#1\undefined}
}%
\providecommand \@ifnum [1]{%
 \ifnum #1\expandafter \@firstoftwo
 \else \expandafter \@secondoftwo
 \fi
}%
\providecommand \@ifx [1]{%
 \ifx #1\expandafter \@firstoftwo
 \else \expandafter \@secondoftwo
 \fi
}%
\providecommand \natexlab [1]{#1}%
\providecommand \enquote  [1]{``#1''}%
\providecommand \bibnamefont  [1]{#1}%
\providecommand \bibfnamefont [1]{#1}%
\providecommand \citenamefont [1]{#1}%
\providecommand \href@noop [0]{\@secondoftwo}%
\providecommand \href [0]{\begingroup \@sanitize@url \@href}%
\providecommand \@href[1]{\@@startlink{#1}\@@href}%
\providecommand \@@href[1]{\endgroup#1\@@endlink}%
\providecommand \@sanitize@url [0]{\catcode `\\12\catcode `\$12\catcode
  `\&12\catcode `\#12\catcode `\^12\catcode `\_12\catcode `\%12\relax}%
\providecommand \@@startlink[1]{}%
\providecommand \@@endlink[0]{}%
\providecommand \url  [0]{\begingroup\@sanitize@url \@url }%
\providecommand \@url [1]{\endgroup\@href {#1}{\urlprefix }}%
\providecommand \urlprefix  [0]{URL }%
\providecommand \Eprint [0]{\href }%
\providecommand \doibase [0]{http://dx.doi.org/}%
\providecommand \selectlanguage [0]{\@gobble}%
\providecommand \bibinfo  [0]{\@secondoftwo}%
\providecommand \bibfield  [0]{\@secondoftwo}%
\providecommand \translation [1]{[#1]}%
\providecommand \BibitemOpen [0]{}%
\providecommand \bibitemStop [0]{}%
\providecommand \bibitemNoStop [0]{.\EOS\space}%
\providecommand \EOS [0]{\spacefactor3000\relax}%
\providecommand \BibitemShut  [1]{\csname bibitem#1\endcsname}%
\let\auto@bib@innerbib\@empty
\bibitem [{\citenamefont {Vishveshwara}(1970)}]{Vishveshwara:1970zz}%
  \BibitemOpen
  \bibfield  {author} {\bibinfo {author} {\bibfnamefont {C.~V.}\ \bibnamefont
  {Vishveshwara}},\ }\href {\doibase 10.1038/227936a0} {\bibfield  {journal}
  {\bibinfo  {journal} {Nature}\ }\textbf {\bibinfo {volume} {227}},\ \bibinfo
  {pages} {936} (\bibinfo {year} {1970})}\BibitemShut {NoStop}%
\bibitem [{\citenamefont {Misner}\ \emph {et~al.}(1973)\citenamefont {Misner},
  \citenamefont {Thorne},\ and\ \citenamefont {Wheeler}}]{Misner:1974qy}%
  \BibitemOpen
  \bibfield  {author} {\bibinfo {author} {\bibfnamefont {C.~W.}\ \bibnamefont
  {Misner}}, \bibinfo {author} {\bibfnamefont {K.~S.}\ \bibnamefont {Thorne}},
  \ and\ \bibinfo {author} {\bibfnamefont {J.~A.}\ \bibnamefont {Wheeler}},\
  }\href@noop {} {\emph {\bibinfo {title} {{Gravitation}}}}\ (\bibinfo
  {publisher} {W. H. Freeman},\ \bibinfo {address} {San Francisco},\ \bibinfo
  {year} {1973})\BibitemShut {NoStop}%
\bibitem [{\citenamefont {Dreyer}\ \emph {et~al.}(2004)\citenamefont {Dreyer},
  \citenamefont {Kelly}, \citenamefont {Krishnan}, \citenamefont {Finn},
  \citenamefont {Garrison},\ and\ \citenamefont
  {Lopez-Aleman}}]{Dreyer:2003bv}%
  \BibitemOpen
  \bibfield  {author} {\bibinfo {author} {\bibfnamefont {O.}~\bibnamefont
  {Dreyer}}, \bibinfo {author} {\bibfnamefont {B.~J.}\ \bibnamefont {Kelly}},
  \bibinfo {author} {\bibfnamefont {B.}~\bibnamefont {Krishnan}}, \bibinfo
  {author} {\bibfnamefont {L.~S.}\ \bibnamefont {Finn}}, \bibinfo {author}
  {\bibfnamefont {D.}~\bibnamefont {Garrison}}, \ and\ \bibinfo {author}
  {\bibfnamefont {R.}~\bibnamefont {Lopez-Aleman}},\ }\href {\doibase
  10.1088/0264-9381/21/4/003} {\bibfield  {journal} {\bibinfo  {journal}
  {Class. Quant. Grav.}\ }\textbf {\bibinfo {volume} {21}},\ \bibinfo {pages}
  {787} (\bibinfo {year} {2004})},\ \Eprint
  {http://arxiv.org/abs/gr-qc/0309007} {arXiv:gr-qc/0309007} \BibitemShut
  {NoStop}%
\bibitem [{\citenamefont {Berti}\ \emph {et~al.}(2007)\citenamefont {Berti},
  \citenamefont {Cardoso}, \citenamefont {Cardoso},\ and\ \citenamefont
  {Cavaglia}}]{Berti:2007zu}%
  \BibitemOpen
  \bibfield  {author} {\bibinfo {author} {\bibfnamefont {E.}~\bibnamefont
  {Berti}}, \bibinfo {author} {\bibfnamefont {J.}~\bibnamefont {Cardoso}},
  \bibinfo {author} {\bibfnamefont {V.}~\bibnamefont {Cardoso}}, \ and\
  \bibinfo {author} {\bibfnamefont {M.}~\bibnamefont {Cavaglia}},\ }\href
  {\doibase 10.1103/PhysRevD.76.104044} {\bibfield  {journal} {\bibinfo
  {journal} {Phys. Rev. D}\ }\textbf {\bibinfo {volume} {76}},\ \bibinfo
  {pages} {104044} (\bibinfo {year} {2007})},\ \Eprint
  {http://arxiv.org/abs/0707.1202} {arXiv:0707.1202 [gr-qc]} \BibitemShut
  {NoStop}%
\bibitem [{\citenamefont {Gossan}\ \emph {et~al.}(2012)\citenamefont {Gossan},
  \citenamefont {Veitch},\ and\ \citenamefont {Sathyaprakash}}]{Gossan:2011ha}%
  \BibitemOpen
  \bibfield  {author} {\bibinfo {author} {\bibfnamefont {S.}~\bibnamefont
  {Gossan}}, \bibinfo {author} {\bibfnamefont {J.}~\bibnamefont {Veitch}}, \
  and\ \bibinfo {author} {\bibfnamefont {B.~S.}\ \bibnamefont
  {Sathyaprakash}},\ }\href {\doibase 10.1103/PhysRevD.85.124056} {\bibfield
  {journal} {\bibinfo  {journal} {Phys. Rev. D}\ }\textbf {\bibinfo {volume}
  {85}},\ \bibinfo {pages} {124056} (\bibinfo {year} {2012})},\ \Eprint
  {http://arxiv.org/abs/1111.5819} {arXiv:1111.5819 [gr-qc]} \BibitemShut
  {NoStop}%
\bibitem [{\citenamefont {Cardoso}\ and\ \citenamefont
  {Pani}(2019)}]{Cardoso:2019rvt}%
  \BibitemOpen
  \bibfield  {author} {\bibinfo {author} {\bibfnamefont {V.}~\bibnamefont
  {Cardoso}}\ and\ \bibinfo {author} {\bibfnamefont {P.}~\bibnamefont {Pani}},\
  }\href {\doibase 10.1007/s41114-019-0020-4} {\bibfield  {journal} {\bibinfo
  {journal} {Living Rev. Rel.}\ }\textbf {\bibinfo {volume} {22}},\ \bibinfo
  {pages} {4} (\bibinfo {year} {2019})},\ \Eprint
  {http://arxiv.org/abs/1904.05363} {arXiv:1904.05363 [gr-qc]} \BibitemShut
  {NoStop}%
\bibitem [{\citenamefont {Carballo-Rubio}\ \emph {et~al.}(2018)\citenamefont
  {Carballo-Rubio}, \citenamefont {Di~Filippo}, \citenamefont {Liberati},\ and\
  \citenamefont {Visser}}]{Carballo-Rubio:2018jzw}%
  \BibitemOpen
  \bibfield  {author} {\bibinfo {author} {\bibfnamefont {R.}~\bibnamefont
  {Carballo-Rubio}}, \bibinfo {author} {\bibfnamefont {F.}~\bibnamefont
  {Di~Filippo}}, \bibinfo {author} {\bibfnamefont {S.}~\bibnamefont
  {Liberati}}, \ and\ \bibinfo {author} {\bibfnamefont {M.}~\bibnamefont
  {Visser}},\ }\href {\doibase 10.1103/PhysRevD.98.124009} {\bibfield
  {journal} {\bibinfo  {journal} {Phys. Rev. D}\ }\textbf {\bibinfo {volume}
  {98}},\ \bibinfo {pages} {124009} (\bibinfo {year} {2018})},\ \Eprint
  {http://arxiv.org/abs/1809.08238} {arXiv:1809.08238 [gr-qc]} \BibitemShut
  {NoStop}%
\bibitem [{\citenamefont {Raposo}\ \emph {et~al.}(2019)\citenamefont {Raposo},
  \citenamefont {Pani},\ and\ \citenamefont {Emparan}}]{Raposo:2018xkf}%
  \BibitemOpen
  \bibfield  {author} {\bibinfo {author} {\bibfnamefont {G.}~\bibnamefont
  {Raposo}}, \bibinfo {author} {\bibfnamefont {P.}~\bibnamefont {Pani}}, \ and\
  \bibinfo {author} {\bibfnamefont {R.}~\bibnamefont {Emparan}},\ }\href
  {\doibase 10.1103/PhysRevD.99.104050} {\bibfield  {journal} {\bibinfo
  {journal} {Phys. Rev. D}\ }\textbf {\bibinfo {volume} {99}},\ \bibinfo
  {pages} {104050} (\bibinfo {year} {2019})},\ \Eprint
  {http://arxiv.org/abs/1812.07615} {arXiv:1812.07615 [gr-qc]} \BibitemShut
  {NoStop}%
\bibitem [{\citenamefont {Abbott}\ \emph {et~al.}(2016)\citenamefont {Abbott}
  \emph {et~al.}}]{TheLIGOScientific:2016src}%
  \BibitemOpen
  \bibfield  {author} {\bibinfo {author} {\bibfnamefont {B.~P.}\ \bibnamefont
  {Abbott}} \emph {et~al.} (\bibinfo {collaboration} {LIGO Scientific,
  Virgo}),\ }\href {\doibase 10.1103/PhysRevLett.116.221101} {\bibfield
  {journal} {\bibinfo  {journal} {Phys. Rev. Lett.}\ }\textbf {\bibinfo
  {volume} {116}},\ \bibinfo {pages} {221101} (\bibinfo {year} {2016})},\
  \bibinfo {note} {[Erratum: Phys.Rev.Lett. 121, 129902 (2018)]},\ \Eprint
  {http://arxiv.org/abs/1602.03841} {arXiv:1602.03841 [gr-qc]} \BibitemShut
  {NoStop}%
\bibitem [{\citenamefont {Ghosh}\ \emph {et~al.}(2018)\citenamefont {Ghosh},
  \citenamefont {Johnson-Mcdaniel}, \citenamefont {Ghosh}, \citenamefont
  {Mishra}, \citenamefont {Ajith}, \citenamefont {Del~Pozzo}, \citenamefont
  {Berry}, \citenamefont {Nielsen},\ and\ \citenamefont
  {London}}]{Ghosh:2017gfp}%
  \BibitemOpen
  \bibfield  {author} {\bibinfo {author} {\bibfnamefont {A.}~\bibnamefont
  {Ghosh}}, \bibinfo {author} {\bibfnamefont {N.~K.}\ \bibnamefont
  {Johnson-Mcdaniel}}, \bibinfo {author} {\bibfnamefont {A.}~\bibnamefont
  {Ghosh}}, \bibinfo {author} {\bibfnamefont {C.~K.}\ \bibnamefont {Mishra}},
  \bibinfo {author} {\bibfnamefont {P.}~\bibnamefont {Ajith}}, \bibinfo
  {author} {\bibfnamefont {W.}~\bibnamefont {Del~Pozzo}}, \bibinfo {author}
  {\bibfnamefont {C.~P.~L.}\ \bibnamefont {Berry}}, \bibinfo {author}
  {\bibfnamefont {A.~B.}\ \bibnamefont {Nielsen}}, \ and\ \bibinfo {author}
  {\bibfnamefont {L.}~\bibnamefont {London}},\ }\href {\doibase
  10.1088/1361-6382/aa972e} {\bibfield  {journal} {\bibinfo  {journal} {Class.
  Quant. Grav.}\ }\textbf {\bibinfo {volume} {35}},\ \bibinfo {pages} {014002}
  (\bibinfo {year} {2018})},\ \Eprint {http://arxiv.org/abs/1704.06784}
  {arXiv:1704.06784 [gr-qc]} \BibitemShut {NoStop}%
\bibitem [{\citenamefont {Berti}\ \emph {et~al.}(2018)\citenamefont {Berti},
  \citenamefont {Yagi}, \citenamefont {Yang},\ and\ \citenamefont
  {Yunes}}]{Berti:2018vdi}%
  \BibitemOpen
  \bibfield  {author} {\bibinfo {author} {\bibfnamefont {E.}~\bibnamefont
  {Berti}}, \bibinfo {author} {\bibfnamefont {K.}~\bibnamefont {Yagi}},
  \bibinfo {author} {\bibfnamefont {H.}~\bibnamefont {Yang}}, \ and\ \bibinfo
  {author} {\bibfnamefont {N.}~\bibnamefont {Yunes}},\ }\href {\doibase
  10.1007/s10714-018-2372-6} {\bibfield  {journal} {\bibinfo  {journal} {Gen.
  Rel. Grav.}\ }\textbf {\bibinfo {volume} {50}},\ \bibinfo {pages} {49}
  (\bibinfo {year} {2018})},\ \Eprint {http://arxiv.org/abs/1801.03587}
  {arXiv:1801.03587 [gr-qc]} \BibitemShut {NoStop}%
\bibitem [{\citenamefont {Isi}\ and\ \citenamefont {Farr}(2021)}]{Isi:2021iql}%
  \BibitemOpen
  \bibfield  {author} {\bibinfo {author} {\bibfnamefont {M.}~\bibnamefont
  {Isi}}\ and\ \bibinfo {author} {\bibfnamefont {W.~M.}\ \bibnamefont {Farr}},\
  }\href@noop {} {\  (\bibinfo {year} {2021})},\ \Eprint
  {http://arxiv.org/abs/2107.05609} {arXiv:2107.05609 [gr-qc]} \BibitemShut
  {NoStop}%
\bibitem [{\citenamefont {Carullo}\ \emph {et~al.}(2019)\citenamefont
  {Carullo}, \citenamefont {Del~Pozzo},\ and\ \citenamefont
  {Veitch}}]{Carullo:2019flw}%
  \BibitemOpen
  \bibfield  {author} {\bibinfo {author} {\bibfnamefont {G.}~\bibnamefont
  {Carullo}}, \bibinfo {author} {\bibfnamefont {W.}~\bibnamefont {Del~Pozzo}},
  \ and\ \bibinfo {author} {\bibfnamefont {J.}~\bibnamefont {Veitch}},\ }\href
  {\doibase 10.1103/PhysRevD.99.123029} {\bibfield  {journal} {\bibinfo
  {journal} {Phys. Rev. D}\ }\textbf {\bibinfo {volume} {99}},\ \bibinfo
  {pages} {123029} (\bibinfo {year} {2019})},\ \bibinfo {note} {[Erratum:
  Phys.Rev.D 100, 089903 (2019)]},\ \Eprint {http://arxiv.org/abs/1902.07527}
  {arXiv:1902.07527 [gr-qc]} \BibitemShut {NoStop}%
\bibitem [{\citenamefont {Isi}\ \emph {et~al.}(2019)\citenamefont {Isi},
  \citenamefont {Giesler}, \citenamefont {Farr}, \citenamefont {Scheel},\ and\
  \citenamefont {Teukolsky}}]{Isi:2019aib}%
  \BibitemOpen
  \bibfield  {author} {\bibinfo {author} {\bibfnamefont {M.}~\bibnamefont
  {Isi}}, \bibinfo {author} {\bibfnamefont {M.}~\bibnamefont {Giesler}},
  \bibinfo {author} {\bibfnamefont {W.~M.}\ \bibnamefont {Farr}}, \bibinfo
  {author} {\bibfnamefont {M.~A.}\ \bibnamefont {Scheel}}, \ and\ \bibinfo
  {author} {\bibfnamefont {S.~A.}\ \bibnamefont {Teukolsky}},\ }\href {\doibase
  10.1103/PhysRevLett.123.111102} {\bibfield  {journal} {\bibinfo  {journal}
  {Phys. Rev. Lett.}\ }\textbf {\bibinfo {volume} {123}},\ \bibinfo {pages}
  {111102} (\bibinfo {year} {2019})},\ \Eprint
  {http://arxiv.org/abs/1905.00869} {arXiv:1905.00869 [gr-qc]} \BibitemShut
  {NoStop}%
\bibitem [{\citenamefont {Capano}\ \emph {et~al.}(2021)\citenamefont {Capano},
  \citenamefont {Cabero}, \citenamefont {Westerweck}, \citenamefont {Abedi},
  \citenamefont {Kastha}, \citenamefont {Nitz}, \citenamefont {Nielsen},\ and\
  \citenamefont {Krishnan}}]{Capano:2021etf}%
  \BibitemOpen
  \bibfield  {author} {\bibinfo {author} {\bibfnamefont {C.~D.}\ \bibnamefont
  {Capano}}, \bibinfo {author} {\bibfnamefont {M.}~\bibnamefont {Cabero}},
  \bibinfo {author} {\bibfnamefont {J.}~\bibnamefont {Westerweck}}, \bibinfo
  {author} {\bibfnamefont {J.}~\bibnamefont {Abedi}}, \bibinfo {author}
  {\bibfnamefont {S.}~\bibnamefont {Kastha}}, \bibinfo {author} {\bibfnamefont
  {A.~H.}\ \bibnamefont {Nitz}}, \bibinfo {author} {\bibfnamefont {A.~B.}\
  \bibnamefont {Nielsen}}, \ and\ \bibinfo {author} {\bibfnamefont
  {B.}~\bibnamefont {Krishnan}},\ }\href@noop {} {\  (\bibinfo {year}
  {2021})},\ \Eprint {http://arxiv.org/abs/2105.05238} {arXiv:2105.05238
  [gr-qc]} \BibitemShut {NoStop}%
\bibitem [{\citenamefont {LSC}(2016)}]{aPlusWP}%
  \BibitemOpen
  \bibfield  {author} {\bibinfo {author} {\bibnamefont {LSC}}\ }(\bibinfo
  {year} {2016})\BibitemShut {NoStop}%
\bibitem [{\citenamefont {Kuns}\ \emph {et~al.}(2020)\citenamefont {Kuns},
  \citenamefont {Srivastava}, \citenamefont {Hall}, \citenamefont {Evans},\
  and\ \citenamefont {Ballmer}}]{AdV_KAGRA_plus}%
  \BibitemOpen
  \bibfield  {author} {\bibinfo {author} {\bibfnamefont {K.}~\bibnamefont
  {Kuns}}, \bibinfo {author} {\bibfnamefont {V.}~\bibnamefont {Srivastava}},
  \bibinfo {author} {\bibfnamefont {E.}~\bibnamefont {Hall}}, \bibinfo {author}
  {\bibfnamefont {M.}~\bibnamefont {Evans}}, \ and\ \bibinfo {author}
  {\bibfnamefont {S.}~\bibnamefont {Ballmer}}\ }(\bibinfo {year}
  {2020})\BibitemShut {NoStop}%
\bibitem [{\citenamefont {Adhikari}\ \emph {et~al.}(2019)\citenamefont
  {Adhikari} \emph {et~al.}}]{Adhikari:2019zpy}%
  \BibitemOpen
  \bibfield  {author} {\bibinfo {author} {\bibfnamefont {R.~X.}\ \bibnamefont
  {Adhikari}} \emph {et~al.},\ }\href {\doibase 10.1088/1361-6382/ab3cff}
  {\bibfield  {journal} {\bibinfo  {journal} {Class. Quant. Grav.}\ }\textbf
  {\bibinfo {volume} {36}},\ \bibinfo {pages} {245010} (\bibinfo {year}
  {2019})},\ \Eprint {http://arxiv.org/abs/1905.02842} {arXiv:1905.02842
  [astro-ph.HE]} \BibitemShut {NoStop}%
\bibitem [{\citenamefont {Punturo}\ \emph {et~al.}(2010)\citenamefont {Punturo}
  \emph {et~al.}}]{Punturo:2010zz}%
  \BibitemOpen
  \bibfield  {author} {\bibinfo {author} {\bibfnamefont {M.}~\bibnamefont
  {Punturo}} \emph {et~al.},\ }\href {\doibase 10.1088/0264-9381/27/19/194002}
  {\bibfield  {journal} {\bibinfo  {journal} {Class. Quant. Grav.}\ }\textbf
  {\bibinfo {volume} {27}},\ \bibinfo {pages} {194002} (\bibinfo {year}
  {2010})}\BibitemShut {NoStop}%
\bibitem [{\citenamefont {Abbott}\ \emph {et~al.}(2017)\citenamefont {Abbott}
  \emph {et~al.}}]{LIGOScientific:2016wof}%
  \BibitemOpen
  \bibfield  {author} {\bibinfo {author} {\bibfnamefont {B.~P.}\ \bibnamefont
  {Abbott}} \emph {et~al.} (\bibinfo {collaboration} {LIGO Scientific}),\
  }\href {\doibase 10.1088/1361-6382/aa51f4} {\bibfield  {journal} {\bibinfo
  {journal} {Class. Quant. Grav.}\ }\textbf {\bibinfo {volume} {34}},\ \bibinfo
  {pages} {044001} (\bibinfo {year} {2017})},\ \Eprint
  {http://arxiv.org/abs/1607.08697} {arXiv:1607.08697 [astro-ph.IM]}
  \BibitemShut {NoStop}%
\bibitem [{\citenamefont {Ackley}\ \emph {et~al.}(2020)\citenamefont {Ackley}
  \emph {et~al.}}]{Ackley:2020atn}%
  \BibitemOpen
  \bibfield  {author} {\bibinfo {author} {\bibfnamefont {K.}~\bibnamefont
  {Ackley}} \emph {et~al.},\ }\href {\doibase 10.1017/pasa.2020.39} {\bibfield
  {journal} {\bibinfo  {journal} {Publ. Astron. Soc. Austral.}\ }\textbf
  {\bibinfo {volume} {37}},\ \bibinfo {pages} {e047} (\bibinfo {year}
  {2020})},\ \Eprint {http://arxiv.org/abs/2007.03128} {arXiv:2007.03128
  [astro-ph.HE]} \BibitemShut {NoStop}%
\bibitem [{\citenamefont {Amaro-Seoane}\ \emph {et~al.}(2017)\citenamefont
  {Amaro-Seoane} \emph {et~al.}}]{LISA:2017pwj}%
  \BibitemOpen
  \bibfield  {author} {\bibinfo {author} {\bibfnamefont {P.}~\bibnamefont
  {Amaro-Seoane}} \emph {et~al.} (\bibinfo {collaboration} {LISA}),\
  }\href@noop {} {\  (\bibinfo {year} {2017})},\ \Eprint
  {http://arxiv.org/abs/1702.00786} {arXiv:1702.00786 [astro-ph.IM]}
  \BibitemShut {NoStop}%
\bibitem [{\citenamefont {Kamaretsos}\ \emph {et~al.}(2012)\citenamefont
  {Kamaretsos}, \citenamefont {Hannam}, \citenamefont {Husa},\ and\
  \citenamefont {Sathyaprakash}}]{Kamaretsos:2011um}%
  \BibitemOpen
  \bibfield  {author} {\bibinfo {author} {\bibfnamefont {I.}~\bibnamefont
  {Kamaretsos}}, \bibinfo {author} {\bibfnamefont {M.}~\bibnamefont {Hannam}},
  \bibinfo {author} {\bibfnamefont {S.}~\bibnamefont {Husa}}, \ and\ \bibinfo
  {author} {\bibfnamefont {B.~S.}\ \bibnamefont {Sathyaprakash}},\ }\href
  {\doibase 10.1103/PhysRevD.85.024018} {\bibfield  {journal} {\bibinfo
  {journal} {Phys. Rev. D}\ }\textbf {\bibinfo {volume} {85}},\ \bibinfo
  {pages} {024018} (\bibinfo {year} {2012})},\ \Eprint
  {http://arxiv.org/abs/1107.0854} {arXiv:1107.0854 [gr-qc]} \BibitemShut
  {NoStop}%
\bibitem [{\citenamefont {Baibhav}\ \emph {et~al.}(2018)\citenamefont
  {Baibhav}, \citenamefont {Berti}, \citenamefont {Cardoso},\ and\
  \citenamefont {Khanna}}]{Baibhav:2017jhs}%
  \BibitemOpen
  \bibfield  {author} {\bibinfo {author} {\bibfnamefont {V.}~\bibnamefont
  {Baibhav}}, \bibinfo {author} {\bibfnamefont {E.}~\bibnamefont {Berti}},
  \bibinfo {author} {\bibfnamefont {V.}~\bibnamefont {Cardoso}}, \ and\
  \bibinfo {author} {\bibfnamefont {G.}~\bibnamefont {Khanna}},\ }\href
  {\doibase 10.1103/PhysRevD.97.044048} {\bibfield  {journal} {\bibinfo
  {journal} {Phys. Rev. D}\ }\textbf {\bibinfo {volume} {97}},\ \bibinfo
  {pages} {044048} (\bibinfo {year} {2018})},\ \Eprint
  {http://arxiv.org/abs/1710.02156} {arXiv:1710.02156 [gr-qc]} \BibitemShut
  {NoStop}%
\bibitem [{\citenamefont {London}(2020)}]{London:2018gaq}%
  \BibitemOpen
  \bibfield  {author} {\bibinfo {author} {\bibfnamefont {L.~T.}\ \bibnamefont
  {London}},\ }\href {\doibase 10.1103/PhysRevD.102.084052} {\bibfield
  {journal} {\bibinfo  {journal} {Phys. Rev. D}\ }\textbf {\bibinfo {volume}
  {102}},\ \bibinfo {pages} {084052} (\bibinfo {year} {2020})},\ \Eprint
  {http://arxiv.org/abs/1801.08208} {arXiv:1801.08208 [gr-qc]} \BibitemShut
  {NoStop}%
\bibitem [{\citenamefont {Ota}\ and\ \citenamefont
  {Chirenti}(2020)}]{Ota:2019bzl}%
  \BibitemOpen
  \bibfield  {author} {\bibinfo {author} {\bibfnamefont {I.}~\bibnamefont
  {Ota}}\ and\ \bibinfo {author} {\bibfnamefont {C.}~\bibnamefont {Chirenti}},\
  }\href {\doibase 10.1103/PhysRevD.101.104005} {\bibfield  {journal} {\bibinfo
   {journal} {Phys. Rev. D}\ }\textbf {\bibinfo {volume} {101}},\ \bibinfo
  {pages} {104005} (\bibinfo {year} {2020})},\ \Eprint
  {http://arxiv.org/abs/1911.00440} {arXiv:1911.00440 [gr-qc]} \BibitemShut
  {NoStop}%
\bibitem [{\citenamefont {Giesler}\ \emph {et~al.}(2019)\citenamefont
  {Giesler}, \citenamefont {Isi}, \citenamefont {Scheel},\ and\ \citenamefont
  {Teukolsky}}]{Giesler:2019uxc}%
  \BibitemOpen
  \bibfield  {author} {\bibinfo {author} {\bibfnamefont {M.}~\bibnamefont
  {Giesler}}, \bibinfo {author} {\bibfnamefont {M.}~\bibnamefont {Isi}},
  \bibinfo {author} {\bibfnamefont {M.~A.}\ \bibnamefont {Scheel}}, \ and\
  \bibinfo {author} {\bibfnamefont {S.}~\bibnamefont {Teukolsky}},\ }\href
  {\doibase 10.1103/PhysRevX.9.041060} {\bibfield  {journal} {\bibinfo
  {journal} {Phys. Rev. X}\ }\textbf {\bibinfo {volume} {9}},\ \bibinfo {pages}
  {041060} (\bibinfo {year} {2019})},\ \Eprint
  {http://arxiv.org/abs/1903.08284} {arXiv:1903.08284 [gr-qc]} \BibitemShut
  {NoStop}%
\bibitem [{\citenamefont {Bhagwat}\ \emph {et~al.}(2020)\citenamefont
  {Bhagwat}, \citenamefont {Forteza}, \citenamefont {Pani},\ and\ \citenamefont
  {Ferrari}}]{Bhagwat:2019dtm}%
  \BibitemOpen
  \bibfield  {author} {\bibinfo {author} {\bibfnamefont {S.}~\bibnamefont
  {Bhagwat}}, \bibinfo {author} {\bibfnamefont {X.~J.}\ \bibnamefont
  {Forteza}}, \bibinfo {author} {\bibfnamefont {P.}~\bibnamefont {Pani}}, \
  and\ \bibinfo {author} {\bibfnamefont {V.}~\bibnamefont {Ferrari}},\ }\href
  {\doibase 10.1103/PhysRevD.101.044033} {\bibfield  {journal} {\bibinfo
  {journal} {Phys. Rev. D}\ }\textbf {\bibinfo {volume} {101}},\ \bibinfo
  {pages} {044033} (\bibinfo {year} {2020})},\ \Eprint
  {http://arxiv.org/abs/1910.08708} {arXiv:1910.08708 [gr-qc]} \BibitemShut
  {NoStop}%
\bibitem [{\citenamefont {Jim\'enez~Forteza}\ \emph {et~al.}(2020)\citenamefont
  {Jim\'enez~Forteza}, \citenamefont {Bhagwat}, \citenamefont {Pani},\ and\
  \citenamefont {Ferrari}}]{Forteza:2020cve}%
  \BibitemOpen
  \bibfield  {author} {\bibinfo {author} {\bibfnamefont {X.}~\bibnamefont
  {Jim\'enez~Forteza}}, \bibinfo {author} {\bibfnamefont {S.}~\bibnamefont
  {Bhagwat}}, \bibinfo {author} {\bibfnamefont {P.}~\bibnamefont {Pani}}, \
  and\ \bibinfo {author} {\bibfnamefont {V.}~\bibnamefont {Ferrari}},\ }\href
  {\doibase 10.1103/PhysRevD.102.044053} {\bibfield  {journal} {\bibinfo
  {journal} {Phys. Rev. D}\ }\textbf {\bibinfo {volume} {102}},\ \bibinfo
  {pages} {044053} (\bibinfo {year} {2020})},\ \Eprint
  {http://arxiv.org/abs/2005.03260} {arXiv:2005.03260 [gr-qc]} \BibitemShut
  {NoStop}%
\bibitem [{\citenamefont {Hughes}\ \emph {et~al.}(2019)\citenamefont {Hughes},
  \citenamefont {Apte}, \citenamefont {Khanna},\ and\ \citenamefont
  {Lim}}]{Hughes:2019zmt}%
  \BibitemOpen
  \bibfield  {author} {\bibinfo {author} {\bibfnamefont {S.~A.}\ \bibnamefont
  {Hughes}}, \bibinfo {author} {\bibfnamefont {A.}~\bibnamefont {Apte}},
  \bibinfo {author} {\bibfnamefont {G.}~\bibnamefont {Khanna}}, \ and\ \bibinfo
  {author} {\bibfnamefont {H.}~\bibnamefont {Lim}},\ }\href {\doibase
  10.1103/PhysRevLett.123.161101} {\bibfield  {journal} {\bibinfo  {journal}
  {Phys. Rev. Lett.}\ }\textbf {\bibinfo {volume} {123}},\ \bibinfo {pages}
  {161101} (\bibinfo {year} {2019})},\ \Eprint
  {http://arxiv.org/abs/1901.05900} {arXiv:1901.05900 [gr-qc]} \BibitemShut
  {NoStop}%
\bibitem [{\citenamefont {Lim}\ \emph {et~al.}(2019)\citenamefont {Lim},
  \citenamefont {Khanna}, \citenamefont {Apte},\ and\ \citenamefont
  {Hughes}}]{Lim:2019xrb}%
  \BibitemOpen
  \bibfield  {author} {\bibinfo {author} {\bibfnamefont {H.}~\bibnamefont
  {Lim}}, \bibinfo {author} {\bibfnamefont {G.}~\bibnamefont {Khanna}},
  \bibinfo {author} {\bibfnamefont {A.}~\bibnamefont {Apte}}, \ and\ \bibinfo
  {author} {\bibfnamefont {S.~A.}\ \bibnamefont {Hughes}},\ }\href {\doibase
  10.1103/PhysRevD.100.084032} {\bibfield  {journal} {\bibinfo  {journal}
  {Phys. Rev. D}\ }\textbf {\bibinfo {volume} {100}},\ \bibinfo {pages}
  {084032} (\bibinfo {year} {2019})},\ \Eprint
  {http://arxiv.org/abs/1901.05902} {arXiv:1901.05902 [gr-qc]} \BibitemShut
  {NoStop}%
\bibitem [{\citenamefont {Dhani}(2021)}]{Dhani:2020nik}%
  \BibitemOpen
  \bibfield  {author} {\bibinfo {author} {\bibfnamefont {A.}~\bibnamefont
  {Dhani}},\ }\href {\doibase 10.1103/PhysRevD.103.104048} {\bibfield
  {journal} {\bibinfo  {journal} {Phys. Rev. D}\ }\textbf {\bibinfo {volume}
  {103}},\ \bibinfo {pages} {104048} (\bibinfo {year} {2021})},\ \Eprint
  {http://arxiv.org/abs/2010.08602} {arXiv:2010.08602 [gr-qc]} \BibitemShut
  {NoStop}%
\bibitem [{\citenamefont {Forteza}\ and\ \citenamefont
  {Mourier}(2021)}]{Forteza:2021wfq}%
  \BibitemOpen
  \bibfield  {author} {\bibinfo {author} {\bibfnamefont {X.~J.}\ \bibnamefont
  {Forteza}}\ and\ \bibinfo {author} {\bibfnamefont {P.}~\bibnamefont
  {Mourier}},\ }\href@noop {} {\  (\bibinfo {year} {2021})},\ \Eprint
  {http://arxiv.org/abs/2107.11829} {arXiv:2107.11829 [gr-qc]} \BibitemShut
  {NoStop}%
\bibitem [{\citenamefont {Buonanno}\ \emph {et~al.}(2007)\citenamefont
  {Buonanno}, \citenamefont {Cook},\ and\ \citenamefont
  {Pretorius}}]{Buonanno:2006ui}%
  \BibitemOpen
  \bibfield  {author} {\bibinfo {author} {\bibfnamefont {A.}~\bibnamefont
  {Buonanno}}, \bibinfo {author} {\bibfnamefont {G.~B.}\ \bibnamefont {Cook}},
  \ and\ \bibinfo {author} {\bibfnamefont {F.}~\bibnamefont {Pretorius}},\
  }\href {\doibase 10.1103/PhysRevD.75.124018} {\bibfield  {journal} {\bibinfo
  {journal} {Phys. Rev. D}\ }\textbf {\bibinfo {volume} {75}},\ \bibinfo
  {pages} {124018} (\bibinfo {year} {2007})},\ \Eprint
  {http://arxiv.org/abs/gr-qc/0610122} {arXiv:gr-qc/0610122} \BibitemShut
  {NoStop}%
\bibitem [{\citenamefont {Pan}\ \emph {et~al.}(2014)\citenamefont {Pan},
  \citenamefont {Buonanno}, \citenamefont {Taracchini}, \citenamefont {Kidder},
  \citenamefont {Mrou\'e}, \citenamefont {Pfeiffer}, \citenamefont {Scheel},\
  and\ \citenamefont {Szil\'agyi}}]{Pan:2013rra}%
  \BibitemOpen
  \bibfield  {author} {\bibinfo {author} {\bibfnamefont {Y.}~\bibnamefont
  {Pan}}, \bibinfo {author} {\bibfnamefont {A.}~\bibnamefont {Buonanno}},
  \bibinfo {author} {\bibfnamefont {A.}~\bibnamefont {Taracchini}}, \bibinfo
  {author} {\bibfnamefont {L.~E.}\ \bibnamefont {Kidder}}, \bibinfo {author}
  {\bibfnamefont {A.~H.}\ \bibnamefont {Mrou\'e}}, \bibinfo {author}
  {\bibfnamefont {H.~P.}\ \bibnamefont {Pfeiffer}}, \bibinfo {author}
  {\bibfnamefont {M.~A.}\ \bibnamefont {Scheel}}, \ and\ \bibinfo {author}
  {\bibfnamefont {B.}~\bibnamefont {Szil\'agyi}},\ }\href {\doibase
  10.1103/PhysRevD.89.084006} {\bibfield  {journal} {\bibinfo  {journal} {Phys.
  Rev. D}\ }\textbf {\bibinfo {volume} {89}},\ \bibinfo {pages} {084006}
  (\bibinfo {year} {2014})},\ \Eprint {http://arxiv.org/abs/1307.6232}
  {arXiv:1307.6232 [gr-qc]} \BibitemShut {NoStop}%
\bibitem [{\citenamefont {Taracchini}\ \emph {et~al.}(2014)\citenamefont
  {Taracchini} \emph {et~al.}}]{Taracchini:2013rva}%
  \BibitemOpen
  \bibfield  {author} {\bibinfo {author} {\bibfnamefont {A.}~\bibnamefont
  {Taracchini}} \emph {et~al.},\ }\href {\doibase 10.1103/PhysRevD.89.061502}
  {\bibfield  {journal} {\bibinfo  {journal} {Phys. Rev. D}\ }\textbf {\bibinfo
  {volume} {89}},\ \bibinfo {pages} {061502} (\bibinfo {year} {2014})},\
  \Eprint {http://arxiv.org/abs/1311.2544} {arXiv:1311.2544 [gr-qc]}
  \BibitemShut {NoStop}%
\bibitem [{\citenamefont {Babak}\ \emph {et~al.}(2017)\citenamefont {Babak},
  \citenamefont {Taracchini},\ and\ \citenamefont {Buonanno}}]{Babak:2016tgq}%
  \BibitemOpen
  \bibfield  {author} {\bibinfo {author} {\bibfnamefont {S.}~\bibnamefont
  {Babak}}, \bibinfo {author} {\bibfnamefont {A.}~\bibnamefont {Taracchini}}, \
  and\ \bibinfo {author} {\bibfnamefont {A.}~\bibnamefont {Buonanno}},\ }\href
  {\doibase 10.1103/PhysRevD.95.024010} {\bibfield  {journal} {\bibinfo
  {journal} {Phys. Rev. D}\ }\textbf {\bibinfo {volume} {95}},\ \bibinfo
  {pages} {024010} (\bibinfo {year} {2017})},\ \Eprint
  {http://arxiv.org/abs/1607.05661} {arXiv:1607.05661 [gr-qc]} \BibitemShut
  {NoStop}%
\bibitem [{\citenamefont {Teukolsky}(1973)}]{Teukolsky:1973ha}%
  \BibitemOpen
  \bibfield  {author} {\bibinfo {author} {\bibfnamefont {S.~A.}\ \bibnamefont
  {Teukolsky}},\ }\href {\doibase 10.1086/152444} {\bibfield  {journal}
  {\bibinfo  {journal} {Astrophys. J.}\ }\textbf {\bibinfo {volume} {185}},\
  \bibinfo {pages} {635} (\bibinfo {year} {1973})}\BibitemShut {NoStop}%
\bibitem [{\citenamefont {Berti}\ \emph {et~al.}(2006)\citenamefont {Berti},
  \citenamefont {Cardoso},\ and\ \citenamefont {Casals}}]{Berti:2005gp}%
  \BibitemOpen
  \bibfield  {author} {\bibinfo {author} {\bibfnamefont {E.}~\bibnamefont
  {Berti}}, \bibinfo {author} {\bibfnamefont {V.}~\bibnamefont {Cardoso}}, \
  and\ \bibinfo {author} {\bibfnamefont {M.}~\bibnamefont {Casals}},\ }\href
  {\doibase 10.1103/PhysRevD.73.109902} {\bibfield  {journal} {\bibinfo
  {journal} {Phys. Rev. D}\ }\textbf {\bibinfo {volume} {73}},\ \bibinfo
  {pages} {024013} (\bibinfo {year} {2006})},\ \bibinfo {note} {[Erratum:
  Phys.Rev.D 73, 109902 (2006)]},\ \Eprint {http://arxiv.org/abs/gr-qc/0511111}
  {arXiv:gr-qc/0511111} \BibitemShut {NoStop}%
\bibitem [{\citenamefont {Berti}\ and\ \citenamefont
  {Klein}(2014)}]{Berti:2014fga}%
  \BibitemOpen
  \bibfield  {author} {\bibinfo {author} {\bibfnamefont {E.}~\bibnamefont
  {Berti}}\ and\ \bibinfo {author} {\bibfnamefont {A.}~\bibnamefont {Klein}},\
  }\href {\doibase 10.1103/PhysRevD.90.064012} {\bibfield  {journal} {\bibinfo
  {journal} {Phys. Rev. D}\ }\textbf {\bibinfo {volume} {90}},\ \bibinfo
  {pages} {064012} (\bibinfo {year} {2014})},\ \Eprint
  {http://arxiv.org/abs/1408.1860} {arXiv:1408.1860 [gr-qc]} \BibitemShut
  {NoStop}%
\bibitem [{\citenamefont {Kelly}\ and\ \citenamefont
  {Baker}(2013)}]{Kelly:2012nd}%
  \BibitemOpen
  \bibfield  {author} {\bibinfo {author} {\bibfnamefont {B.~J.}\ \bibnamefont
  {Kelly}}\ and\ \bibinfo {author} {\bibfnamefont {J.~G.}\ \bibnamefont
  {Baker}},\ }\href {\doibase 10.1103/PhysRevD.87.084004} {\bibfield  {journal}
  {\bibinfo  {journal} {Phys. Rev. D}\ }\textbf {\bibinfo {volume} {87}},\
  \bibinfo {pages} {084004} (\bibinfo {year} {2013})},\ \Eprint
  {http://arxiv.org/abs/1212.5553} {arXiv:1212.5553 [gr-qc]} \BibitemShut
  {NoStop}%
\bibitem [{\citenamefont {Mehta}\ \emph {et~al.}(2019)\citenamefont {Mehta},
  \citenamefont {Tiwari}, \citenamefont {Johnson-McDaniel}, \citenamefont
  {Mishra}, \citenamefont {Varma},\ and\ \citenamefont
  {Ajith}}]{Mehta:2019wxm}%
  \BibitemOpen
  \bibfield  {author} {\bibinfo {author} {\bibfnamefont {A.~K.}\ \bibnamefont
  {Mehta}}, \bibinfo {author} {\bibfnamefont {P.}~\bibnamefont {Tiwari}},
  \bibinfo {author} {\bibfnamefont {N.~K.}\ \bibnamefont {Johnson-McDaniel}},
  \bibinfo {author} {\bibfnamefont {C.~K.}\ \bibnamefont {Mishra}}, \bibinfo
  {author} {\bibfnamefont {V.}~\bibnamefont {Varma}}, \ and\ \bibinfo {author}
  {\bibfnamefont {P.}~\bibnamefont {Ajith}},\ }\href {\doibase
  10.1103/PhysRevD.100.024032} {\bibfield  {journal} {\bibinfo  {journal}
  {Phys. Rev. D}\ }\textbf {\bibinfo {volume} {100}},\ \bibinfo {pages}
  {024032} (\bibinfo {year} {2019})},\ \Eprint
  {http://arxiv.org/abs/1902.02731} {arXiv:1902.02731 [gr-qc]} \BibitemShut
  {NoStop}%
\bibitem [{\citenamefont {London}\ \emph {et~al.}(2014)\citenamefont {London},
  \citenamefont {Shoemaker},\ and\ \citenamefont {Healy}}]{London:2014cma}%
  \BibitemOpen
  \bibfield  {author} {\bibinfo {author} {\bibfnamefont {L.}~\bibnamefont
  {London}}, \bibinfo {author} {\bibfnamefont {D.}~\bibnamefont {Shoemaker}}, \
  and\ \bibinfo {author} {\bibfnamefont {J.}~\bibnamefont {Healy}},\ }\href
  {\doibase 10.1103/PhysRevD.90.124032} {\bibfield  {journal} {\bibinfo
  {journal} {Phys. Rev. D}\ }\textbf {\bibinfo {volume} {90}},\ \bibinfo
  {pages} {124032} (\bibinfo {year} {2014})},\ \bibinfo {note} {[Erratum:
  Phys.Rev.D 94, 069902 (2016)]},\ \Eprint {http://arxiv.org/abs/1404.3197}
  {arXiv:1404.3197 [gr-qc]} \BibitemShut {NoStop}%
\bibitem [{\citenamefont {Cook}(2020)}]{Cook:2020otn}%
  \BibitemOpen
  \bibfield  {author} {\bibinfo {author} {\bibfnamefont {G.~B.}\ \bibnamefont
  {Cook}},\ }\href {\doibase 10.1103/PhysRevD.102.024027} {\bibfield  {journal}
  {\bibinfo  {journal} {Phys. Rev. D}\ }\textbf {\bibinfo {volume} {102}},\
  \bibinfo {pages} {024027} (\bibinfo {year} {2020})},\ \Eprint
  {http://arxiv.org/abs/2004.08347} {arXiv:2004.08347 [gr-qc]} \BibitemShut
  {NoStop}%
\bibitem [{\citenamefont {{SXS Collaboration}}()}]{SXS}%
  \BibitemOpen
  \bibfield  {author} {\bibinfo {author} {\bibnamefont {{SXS Collaboration}}},\
  }\href@noop {} {\enquote {\bibinfo {title} {{SXS Gravitational Waveform
  Database}},}\ }\bibinfo {howpublished}
  {\url{http://www.black-holes.org/waveforms/}}\BibitemShut {NoStop}%
\bibitem [{\citenamefont {Bhagwat}\ and\ \citenamefont
  {Pacilio}(2021)}]{Bhagwat:2021kfa}%
  \BibitemOpen
  \bibfield  {author} {\bibinfo {author} {\bibfnamefont {S.}~\bibnamefont
  {Bhagwat}}\ and\ \bibinfo {author} {\bibfnamefont {C.}~\bibnamefont
  {Pacilio}},\ }\href {\doibase 10.1103/PhysRevD.104.024030} {\bibfield
  {journal} {\bibinfo  {journal} {Phys. Rev. D}\ }\textbf {\bibinfo {volume}
  {104}},\ \bibinfo {pages} {024030} (\bibinfo {year} {2021})},\ \Eprint
  {http://arxiv.org/abs/2101.07817} {arXiv:2101.07817 [gr-qc]} \BibitemShut
  {NoStop}%
\end{thebibliography}%
\end{document}